\documentclass[12pt, a4paper, twoside, openright]{book}

\usepackage{vuwthesis} 

\usepackage{palatino} 

\usepackage[nottoc]{tocbibind} 

\usepackage[titletoc]{appendix}

\usepackage{amsmath, amsthm, amssymb, physics, xcolor, tensor, mathrsfs, bm, subcaption}

\usepackage[hidelinks, linktoc=all]{hyperref}

\usepackage{graphicx, float, array, tabu}
\graphicspath{ {Figures/} }

\usepackage[normalem]{ulem}

\newtheorem{definition}{Definition}

\theoremstyle{definition}
\newtheorem{theorem}{Theorem}

\theoremstyle{definition}
\newtheorem{postulate}{Postulate}

\theoremstyle{definition}

\newcommand{\blue}[1]{{\color{blue}#1}}

\newcommand{\M}{\mathcal{M}}
\newcommand{\RR}{\mathbb{R}}
\newcommand{\connec}[2]{C\indices{^{#1}_{#2}}}
\newcommand{\chris}[2]{\Gamma\indices{^{#1}_{#2}}}
\newcommand{\R}[2]{R\indices{^{#1}_{#2}}}
\newcommand{\LL}{\mathcal{L}}
\newcommand{\g}{\mathbf{g}}
\newcommand{\T}{\mathscr{T}}
\renewcommand{\O}{\mathcal{O}}
\newcommand{\diag}{\text{diag}}
\newcommand{\HH}{\mathscr{H}}

\newcommand{\eq}[1]{equation{#1}}			
\newcommand{\bb}[1]{\mathbb{#1}}			
\newcommand{\com}[2]{\left[#1, #2\right]}		
\newcommand{\w}[1]{\vb{w}_{#1}}			
\newcommand{\ii}{\vb{i}}					
\newcommand{\jj}{\vb{j}}					
\newcommand{\kk}{\vb{k}}					
\newcommand{\q}{\vb{q}}					
\renewcommand{\v}{\vec{v}}				
\newcommand{\e}{\mathrm{e}}	

\newcommand{\gp}{1-\frac{2m_+ \e^{-a_+/R}}{R}}
\newcommand{\gm}{1-\frac{2m_- \e^{-a_-/R}}{R}}
\newcommand{\gpm}{1-\frac{2m_\pm \e^{-a_\pm/R}}{R}}
\renewcommand{\P}{\mathcal{P}}
\newcommand{\V}{\mathcal{V}}

\renewcommand{\ggg}{1-\frac{2m \, \e^{-a/R}}{R}}

\renewcommand{\d}{\mathrm{d}}
\def\sign{{\mathrm{sign}}}
\renewcommand{\r}{\hat{r}}
\renewcommand{\t}{\hat{t}}
\renewcommand{\th}{\hat{\theta}}
\newcommand{\ph}{\hat{\phi}}
\def\H{{\scriptscriptstyle{\mathrm{H}}}}

\begin{document}

\frontmatter


\title{
Mimicking Black Holes in General Relativity
}
\author{Thomas Berry}

\subject{Mathematics}

\addcontentsline{toc}{chapter}{Abstract \hspace{28em}i}	
\abstract{
The central theme of this thesis is the study and analysis of black hole mimickers. 
The concept of a black hole mimicker is introduced, and various mimicker spacetime models are examined within the framework of classical general relativity. 
The mimickers examined fall into the classes of regular black holes and traversable wormholes under spherical symmetry.
The regular black holes examined can be further categorised as static spacetimes, however the traversable wormhole is allowed to have a dynamic (non-static) throat.
Astrophysical observables are calculated for a recently proposed regular black hole model containing an exponential suppression of the Misner--Sharp quasi-local mass.
This same regular black hole model is then used to construct a wormhole via the ``cut-and-paste'' technique. 
The resulting wormhole is then analysed within the Darmois-Israel thin-shell formalism, and a linearised stability analysis of the (dynamic) wormhole throat is undertaken. 
Yet another regular black hole model spacetime is proposed, extending a previous work which attempted to construct a regular black hole through a quantum ``deformation'' of the Schwarzschild spacetime. 
The resulting spacetime is again analysed within the framework of classical general relativity. 

In addition to the study of black hole mimickers, I start with a brief overview of the theory of special relativity where a new and novel result is presented for the combination of relativistic velocities in general directions using quaternions. 
This is succeed by an introduction to concepts in differential geometry needed for the successive introduction to the theory of general relativity. 
A thorough discussion of the concept of spacetime singularities is then provided, before analysing the specific black hole mimickers discussed above. 
}

\mscthesisonly


\maketitle

\chapter{Acknowledgments}\label{C:ack} 
\chaptermark{}

I would like to thank everyone who has supported me throughout my degree and in completing this thesis.
During this time I have been lucky enough to be supervised by Professor Matt Visser, whose tutelage, support, and insight has been invaluable. 
The amount of time you have dedicated as a mentor to me has not gone unnoticed, and no doubt the many priceless lessons and skills you have taught me will stay with me for the rest of my life.

Along this line, I would also like to thank Alex Simpson.
You've been a great friend, and I look forward to the many future collaborations bound to come. 
I am also thankful for the work you have put into our co-authored papers over the past year, many of which have ended up in this thesis in some capacity. 

I would also like to thank Francisco Lobo.
Particularly for your insight and input into chapter \ref{C:thin-shell-wormhole}.

Special thanks goes to my partner Cassidy and to my family.
Your endless support and guidance throughout my time at university has made this thesis possible.

\vspace{1em}

\noindent This work was directly supported by a Victoria University of Wellington Master's by Thesis Scholarship, and also indirectly supported by the Marsden fund, via a grant administered by the Royal Society of New Zealand.

\tableofcontents

\listoffigures


\mainmatter



\chapter{Introduction}\label{C:Intro}

General relativity is currently our best theoretical model for gravity and has made many predications that have been verified by astronomical observations.
Perhaps one of the best well-known predictions made by the theory is that of black holes.
However, within the framework of classical general relativity, black holes contain a spacetime singularity at their core.
As shown by the Penrose singularity theorems, under suitable physically-reasonable assumptions these singularities are unavoidable consequences of theory once a trapped surface forms.
These singularities present a host of issues from a physical standpoint.
Critically, the theory of general relativity ceases to be predictive at the singularity.
This is clearly an issue, as one of the most important properties of any physical theory is that it can make accurate predictions about the universe we live in. 

In spite of this, in many cases singularities are not an issue for every-day physics as they are hidden by an event horizon, and as such physics outside of the black hole is often unaffected. 
However, there are regimes where the presence of a singularity could be detected (at least, in theory).
Examples of such are the information loss paradox, or during the very last stages of a black hole's evaporation.

Although the issues surrounding singularities have not been resolved, there is convincing astronomical data which suggests that general relativity is extremely accurate in its description of black holes.
Thus, we either have to accept the reality of spacetime singularities at the centre of black holes, thereby accepting that some of out most basic notions of physics no longer hold, or we have to accept that physical black holes are different to their mathematical counterparts.
A proposed solution to this issue is the concept of black hole `mimickers'.
These are objects that are sufficiently similar to black holes so that they agree with astronomical observations but, crucially, do not contain singularities at their cores.

It is commonly believed that singularities will not be present in a consistent theory of quantum gravity.
However, it is unlikely that such a theory will be achieved any time in the near future, and so black hole mimickers provide an effective, classical approach to the resolution of singularities in general relativity.
In this thesis, I investigate a variety of black hole mimickers within the scope of classical general relativity and discuss their validity as real, physical alternatives to black holes as predicted by classical general relativity.

In chapter \ref{C:SR}, the reader is reminded about some of the main results of the theory of special relativity, providing a starting point for discussing the more general theory in later chapters. 
Familiarity with the theory of special relativity is assumed herein.

Chapter \ref{C:comb-veloc} then studies the special-relativistic combination of velocities using the quaternion number system.
A new and novel result for combining relativistic velocities is proposed and thoroughly investigated within the framework of the theory of special relativity. 

In chapter \ref{C:GR-maths-background}, we move on to discuss the theory of general relativity. 
The mathematical framework needed to understand the results in later chapters is provided and discussed alongside the key postulates which lead Einstein to the development of the theory\footnote{The term ``postulate'' is often used to simply mean a collection of very good experimental evidence. Einstein's postulates have a very solid experimental foundation.}.
This chapter is intended as a review of classical general relativity, and as such no new results will be presented. 

Chapter \ref{C:BH-mimickers} discusses, in detail, singularities in general relativity.
This includes a rigorous definition of a singularity, as well as a general discussion of the Penrose singularity theorems. 
This is then used to introduce the idea of black hole mimickers, including regular black hole and traversable wormhole spacetimes.

In chapter \ref{C:Mink-core}, a specific regular black hole model is analysed within the context of classical general relativity. 
Specifically, the location of timelike and null circular geodesics are investigated in detail, the spin-dependent Regge-Wheeler potential is calculated, and a first-order WKB approximation of the quasi normal modes is completed.
The novel regular black hole under investigation results in a far richer phenomenology than standard (non-regular) black holes.

In chapter \ref{C:thin-shell-wormhole}, the same regular black hole spacetime is used to construct a thin-shell traversable wormhole via the ``cut-and-paste'' technique, there\-by constructing yet another black hole mimicker.
A linearised stability analysis is conducted for the wormhole throat and a series of specific examples are investigated wherein the spacetime parameters are changed (and allowed to be different) between the two manifolds used in the cut-and-paste thin-shell construction.
Again, the novelty of the spacetime results in a rich phenomenology of potential interest to observational astronomers and astrophysicists. 

Chapter \ref{C:QMS} introduces a family of regular black hole spacetimes, which is analysed within the framework of classical general relativity. 
The family of regular black hole spacetimes were inspired by a (non-regular) black hole spacetime which arises as a quantum modification to the Schwarzschild black hole.

Finally, in chapter \ref{C:conclusions} we provide a brief summary of the main results in this thesis and provide an outlook on the future of the field and avenues of potential future research.

\chapter{Special relativity}\label{C:SR}

Before we introduce Einsteins theory of general relativity, we will provide a brief overview of his theory of Special relativity, and provide some new results on the combination of relativistic velocities. 
Note that this is not intended to be a complete overview of the theory, and many results will be assumed to be prior knowledge to the reader.

Special relativity is where one typically first encounters the notion of spacetime: one time dimension \( t \), and three space dimensions \( (x,y,z) \) combined into one four-dimensional space representing the collective set of points \( \{(t,x,y,z)\} \).
In Newtonian mechanics, there is no limit on how fast an object may move through space, and notions such as `the length of an object', or `how fast a clock ticks' are the same no matter who makes the measurements. 
In special relativity, however, the situation is much different -- the `length of an object' or `how fast a clock ticks' is dependent on the relative speed of the observer making the measurements. 

The theory of special relativity is built from two ingredients: 
\begin{enumerate}
    \item[(1)] Minkowski space: the mathematical `space' representing spacetime in which all observers move along their `worldlines'.
    \item[(2)] Einstein's postulates: the `laws' of physics, or mathematical `axioms' (i.e. summary of experimental evidence) which we use to derive equations of motion governing how objects move throughout Mink\-owski space. 
\end{enumerate}
With these two ingredients, we can completely describe the kinematics of an object, no matter how fast it is travelling, so long as we do not consider the effects of gravity.
Bringing gravity into the picture induces many additional complications which will be captured by the full theory of general relativity discussed in chapter \ref{C:GR-maths-background}.
That is not to say that special relativity is not a good, or useful theory.
In many cases, one can ignore the effects of gravity and work completely inside the framework of special relativity.
In fact, this is built into general relativity at a fundamental level: so long as you are working in sufficiently small areas of spacetime, you need not worry about any gravitational effects.
Thus, for now, let us assume that we are working only within the framework of special relativity and worry about gravitational effects in later chapters.

\section{Minkowski space and Einstein's postulates}

Special relativity is a theory of how objects move throughout spacetime.
Thus, first of all, we need to define a notion of what we mean by `spacetime'.

\begin{definition}\label{def:Minkowski-space}
Minkowski spacetime is the pair \( (\RR^4, \eta) \), where \( \eta \) is the quadratic form given in matrix representation by
\begin{equation}
\eta = 
    \begin{pmatrix}
    -1 & 0 & 0 & 0 \\
    0 & 1 & 0 & 0 \\
    0 & 0 & 1 & 0 \\
    0 & 0 & 0 & 1
    \end{pmatrix}.
\label{eq:eta}
\end{equation}
\end{definition}

As one would expect, Minkowski spacetime is four-dimensional, and so we can use it to construct a theory of spacetime.
At this point, we need introduce our laws of physics from which we can deduce equations of motion. 
These laws are given by Einstein's postulates:

\begin{postulate}[Principle of relativity\footnote{It is worth noting that postulate \ref{P:SR1} holds for mechanical processes even in Galilean relativity. It is postulate \ref{P:SR2} which is unique to special relativity.}]\label{P:SR1}
The laws of physics are the same in all inertial frames of reference.
\end{postulate}

\begin{postulate}[Invariance of \( c \)]\label{P:SR2}
The speed of light has the same value in all inertial frames of reference. 
\end{postulate}

Like all laws of physics, these postulates have been rigorously experimentally tested, and so far have proven to be an accurate reflection of how the universe works. 
Details of the experimental verification of special relativity can be found in Ref. \cite{Mattingly:2005}.

We first define a four-vector \( X \) which has components $X^\mu = (x^0, \vec{x}) = (ct, x, y, z)$ (note the components have dimension length).
The four-vector \( X \) represents an event in spacetime  (i.e. a point in Minkowski space).
Suppose now that we have two events \( X_1 \) and \( X_2 \) such that a light ray passes from event 1 to event 2.
From postulate \ref{P:SR2}, the difference between these two events is \( \Delta X = X_2 - X_1 = (c\Delta t, \Delta\vec{x})^T \).
Since a light ray connects the two events, we have that \( \abs{\Delta\vec{x}}/\abs{\Delta t} = c \).
That is, \( \abs{\Delta\vec{x}}^2 - (x^0)^2 = 0 \).
We can write this as matrix multiplication in terms of the quadratic form \( \eta \) as \( \eta(\Delta X, \Delta X) \equiv (\Delta X)^T \eta \Delta X = 0 \). 

\begin{figure}[t]
\centering
    \includegraphics[scale=0.45]{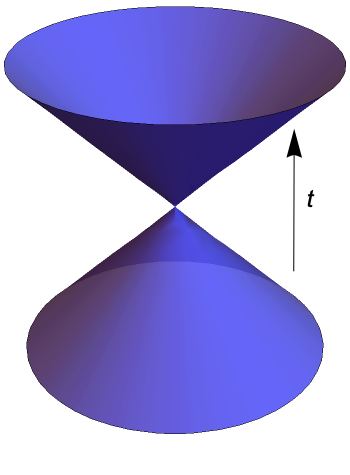}
    \caption[Light cone.]{The light cone of an event in spacetime. Time runs vertically up the page.}
\label{fig:light-cone}
\end{figure}

This is our first notion of the idea of \emph{causality}: as nothing can travel faster than a light signal, event 1 can only \emph{cause} event 2 if at least a light signal can join the two points in spacetime (i.e. if \( \eta(\Delta X, \Delta X) = 0 \)).
Note also that if \( \eta(\Delta X, \Delta X) = 0 \), then we also have \( \eta(\Delta (-X), \Delta (-X)) = 0 \), and so (at least in this purely mathematical framework) if event 1 can cause event 2, so too can event 2 cause event 1.
In most \emph{physical} situations, however, we will require that time flows in the positive direction.
Then we will have a well-defined time-ordering of events 1 and 2, thereby removing any ambiguity about which event `caused' the other.
The set of points 
\begin{equation}
C(X_1) = \{ X_2 \mid \eta(\Delta X, \Delta X) = 0 \},
\label{eq:light-cone}
\end{equation}
forms the surface of a double-cone with apex at \( X_1 \), called the ``light-cone'' of \( X_1 \) (see figure \ref{fig:light-cone}).
Any event \( X_2 \) that can be reached from \( X_1 \) by a signal travelling slower than the speed of light will lie inside the surface of the light-cone.
That is \( X_2 \) is in the future of \( X_1 \).
(Note, if \( X_2 \) is in the past of \( X_1 \), it will lie in its past light-cone).
Conversely, any event that cannot be reached by \( X_1 \) without travelling faster than the speed of light will lie outside of the light-cone of event 1.
That is, event 2 is \emph{not} in the future of event 1.

We can place the separation of two events into three distinct classes:
\begin{enumerate}
    \item[(1)] Timelike separated: event 1 can reach event 2 by slower-than-light travel; \( \eta(\Delta X, \Delta X) < 0 \).
    \item[(2)] Lightlike (null) separated: event 1 can only reach event 2 by speed of light travel; \( \eta(\Delta X, \Delta X) = 0 \).
    \item[(3)] Spacelike separated: event 1 cannot reach event 2 without travelling faster than the speed of light; \( \eta(\Delta X, \Delta X) > 0 \).
\end{enumerate}

\section{Lorentz transformations}

\begin{figure}[t]
\centering
    \includegraphics[scale=0.5]{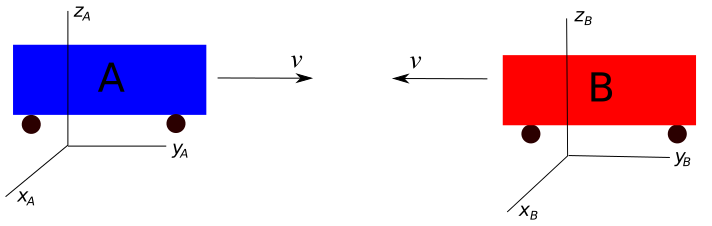}
    \caption[Co-moving observers.]{Two observers travelling towards each other in the non-relativistic case. Each observer sees the other travelling towards them at speed \( 2v \).}
\label{fig:G-trans}
\end{figure}

Postulate \ref{P:SR2} leads to some interesting results that disagree with Newtonian kinematics.
Consider, for example, what happens if two observers \( A \) and \( B \) are travelling toward each other, each with a speed \( v \) as measured by some third observer (see figure \ref{fig:G-trans}). 
According to \( A \), they will see \( B \) moving toward them at a speed \( v+v=2v \); and vice versa for \( B \).
This is a perfectly acceptable scenario within the framework of Newtonian kinematics.
Now suppose that the velocities \( v \) approach a reasonable fraction of the speed of light, say \( v = 0.75c \).
We know from postulate \ref{P:SR2} that nothing can travel faster than \( v = c \), and so A cannot possibly see B travelling toward them at a speed \( 0.75c+0.75c=1.5c \).
This suggests that the simple non-relativistic Galilean transformations of Newtonian kinematics no longer work at relativistic speeds, and that we need a new set of transformations which incorporate postulate \ref{P:SR2} in a fundamental way.
These are known as the Lorentz transformations.

One of the biggest differences between Lorentz transformations and the Galilean transformations of Newtonian kinematics is that the Lorentz transformations also transform the time coordinate between the two reference frames.
For the case of co-linear velocities (along the \( x \)-axis), they take the particularly simple form\footnote{Derivations can be found in any of the standard introductory textbooks on special relativity.}
\begin{align}
ct' &= \gamma(ct-\beta x),	\notag \\
x' &= \gamma(x-\beta ct), \notag \\
y' &= y, \notag \\
z' &= z;
\label{eq:LTs-linear}
\end{align}
where \( \beta=v/c \), \( \gamma=1/\sqrt{1-\beta^2} \), and the primed and un-primed coordinates represent the coordinates in the different frames.
Note that so long as we are only working with two coordinate frames, we can always define the coordinate axes in such a way as to ensure that they are co-linear, and so equations \eqref{eq:LTs-linear} hold.
This does become an issue, however, if one is dealing with more than two coordinate systems.
In such a case, a more complicated form of the Lorentz transformations are needed:
\begin{align}
ct' &= \gamma\left(ct - \vec{\beta}\cdot\vec{r_\parallel}\right), \notag \\
\vec{r}'_\parallel &= \gamma(\vec{r}_\parallel - \vec{\beta}ct), \notag \\
\vec{r}'_\perp &= \vec{r}_\perp,
\label{eq:LTs-full}
\end{align}
where \( \vec{r} = \vec{r}_\parallel + \vec{r}_\perp \) is the position vector connecting the two frames, decomposed into its components parallel and perpendicular to the relative velocity \( \vec{\beta} = \vec{v}/c \).
The Lorentz transformations \eqref{eq:LTs-full} are known as Lorentz boosts, or pure Lorentz transformations, and form a subset of the group of all Lorentz transformations, the Lorentz group. 

Mathematically, the Lorentz group is isomorphic to \( \text{O}(1,3) \), the orthogonal group of one time and three space dimensions that preserves the space-time interval
\begin{equation}
\eta(\Delta X, \Delta X) = -t^2+x^2+y^2+z^2.
\label{eq;stinterval}
\end{equation}
Note that, here and hereafter, we adopt units where the speed of light is set to unity.
It is clear from this description that rotations of space-time are included in the Lorentz group, as well as the more familiar Lorentz boosts \eqref{eq:LTs-full}.
In fact, the pure Lorentz transformations do not even form a subgroup of the Lorentz group as, in general, the composition of two boosts \( B_1 \) and \( B_2 \) is not another boost but in fact a boost and a rotation \( B_{12}R_{12} = B_1 B_2\); whilst \( B_{21} R_{21} = B_2 B_1\).
This rotation, known as the Wigner rotation, was first discovered by Llewellyn Thomas in 1926 whilst trying to describe the Zeeman effect from a relativistic view-point~\cite{thomas:1926}, and was more fully analysed by Eugene Wigner in 1939 \cite{wigner:1939}.
(For more recent discussions see~\cite{ferraro:1999, fisher:1972, malykin:2006, visser_odonnell:2011, ritus:2007}.)

It is well--known that the composition of Lorentz transformations is non-commutative.
That is, applying two successive boosts \( B_1 \) and \( B_2 \) in different orders results in the same final boost, \( B_{12} = B_{21} \), but different rotations, \( R_{12} \neq R_{21} \).
In the context of the combination of two velocities \( \v_1 \) and \( \v_2 \), this means that the final speed is the same no matter the order we combine the velocities, \( \norm{ \v_1 \oplus \v_2 } = \norm{ \v_2 \oplus \v_1 } \), but the final directions they point in are different $\hat v_{1\oplus2} \neq \hat v_{2\oplus 1}$.
Although not immediately obvious, the angle between \( \v_1 \oplus \v_2 \) and \( \v_2 \oplus \v_1 \) is in fact the Wigner angle \( \Omega \) \cite{visser_odonnell:2011}.
%

In the following chapter, we will provide a way of using Lorentz transformations to derive results for the combination of relativistic velocities using a very special representation of the Lorentz group: the quaternions.
\chapter{Combination of velocities using quaternions}\label{C:comb-veloc}
\chaptermark{Combination of velocities\dots}

Hamilton first described the quaternions in the mid-1800s, primarily 
with a view to finding algebraically simple ways to handle 3-dimensional rotations.
With the advent of special relativity in 1905, and noting the manifestly 4-dimensional nature of quaternions once one adds a real part, multiple authors have tried to interpret special relativity in an intrinsically  quaternionic fashion \cite{dirac:1944, girard:1984, mocanu:1992, rastall:1964, silberstein:1912, silberstein:1914, ungar:1989}.

Despite technical success in applying quaternions to  special relativity, the use of quaternions in this subject has never really gained all that much traction in the physics community.
Perhaps one of the reasons for this is that there are a number of sub-optimal notational choices in Silberstein's original work~\cite{silberstein:1912, silberstein:1914}, and the fact that there is no generally accepted way of using quaternions to represent Lorentz transformations, with many different authors employing their own quite distinct methods \cite{DeLeo:1996, dirac:1944, girard:1984, mocanu:1992, rastall:1964, silberstein:1912, silberstein:1914, ungar:1989}.
Even in more recent, post-millennial, articles on ``quaternionic special relativity'' there is considerable disagreement on notational choices \cite{Friedman:2005a, Friedman:2005b, Greiter:2003, Yefremov:2016}.

In this chapter, we shall introduce what we feel is a particularly simple and straightforward method for combining relativistic 3-velocities using quaternions.
All of the interesting features due to non-commutativity properties of non-collinear boosts are implicitly and rather efficiently dealt with by the non-commutative algebra of quaternions.
The method is based on an extension of an analysis by Giust, Vigoureux, and Lages \cite{vigoureuxetal:2009,vigoureuxetal:2008}, who (because they were working with the usual complex numbers) were essentially limited to motion in 2-space; their formalism is not really well-adapted to general motions in 3-space.
Related constructions can also be found in references \cite{Friedman:2005a, Friedman:2005b}.

One could instead try to deal with the non-commutativity of the Lore\-ntz transformations by adapting the general formalism of the Baker--Camp\-bell--Hausdorff (BCH) theorem~\cite{BCH1,BCH2,Van-Brunt:2015a,Van-Brunt:2015b,Van-Brunt:2015c}. Unfortunately, the general BCH formalism applied to this problem very quickly becomes intractable, and we have found that the specifics of the quaternion formalism yield much more useful and tractable results.
Similarly, since the full symmetry group of the Maxwell equations is the conformal extension of the Poincare group, it is sometimes useful, (when looking at pure electromagnetic effects), to work with this conformal extension. However physical observers, (physical clocks and physical rulers), break the conformal invariance, and to even meaningfully define 3-velocities one needs to restrict attention to the Poincare group. We shall go even further and take translation invariance (spatial and temporal homogeneity) for granted, and focus more specifically on the Lorentz group.

\section{Quaternions}		\label{sec;quaternions}

The quaternions are numbers that can be written in the form \( a+b\,\ii+c\,\jj+d\,\kk \), where \( a, \) \(b,\) \( c, \) and \( d \) are real numbers; and \( \ii, \) \( \jj, \) and \( \kk \) are the quaternion units which satisfy the famous relation 
\begin{equation}
    \ii^2 = \jj^2 = \kk^2 = \ii\jj\kk = -1.
\end{equation}
They form a four--dimensional number system that is generally treated as an extension of the complex numbers.
We shall define the quaternion conjugate of the quaternion \( \vb{q} = a + b\,\ii + c\,\jj + d\,\kk \) to be \( \vb{q}^\star = a - b\,\ii - c\,\jj - d\,\kk \), and define the norm of \( \vb{q} \) to be \( \vb{q q}^\star = \abs{\vb{q}}^2 = a^2+b^2+c^2+d^2 \in \bb{R} \).
This allows us to evaluate the quaternion inverse as $\q^{-1} = \q^\star/|\q|^2$. 

Trying to define a ``norm'' as \( \vb{q}^2 = a^2 - b^2 - c^2 - d^2 \), while superficially more ``relativistic'', violates the usual mathematical definition of ``norm'', and furthermore is not useful when it comes to evaluating the quaternion inverse $\vb{q}^{-1}$.

For current purposes we  focus our attention on pure quaternions.
That is, quaternions of the form \( a\,\ii + b\,\jj + c\,\kk \).
Many quaternion operations become much simpler when we are dealing with pure quaternions.
For example, the product of two pure quaternions \( \vb{p} \) and \( \vb{q} \) is given by \( \vb{p}\vb{q} = -\vec{p}\cdot\vec{q} + (\vec{p}\cross\vec{q})\cdot(\ii, \jj, \kk) \), where, in general, we shall set \( \vb{v} = \vec{v} \cdot (\ii,\jj,\kk) \).
From this, we obtain the useful relations 
\begin{equation}
    \com{\vb{p}}{\vb{q}} = 2 ( \vec{p}\cross\vec{q}) \cdot (\ii, \jj, \kk), \qq{and} \{ \vb{p},\vb{q} \} = -2\, \vec{p}\cdot\vec{q}.
\label{eq;pureqrelats}
\end{equation}
A notable consequence of \eqref{eq;pureqrelats} is \( \vb{q}^2 = -\vec{q}\cdot\vec{q} = -q^2 =  -\abs{\vb{q}}^2 \).
There is a natural isomorphism between the space of pure quaternions and \( \bb{R}^3 \) given by 
\begin{equation}
    \ii \mapsto \hat{x}, \quad \jj \mapsto \hat{y}, \quad \kk \mapsto \hat{z};
\label{eq;natiso}
\end{equation}
where \( \hat{x}, \, \hat{y}, \) and \( \hat{z} \) are the standard unit vectors in \( \bb{R}^3 \).

One of the most common uses for quaternions today (2021) is in the computer graphics community, where they are used to compactly and efficiently generate rotations in 3-space.
Indeed, if \( \vb{q} = \cos(\theta/2) + \vu{u}\sin(\theta/2) \) is an arbitrary unit quaternion and \( \vb{v} \) is the image of a vector in \( \bb{R}^3 \) under the isomorphism \eqref{eq;natiso}, then the mapping \( \vb{v} \mapsto \vb{q}\vb{v}\vb{q}^{-1} \) rotates \( \vb{v} \) through an angle \( \theta \) about the axis defined by \( \vu{u} \).
The mapping \( \vb{v} \mapsto \vb{q}\vb{v}\vb{q}^{-1} \) is called \textit{quaternion conjugation} by \( \vb{q} \).
Furthermore, an extension of the quaternions, the dual quaternions, are used in the field of theoretical robot kinematics, due to their ability to efficiently handle rotations and translations of vectors \cite{Berry:robots}.

\section{Combining two 3-velocities}
In the paper by Giust, Vigoureux, and Lages \cite{vigoureuxetal:2009}, see also \cite{vigoureuxetal:2008}, (and the somewhat related discussion in reference \cite{Friedman:2005a}), a method is developed to compactly combine relativistic velocities in two space dimensions, and by extension, coplanar relativistic velocities in 3 space dimensions.
In the following subsection, we first provide a short summary of their approach, and then in the next subsection extend their method to general non-coplanar 3-velocities.

\subsection{Velocities in the ($x$,$y$)-plane} \label{sec:3-velocities}

The success of this Giust, Vigoureux, and Lages approach relies on the angle addition formula for the hyperbolic tangent function,
\begin{equation}
	\tanh(\xi_1 + \xi_2) = \frac{ \tanh{\xi_1} + \tanh{\xi_2} }{ 1 + \tanh{\xi_1}\tanh{\xi_2} }.
\label{eq;tanhadd}
\end{equation}
The tanh function is a natural choice for combining relativistic velocities since it is limited to the interval \( \left[-1 ,1\right] \).
Indeed, using the rapidity \( \xi \) defined by \( v=  \tanh(\xi)  \), we can easily combine collinear relativistic speeds using \eq{} \eqref{eq;tanhadd}.
In order to use this for the combination of non-collinear relativistic 2-velocities, we replace each 2-velocity \( \v \) by the complex number 
\begin{equation}
\label{E:w}
	V = \tanh({\xi}/{2}) \; \e^{i\varphi}.
\end{equation}
Here \( \xi \) is the rapidity of the velocity \( \v \), and \( \varphi \) gives the orientation of \( \v \) according to some observer in the plane defined by \( \v_1 \) and \( \v_2 \).
Giust, Vigoureux, and Lages then define the composition law \( \oplus \) for coplanar velocities \( \v_1 \) and \( \v_2 \) by 
\begin{equation}
        	W = \tanh\frac{\xi}{2} \;\e^{i\varphi_{1\oplus2}} = V_1 \oplus V_2 = \frac{V_1 + V_2}{1 + \overline{V_2} \, V_1} = \frac{ \tanh\frac{\xi_1}{2}\,\e^{i\varphi_1} + \tanh\frac{\xi_2}{2}\, \e^{i\varphi_2} }{ 1 + \tanh\frac{\xi_2}{2}\,\e^{-i\varphi_2} \tanh\frac{\xi_1}{2}\,\e^{i\varphi_1} },
\label{eq;oplus2d}
\end{equation}
where \( \overline{V} \) is the standard complex conjugate of \( V \).
By using \( \xi/2 \) instead of \( \xi \) in equations \eqref{E:w} and  \eqref{eq;oplus2d}, we are actually dealing with the ``relativistic half--velocities'', \( \tanh(\xi/2) \),  (sometimes called the ``symmetric velocities''),
where
\begin{equation}
w = \tanh(\xi/2) ; \qquad v = \tanh(\xi) = {2w\over 1+w^2}.
\end{equation}
That is:
\begin{equation}
w = \tanh\left(\frac{1}{2}\tanh^{-1}(v)\right) = \frac{v}{1+\sqrt{1-v^2}}.
\end{equation}
Using \eq{s} \eqref{eq;tanhadd} and \eqref{eq;oplus2d} we can easily retrieve the real velocity from the half-velocity by using \( \oplus \) operator: \( v= \tanh\xi = \tanh{\xi/2} \oplus \tanh{\xi/2} = w\oplus w\).

As an aside, it is worth noting that these half-velocities are often first encountered when working with Loedel diagrams \cite{Loedel-wiki}.
Standard spacetime diagrams (often called Minkowski diagrams) have orthogonal spacetime axes in a given rest frame.
As such, the axes of other reference frames (moving with a relative velocity to the rest frame) form an acute angle. 
This asymmetry between reference frames in a Minkowski diagram is often misleading, as postulate \ref{P:SR1} enforces the equivalence of any two frames of reference.
Loedel diagrams are constructed in a third reference frame travelling at the relative half-velocity of the two initial frames, and so the symmetry between the two frames of reference is manifest. 

In terms of the half velocities, we can write the combined velocity as
\begin{equation}
 w_{1\oplus2} \;\e^{i\varphi_{1\oplus2}} 
 = \frac{ w_1\,\e^{i\varphi_1} + w_2\, \e^{i\varphi_2} }{ 1 + w_1 w_2\,\,\e^{i(\varphi_1-\varphi_2)} }.
\end{equation}

The \( \oplus \) addition law is non-commutative, which is most easily seen by first setting \( \theta = \varphi_2 - \varphi_1 \), then $\Omega = \varphi_{1\oplus2} - \varphi_{2\oplus1}$, and finally observing that the ratio 
\begin{equation}
    \e^{i\Omega/2} = \frac{ 1 + \tanh\frac{\xi_1}{2}\tanh\frac{\xi_2}{2}\e^{i\theta} }{ 1 + \tanh\frac{\xi_1}{2}\tanh\frac{\xi_2}{2}\e^{-i\theta} }
    = {1+w_1w_2 \e^{i\theta}\over 1+ w_1 w_2 \e^{-i\theta}}
\label{eq;noncomplanar}
\end{equation}
is not equal to unity for non--zero \( \theta \), meaning that \( \Omega=\varphi_{1\oplus2} - \varphi_{2\oplus1} \) is non-zero.

The angle \( \Omega=\varphi_{1\oplus2} - \varphi_{2\oplus1} \) is in fact the Wigner angle \( \Omega \), so an expression for this angle can be obtained by taking the real and imaginary parts of \eq{} \eqref{eq;noncomplanar}: 
\begin{equation}
    \tan\frac{\Omega}{2} = \frac{ \tanh\frac{ \xi_1 }{2} \,\tanh\frac{\xi_2}{2}\, \sin\theta}{1+\tanh\frac{\xi_1}{2}\, \tanh\frac{\xi_2}{2} \,\cos\theta} 
    = {w_1 w_2 \sin\theta\over1+w_1w_2\cos\theta}.
\label{eq;wigvig}
\end{equation}
This expression does not explicitly appear in 
reference~\cite{vigoureuxetal:2009} though something functionally equivalent, in the form $\Omega=2\; \hbox{arg}(1+w_1w_2\e^{i\theta})$, appears in reference~\cite{vigoureuxetal:2008}.

The \( \oplus \) law can be applied to any number of coplanar velocities by iteration:
\begin{equation}
    W = (((V_1 \oplus V_{2} )\oplus \dots \oplus V_{n-1} )\oplus V_n).
\end{equation}
Thus, it would be desirable to cleanly extend this formalism to general three-dimensional velocities.
Note that the order of composition is important, as we shall see in more detail below, the $\oplus$ operation is in general \emph{not} associative.

\subsection{General 3-velocities}

We now extend the result of Giust, Vigoureux, and Lages to arbitrary 3-velocities in three dimensions.

\subsubsection{Algorithm}

Suppose we have a velocity \( \v_i \) in the \((x,y)\)-plane, represented by the pure quaternion \( \vb{w}_i = \tanh(\xi_i/2) \vu{n}_i = \tanh(\xi_i/2)\;(\ii \cos\theta_i + \jj \sin\theta_i ) \). 
Using the rules for quaternion multiplication, we can write this as 
\begin{equation}
\vb{w}_i = \tanh(\xi_i/2)\;(\cos\theta_i + \kk \sin\theta_i) \ii.
\end{equation}
The term inside the brackets now looks very similar to what would be a natural extension of the exponential function to the quaternions, \( \e^{\kk\theta} = \cos\theta + \kk\sin\theta \).
To formalise this, we define the exponential of a quaternion \( \vb{q} \) by the power series 
\begin{equation}
    \e^{\vb{q}} = \sum_{k=0}^\infty \frac{\vb{q}^k}{k!}.
\label{eq;expqdefine}
\end{equation}
To calculate an explicit formula for \eq{} \eqref{eq;expqdefine}, we first consider the case of a pure quaternion \( \vb{u} \).
We know from section \ref{sec;quaternions} that for a pure quaternion we have \( \vb{u}^2 = -\abs{\vb{u}}^2 \), and so we find \( \vb{u}^3 = -\abs{\vb{u}}^2\vb{u}, \, \vb{u}^4 = \abs{\vb{u}}^4, \) and so on.
Thus, we can compute directly from the definition \eqref{eq;expqdefine}:
\begin{align}
    \e^{\vb{u}} 
    &= \left(1 - {1\over2!}\abs{\vb{u}}^2 + {1\over4!}\abs{\vb{u}}^4 - \dots\right) + {\vb{u}\over\abs{\vb{u}}}\left( \abs{\vb{u}} - {1\over3!}\abs{\vb{u}}^3 + {1\over5!}\abs{\vb{u}}^5 - \dots\right)	\notag \\
    &= \cos\abs{\vb{u}} + \vu{u}\sin\abs{\vb{u}}.	\label{eq;expquat}
\end{align}
Following the same procedure above, we find the exponential of a pure unit quaternion \( \vu{u} \) and real number \( \phi \) to be
\begin{equation}
	\e^{\vu{u}\phi} = \cos\phi + \vu{u}\sin\phi.
\label{eq;expphiu}
\end{equation}
This nice result reflects the expression for the exponential of a complex number.

We can now extend this result to any arbitrary quaternion \( \vb{q} = a + \vb{u} \) by noting that the real number \( a \) commutes with all the terms in \( \vb{u} \), thereby allowing us to write \( \e^{\vb{q}} = \e^a \e^{\vb{u}} \), where \( \e^{\vb{u}} \) has the same form as \eq{} \eqref{eq;expquat}.
Explicitly,
\begin{equation}
    \e^{\vb{q}} = \e^a (\cos\abs{\vb{u}} + \vu{u} \; \sin\abs{\vb{u}} ).
\label{eq;expqgen}
\end{equation}

The exponential of a quaternion possesses many of the same properties as the exponential of a complex number.
Two particularly useful ones we use below are
\begin{equation}
    \left( \e^{\vu{u}\phi} \right)^\star = \e^{-\vu{u}\phi} = \cos\phi-\vu{u}\sin\phi, \qq{and} \abs{\e^{\vu{u}\phi}}=1.
\end{equation}

Using these results, we are now justified in writing
\begin{equation}
 \vb{w}_i = \tanh(\xi_i/2)\, \e^{\kk\theta_i} \, \ii = w_i \, \e^{\kk\theta_i} \, \ii 
\end{equation}
for our velocity in the \((x,y)\)-plane.

Building on this result, we now find it appropriate to define the \( \oplus \) operator for general 3-velocities, \( \vb{w}_1=w_1 \vb{\hat n_1} \) and \( \vb{w}_2= w_2 \vb{\hat n_2} \), by:
\begin{equation}
	\vb{w}_{1\oplus2} = \vb{w}_1 \oplus \vb{w}_2 = (1 - \vb{w}_1\vb{w}_2)^{-1}(\vb{w}_1 + \vb{w}_2).
\label{eq;oplusgen}
\end{equation}
The usefulness of this definition is best understood by looking at a few examples.

\subsubsection{Example: Parallel velocities}
We consider two parallel velocities \( \v_1 \) and \( \v_2 \) represented by the quaternions
\begin{equation}
    \w{1} = \tanh\frac{\xi_1}{2} \, \vb{\hat n}  \qq{and} \w{2} = \tanh\frac{\xi_2}{2} \, \vb{\hat n} ,
\end{equation}
respectively.
Our composition law \eqref{eq;oplusgen} then gives 
\begin{align}
	\w{1\oplus2} &= \left( 1 + \tanh\frac{\xi_1}{2}\,\tanh\frac{\xi_2}{2} \right)^{-1} \left( \tanh\frac{\xi_1}{2}\,  \vb{\hat n} \,+ \,\tanh\frac{\xi_2}{2}\,  \vb{\hat n}  \right) \notag \\
	&= \frac{ \tanh\frac{\xi_1}{2} + \tanh\frac{\xi_2}{2} }{1 + \tanh\frac{\xi_1}{2} \,\tanh\frac{\xi_2}{2} } \; \vb{\hat n}  	\notag\\
	&= \tanh\left(\frac{\xi_1+\xi_2}{2}\right)  \; \vb{\hat n} , 	
	\label{eq;paraveloc}
\end{align}
which is equivalent to 
\begin{equation}
\w{1\oplus2} = {w_1+w_2\over1+w_1w_2} \; \vb{\hat n},
\end{equation}
and hence, also equivalent to 
the well--known result for the relativistic composition of two parallel velocities, \begin{equation}
    \v_1 \oplus \v_2 = \frac{v_1 + v_2}{1 + v_1 v_2} \; \hat{n}.
\label{eq;paralleladdv}
\end{equation}

\subsubsection{Example: Perpendicular velocities in the $x$--$y$ plane}
We now consider two perpendicular velocities in the $x$--$y$ plane. 
By rotating around the $z$ axis, without loss of generality they can be taken to be given by
\begin{equation}
    \w{1} = w_1 \ii , \quad \w{2} = w_2 \, \jj,
\end{equation}
where we have written \( \tanh(\xi_1/2) = w_1 \) and \( \tanh(\xi_2/2) = w_2 \) for brevity.

Our composition law then gives a combined velocity of 
\begin{equation}
    \vb{w}_{1\oplus2} = (1 - w_1w_2 \ii\jj)^{-1} (w_1\ii + w_2\jj) = \frac{w_1(1-w_2^2) \ii + w_2(1+w_1^2) \jj}{1+w_1^2w_2^2},
\end{equation}
which is definitely not commutative. In contrast the norm is symmetric:
\begin{equation}
    \abs{\vb{w}_{1\oplus2}}^2 = \frac{w_1^2(1-w_2^2)^2 + w_2^2(1+w_1^2)^2}{(1+w_1^2w_2^2)^2} 
    = \frac{w_1^2+w_2^2}{1+w_1^2w_2^2}.
\label{eq;w12perp}
\end{equation}
Here the \( \vb{w}_i \) are the ``relativistic half--velocities'' \( w_i = \tanh(\xi_i/2) \), so the full velocities are
\begin{equation}
    \abs{\vb{v}_i}^2 =  \abs{\vb{w}_i \oplus \vb{w}_i}^2 = \frac{4w_i^2}{(1+w_i^2)^2},
\label{eq;visquared}
\end{equation}
and so give a final speed of
\begin{equation}
    \abs{\vb{v}_{1\oplus2}}^2 = \frac{4\left(w_1^2+w_2^2\right)}{\big(1+w_1^2w_2^2\big) \left[1+\frac{w_1^2+w_2^2}{1+w_1^2w_2^2}\right]^2} = \frac{4\big(w_1^2+w_2^2\big) \big(1+w_1^2w_2^2\big)}{\left[ \big(1+w_1^2\big) \big(1+w_2^2\big) \right]^2}.
\end{equation}
The non-quaternionic result for the composition of two perpendicular velocities is \cite{visser_odonnell:2011}
\begin{equation}
    \norm{\v_{1\oplus2}}^2 = v_1^2 + v_2^2 - v_1^2v_2^2 = 1 - \big(1-v_1^2\big)\big(1-v_2^2\big).
\end{equation}
Thus, we find
\begin{align}
    \norm{\v_{1\oplus2}}^2 &= \frac{4w_1^2}{(1+w_1^2)^2} + \frac{4w_2^2}{(1+w_2^2)^2} - \frac{16w_1^2w_2^2}{(1+w_1^2)^2(1+w_2^2)^2} 	\notag \\
    &= \frac{4(w_1^2+w_2^2)(1+w_1^2w_2^2)}{\left[ (1+w_1^2)(1+w_2^2) \right]^2}.
\end{align}
And so our composition law \( \oplus \) gives the standard result for the composition of two perpendicular velocities in the $x$--$y$ plane.

\subsubsection{Example: Perpendicular velocities in general}

For general perpendicular velocities $\v_{1}$ and $\v_{2}$ the easiest way of proceeding is to simply rotate to point $\v_{1}$ along the $x$-axis and $\v_{2}$ along the $y$-axis, and just copy the argument above. 
If one wishes to be more direct then simply define
\begin{equation}
    \w{1} = w_1 \; \widehat{\w{1}} , \qquad \w{2} = w_2 \; \widehat{\w{2}}; 
    \qquad \widehat{\w{3}} = \widehat{\w{1}} \; \widehat{\w{2}}.
\end{equation}
In view of the mutual orthogonality of the vectors $\hat w_1$, $\hat w_2$, and $\hat w_3$, the unit quaternions $(\widehat{\w{1}},\widehat{\w{2}},\widehat{\w{3}})$ obey exactly the same commutation relations as $(\ii, \jj, \kk)$. Thence
\begin{equation}
    \vb{w}_{1\oplus2} = (1 - w_1w_2 \widehat{\w{1}} \widehat{\w{2}})^{-1} (w_1 \widehat{\w{1}} + w_2 \widehat{\w{2}}) = \frac{w_1(1-w_2^2) \widehat{\w{1}}+ w_2(1+w_1^2) \widehat{\w{2}}}{1+w_1^2w_2^2}.
\end{equation}
This now leads to exactly the same results as above; there was no loss of generality inherent in  working in the $x$--$y$ plane. 

\subsubsection{Example: Reduction to Giust--Vigoureux--Lages result in the $x$--$y$ plane}
It is important to note that our composition law \( \oplus \) reduces to the composition law of Giust, Vigoureux, and Lages when dealing with planar velocities in the $x$--$y$ plane.
As above, we define general velocities in the (\( \ii, \jj \))-plane by \( \w{1} = \tanh(\xi_1/2) \e^{\kk\phi_1} \ii \), and \( \w{2} = \tanh(\xi_2/2) \e^{\kk\phi_2} \ii \), then, using our composition law \eqref{eq;oplusgen}, we find
\begin{equation}
    \w{1\oplus2} = \left(1 - \tanh\frac{\xi_1}{2} \e^{\kk\phi_1}\ii \; \tanh\frac{\xi_2}{2} \e^{\kk\phi_2}\ii \right)^{-1} \left( \tanh\frac{\xi_1}{2} \e^{\kk\phi_1}\ii + \tanh\frac{\xi_2}{2} \e^{\kk\phi_2}\ii \right).
\end{equation}
But, noting that \( \tanh(\xi_2/2) \e^{\kk\phi_2}\ii = \tanh(\xi_2/2) \ii \, \e^{-\kk\phi_2}  \) and $\ii^2=-1$, we can re-write this as 
\begin{equation}
    \w{1\oplus2} = \left(1 + \tanh\frac{\xi_1}{2} \e^{\kk\phi_1} \tanh\frac{\xi_2}{2} \e^{-\kk\phi_2}\ \right)^{-1} \left( \tanh\frac{\xi_1}{2} \e^{\kk\phi_1} + \tanh\frac{\xi_2}{2} \e^{\kk\phi_2}\right)\ii.
\end{equation}
Now, writing 
\begin{equation}
\w{1\oplus2} = \tanh(\xi_{1\oplus2}/2) \, \e^{\kk \phi_{1\oplus2}}\,  \ii
\end{equation}
we can cancel out the trailing $\ii$, to obtain
\begin{multline}
 \tanh{\xi_{1\oplus2}\over2}\; \e^{\kk \phi_{1\oplus2}} =
 \left(1 + \tanh\frac{\xi_1}{2} \e^{\kk\phi_1} \tanh\frac{\xi_2}{2} \e^{-\kk\phi_2}\ \right)^{-1} \times \\
  \left( \tanh\frac{\xi_1}{2} \e^{\kk\phi_1} + \tanh\frac{\xi_2}{2} \e^{\kk\phi_2}\right).
\end{multline}
This expression now only contains $\kk$,  so everything commutes, and we can write
\begin{equation}
w_{1\oplus2}\,\e^{\kk \phi_{1\oplus2}} = 
 {
 w_1\, \e^{\kk\phi_1} + w_2\, \e^{\kk\phi_2}
 \over
 1 + w_1\, \e^{\kk\phi_1}\, w_2\, \e^{-\kk\phi_2}
 }
\end{equation}
which is equivalent to the result of Giust, Vigoureux, and Lages. 

\subsubsection{Example: Composition in general directions}
For general velocities $\v_{1}$ and $\v_{2}$ the easiest way of proceeding is to simply rotate to put $\v_{1}$ and $\v_{2}$  in the  the $x$--$y$ plane, and just copy the Giust--Vigoureux--Lages argument \cite{vigoureuxetal:2009} above. 
If one wishes to be more direct, then simply define
\begin{equation}
    \w{1} = w_1 \; \widehat{\w{1}} , \qquad\qquad \w{2} = w_2 \; \widehat{\w{2}}; 
    \qquad\qquad \widehat{\w{3}} = 
    {[\widehat{\w{1}}, \; \widehat{\w{2}}]\over |\,[{\widehat{\w{1}}, \; \widehat{\w{2}}]\, |}} .
\end{equation}
If $ \widehat{\w{1}}$ is not parallel to  $\widehat{\w{2}}$, then $ \widehat{\w{3}}$ is well defined and perpendicular to both $ \widehat{\w{1}}$ and $ \widehat{\w{2}}$. 
With these definitions one can now write
\begin{equation}
\widehat{\w{2}} = \exp( \phi  \widehat{\w{3}} ) \; \widehat{\w{1}}.
\end{equation}
Then, following the discussion above, we see
\begin{align}
\w{1\oplus2} &=
( 1 + w_1\, w_2\, \e^{-\widehat{\w{3}}\phi})^{-1}
{
( w_1\, \widehat{\w{1}}+ w_2\,\widehat{\w{2}})
 }	\notag \\
 &= 
 ( 1 + w_1\, w_2\, \e^{-\widehat{\w{3}}\phi})^{-1}
 {
 (w_1+ w_2\, \e^{\widehat{\w{3}}\phi}) \widehat{\w{1}}
 }.
\end{align}
From this we can extract
\begin{equation}
w_{1\oplus2}\;\e^{\widehat{\w{3}} \phi_{1\oplus2}} = 
 ( 1 + w_1\, w_2\, \e^{-\widehat{\w{3}}\phi})^{-1}
 {
 (w_1+ w_2\, \e^{\widehat{\w{3}}\phi}) 
 }
 =
 { (w_1+ w_2\, \e^{\widehat{\w{3}}\phi}) \over  ( 1 + w_1\, w_2\, \e^{-\widehat{\w{3}}\phi}) }.
\end{equation}
Finally,
\begin{equation}
w_{1\oplus2}\;\e^{\widehat{\w{3}} \phi_{1\oplus2}} = 
 { (w_1+ w_2\, \e^{\widehat{\w{3}}\phi}) \over  ( 1 + w_1\, w_2\, \e^{-\widehat{\w{3}}\phi}) }.
\end{equation}
This finally is a fully explicit result for general velocities $\v_{1}$ and $\v_{2}$, which is manifestly in agreement with the  Giust--Vigoureux--Lages  results \cite{vigoureuxetal:2009}.

\subsubsection{Uniqueness of the composition law}
Finally, we might note that the expression for the composition law \eqref{eq;oplusgen} is not unique.
For example, by considering the power-series of \( (1-\w{1}\w{2})^{-1} \), we can re-write \eq{} \eqref{eq;oplusgen} as
\begin{equation}
    \w{1\oplus2} = (1-\w{1}\w{2})^{-1} (\w{1}+\w{2}) = \sum_{n=0}^\infty (\w{1}\w{2})^n (\w{1}+\w{2}).
\end{equation}
But, as \( \w{1} \) and \( \w{2} \) are pure quaternions, both \( \w{1}^2 \) and \( \w{2}^2 \) are real numbers, and so commute with \( \w{1} \) and \( \w{2} \).
Thus,
\begin{equation}
    \w{1\oplus2} = \sum_{n=0}^\infty (\w{1}\w{2})^n \w{1} + \sum_{n=0}^\infty (\w{1}\w{2})^n \w{2} = \w{1} \sum_{n=0}^\infty (\w{2}\w{1})^n + \w{2} \sum_{n=0}^\infty (\w{2}\w{1})^n.
\end{equation}
Consequently we find that our composition law can also be written as 
\begin{equation}
    \w{1\oplus2} = (\w{1}+\w{2})\sum_{n=0}^\infty (\w{2}\w{1})^n = (\w{1}+\w{2}) (1 - \w{2}\w{1})^{-1}.
\label{eq;oplusgen2}
\end{equation}
Indeed, one could use either \eq{} \eqref{eq;oplusgen} or \eq{} \eqref{eq;oplusgen2} as the definition of the composition law \( \oplus \).
Nonetheless, we will stick with the convention given in \eqref{eq;oplusgen}.

\subsection{Calculating the Wigner angle}

In this section we obtain an expression for the Wigner angle for general 3-velocities using our composition law \eqref{eq;oplusgen}.
Our calculations are obtained using the result that the Wigner angle is the angle between the velocities \( \w{1\oplus2} \) and \( \w{2\oplus1} \).

We first note
\begin{equation}
|\w{1\oplus2}| = |\w{2\oplus1}| = |1-\w{1}\w{2}|^{-1} |\w{1}+\w{2}| 
= {|w_1 \vb{\hat n}_1+w_2 \vb{\hat n}_2| \over |1- w_1 w_2 \; \vb{\hat n}_1 \vb{\hat n}_2|}. 
\end{equation}
Thus, setting $\cos\theta=\vec n_1 \cdot \vec n_2$ we explicitly verify
\begin{equation}
|\w{1\oplus2}|  = |\w{2\oplus1}|= \sqrt{w_1^2+w_2^2+2w_1 w_2 \cos\theta
 \over 1 + w_1^2 w_2^2 + 2 w_1w_2 \cos\theta}. 
\end{equation}
Now note that because $|\w{1\oplus2}| = |\w{2\oplus1}|$ it follows that  $\left(\w{1\oplus2}\right) \left(\w{2\oplus1}\right)^{-1}$ is a unit norm quaternion. 
In fact, defining
\begin{equation}
\e^{\vb{\Omega}} =\left(\w{1\oplus2}\right) \left(\w{2\oplus1}\right)^{-1},
\end{equation}
we will soon see that the norm of \( \bf{\Omega} \) is the Wigner angle.
Explicitly,
\begin{equation}
\e^{\vb{\Omega}} = \left( (1-\w{1}\w{2})^{-1} (\w{1}+\w{2}) \right)
\left(  (1-\w{2}\w{1})^{-1} (\w{2}+\w{1}) \right)^{-1}.
\end{equation}
But for a product of quaternions we have $(\q_1\q_2)^{-1} = \q_2^{-1} \q_1^{-1}$, and so this reduces to
\begin{equation}
\e^{\vb{\Omega}} =  (1-\w{1}\w{2})^{-1}  (1-\w{2}\w{1}).
\end{equation}
Now
\begin{equation}
\w{1}\w{2} = - w_1w_2\cos\theta + (\vec w_1\times \vec w_2)\cdot(\ii,\jj,\kk).
\end{equation}
Let us define
\begin{equation}
{\hat\Omega} = {(\vec w_1\times \vec w_2)\over |\vec w_1\times \vec w_2|};
\qquad\hbox{so}\qquad
{\hat w_1}\times {\hat w_2}= \sin\theta \;\; {\hat\Omega}.
\end{equation}
Then setting $\vb{\hat\Omega}  = {\hat\Omega}\cdot (\ii,\jj,\kk)$ so that $\vb{\Omega}  =\Omega\, \vb{\hat\Omega}$, we have:
\begin{equation}
\w{1}\w{2} = - w_1w_2 (\cos\theta - \sin\theta\; \vb{\hat\Omega}) 
 =-w_1w_2 \;\e^{-\theta \vb{\hat\Omega}}.
\end{equation}
Consequently the Wigner angle satisfies
\begin{equation}
\e^{\vb{\Omega}} = \e^{\Omega \,\vb{\hat\Omega}} =  
\left(1+w_1w_2\, \e^{-\theta \vb{\hat\Omega}}\right)^{-1}  
\left(1+w_1w_2 \,\e^{\theta \vb{\hat\Omega}}\right)
= {1+w_1w_2 \,\e^{\theta \vb{\hat\Omega}}\over  1+w_1w_2 \,\e^{-\theta \vb{\hat\Omega}}}.
\end{equation} 
Equivalently, 
\begin{equation}
	\e^{\Omega \vb{\hat\Omega}/2} 
	= \frac{1 + w_1w_2\,\e^{\theta\vb{\hat\Omega}}}{\abs{1 + w_1w_2 \, \e^{\theta\vb{\hat\Omega}}}}.
\label{eq;wig4}
\end{equation}
Taking the scalar and vectorial parts of \eq{} \eqref{eq;wig4}, we finally obtain 
\begin{equation}
	\tan\frac{\Omega}{2} = \frac{w_1w_2\sin\theta}{1+w_1w_2\cos\theta} = \frac{\abs{\vec{w}_1\cross \vec{w}_2}}{1+\vec{w}_1 \cdot \vec{w}_2},
\label{eq;tanwighalf}
\end{equation}
as an explicit expression for the Wigner angle \( \Omega \).

The simplicity of \eq{} \eqref{eq;tanwighalf} compared to exisiting formulae for \( \Omega \) in the literature, shows how the composition law \eqref{eq;oplusgen} can lead to much tidier and simpler formulae than other methods allowed for.
This can be seen as the extension of the result \eqref{eq;wigvig} to more general velocities.

We can write \eq{} \eqref{eq;tanwighalf} in a perhaps more familiar (though possibly more tedious) form by first noting that from \eq{} \eqref{eq;visquared} we have
\begin{equation}
    w_i = \frac{1-\sqrt{1-v_i^2}}{v_i} = \frac{\gamma_i-1}{\sqrt{\gamma_i^2-1}} = \sqrt{ \frac{\gamma_i-1}{\gamma_i+1} } = \frac{ \sqrt{\gamma_i^2-1} }{\gamma_i + 1} = \frac{ v_i \gamma_i }{\gamma_i+1},
\end{equation}
and so
\begin{equation}
    \tan\frac{\Omega}{2} = \frac{v_1v_2 \gamma_1\gamma_2 \sin\theta}{(1+\gamma_1)(1+\gamma_2) + v_1v_2\gamma_1\gamma_2\cos\theta}.
\label{eq;tanhalfreal}
\end{equation}

We can check two interesting cases of \eq{} \eqref{eq;tanwighalf} for when \( \theta = 0 \) (parallel velocities) and when \( \theta = \pi/2 \) (perpendicular velocities).
We can see directly that, for parallel velocities, the associated Wigner angle is given by \( \tan(\Omega/2) = 0 \), so that \( \Omega = n\pi \) for \( n \in \bb{Z} \); whilst for perpendicular velocities, the associated Wigner angle is simply given by \( \tan(\Omega/2) = w_1w_2 \).

It is easiest to check our results against the literature using the somewhat messier \eq{} \eqref{eq;tanhalfreal}, in which case parallel velocities again give \( \tan(\Omega/2) = 0 \), whilst perpendicular velocities give 
\begin{equation}
    \tan(\Omega/2) = \frac{v_1v_2\gamma_1\gamma_2}{(1+\gamma_1)(1+\gamma_2)},
\end{equation}
which agrees with the results given in \cite{visser_odonnell:2011}.

\section{Combining three 3-velocities}	\label{sec:three-3-velocities}
Let us now see what happens when we relativistically combine 3 half-velocities.\\
We shall calculate, compare, and contrast $\vb{w}_{(1\oplus2)\oplus3 }$ 
with $\vb{w}_{1\oplus(2\oplus3)}$.

\subsection{Combining 3 half-velocities: $\vb{w}_{(1\oplus2)\oplus3 }$}
Start from our key result
\begin{equation}
\vb{w}_{1\oplus2} = \vb{w}_1 \oplus \vb{w}_2
=(1-\vb{w}_1\vb{w}_2)^{-1} (\vb{w}_1 +\vb{w}_2),
\end{equation}
and iterate it to yield
\begin{equation}
\vb{w}_{(1\oplus2)\oplus3 } 
= \{1-(1-\vb{w}_1\vb{w}_2)^{-1} (\vb{w}_1 +\vb{w}_2)\vb{w}_3 \}^{-1} 
\{(1-\vb{w}_1\vb{w}_2)^{-1} (\vb{w}_1 +\vb{w}_2)+ \vb{w}_3\}.
\end{equation}

It is now a matter of straightforward quaternionic algebra to check that
\begin{align}
\vb{w}_{(1\oplus2)\oplus3 } 
&=
\{(1-\vb{w}_1\vb{w}_2)^{-1} (1-\vb{w}_1\vb{w}_2 -(\vb{w}_1 +\vb{w}_2)\vb{w}_3 )\}^{-1} \times
\notag\\
&\qquad\quad
\{(1-\vb{w}_1\vb{w}_2)^{-1} (\vb{w}_1 +\vb{w}_2)+ \vb{w}_3\}
\notag\\
&=
(1-\vb{w}_1\vb{w}_2 -(\vb{w}_1 +\vb{w}_2)\vb{w}_3 )^{-1} (1-\vb{w}_1\vb{w}_2) \times
\notag\\
&\qquad\quad \{(1-\vb{w}_1\vb{w}_2)^{-1} (\vb{w}_1 +\vb{w}_2)+ \vb{w}_3\}
\nonumber\\
&=
(1-\vb{w}_1\vb{w}_2 -(\vb{w}_1 +\vb{w}_2)\vb{w}_3 )^{-1}  \{(\vb{w}_1 +\vb{w}_2)+ (1-\vb{w}_1\vb{w}_2) \vb{w}_3\}.
\end{align}
Ultimately,
\begin{equation}
\vb{w}_{(1\oplus2)\oplus3 } 
=  \{1-\vb{w}_1\vb{w}_2 -\vb{w}_1\vb{w}_3- \vb{w}_2\vb{w}_3 \}^{-1}  \{\vb{w}_1 +\vb{w}_2+ \vb{w}_3 - \vb{w}_1\vb{w}_2 \vb{w}_3\}.
\label{E:(12)3a}
\end{equation}

An alternative formulation starts from
\begin{equation}
\vb{w}_{1\oplus2} = \vb{w}_1 \oplus \vb{w}_2
= (\vb{w}_1 +\vb{w}_2) (1-\vb{w}_2\vb{w}_1)^{-1},
\end{equation}
which when iterated yields
\begin{equation}
\vb{w}_{(1\oplus2)\oplus3 } 
= 
\{(\vb{w}_1 +\vb{w}_2) (1-\vb{w}_2\vb{w}_1)^{-1} + \vb{w}_3\}
\{1-  \vb{w}_3 (\vb{w}_1 +\vb{w}_2) (1-\vb{w}_2\vb{w}_1)^{-1} \}^{-1}.
\end{equation}
Thus, a little straightforward quaternionic algebra verifies that
\begin{align}
\vb{w}_{(1\oplus2)\oplus3 } 
&=
\{(\vb{w}_1 +\vb{w}_2)  + \vb{w}_3(1-\vb{w}_2\vb{w}_1)\}(1-\vb{w}_2\vb{w}_1)^{-1} \times
\notag\\
&\qquad\qquad \{1-  \vb{w}_3 (\vb{w}_1 +\vb{w}_2) (1-\vb{w}_2\vb{w}_1)^{-1} \}^{-1}
\notag\\
&=
\{(\vb{w}_1 +\vb{w}_2)  + \vb{w}_3(1-\vb{w}_2\vb{w}_1)\}
\{(1-\vb{w}_2\vb{w}_1)-  \vb{w}_3 (\vb{w}_1 +\vb{w}_2) \}^{-1}.
\end{align}
Ultimately,
\begin{equation}
\vb{w}_{(1\oplus2)\oplus3 } 
= 
\{\vb{w}_1 +\vb{w}_2  + \vb{w}_3 -\vb{w}_3\vb{w}_2\vb{w}_1\}
\{1-\vb{w}_2\vb{w}_1-  \vb{w}_3 \vb{w}_1 - \vb{w}_3 \vb{w}_2 \}^{-1}.
\label{E:(12)3b}
\end{equation}
So we have found two equivalent formulae for $\vb{w}_{(1\oplus2)\oplus3}$,
equations (\ref{E:(12)3a}) and (\ref{E:(12)3b}).

\subsection{Combining 3 half-velocities: $\vb{w}_{1\oplus(2\oplus3) }$}
In contrast, the situation for $\vb{w}_{1\oplus(2\oplus3) }$ is considerably more subtle.
Start from the key result that 
\begin{equation}
\vb{w}_{2\oplus3} = \vb{w}_2 \oplus \vb{w}_3
= (1-\vb{w}_2\vb{w}_3)^{-1} (\vb{w}_2 +\vb{w}_3),
\end{equation}
and iterate it to yield
\begin{equation}
\vb{w}_{1\oplus(2\oplus3) } 
=
\{1- \vb{w}_1(1-\vb{w}_2\vb{w}_3)^{-1} (\vb{w}_2 +\vb{w}_3)\}^{-1}
\{ \vb{w}_1 + (1-\vb{w}_2\vb{w}_3)^{-1} (\vb{w}_2 +\vb{w}_3)\}.
\end{equation}
The relevant quaternionic algebra is now a little trickier
\begin{eqnarray}
\vb{w}_{1\oplus(2\oplus3) } 
&=&
\{1- \vb{w}_1(1-\vb{w}_2\vb{w}_3)^{-1} (\vb{w}_2 +\vb{w}_3)\}^{-1}
(1-\vb{w}_2\vb{w}_3)^{-1} \times
\nonumber\\
&&\qquad\quad\{ (1-\vb{w}_2\vb{w}_3)\vb{w}_1 +  (\vb{w}_2 +\vb{w}_3)\}
\nonumber\\
&=&
\{ (1-\vb{w}_2\vb{w}_3)(1- \vb{w}_1(1-\vb{w}_2\vb{w}_3)^{-1} (\vb{w}_2 +\vb{w}_3)\}^{-1} \times
\nonumber\\
&&\qquad\quad
\{ (1-\vb{w}_2\vb{w}_3)\vb{w}_1 +  (\vb{w}_2 +\vb{w}_3)\}
\nonumber\\
&=&
\{ 1-\vb{w}_2\vb{w}_3- (1-\vb{w}_2\vb{w}_3) \vb{w}_1(1-\vb{w}_2\vb{w}_3)^{-1} (\vb{w}_2 +\vb{w}_3)\}^{-1} \times
\nonumber\\
&&\qquad\quad
\{ \vb{w}_1 +  \vb{w}_2 +\vb{w}_3 - \vb{w}_2\vb{w}_3\vb{w}_1\}.
\end{eqnarray}
To proceed we note that
\begin{align}
(1-\vb{w}_2\vb{w}_3) \vb{w}_1(1-\vb{w}_2\vb{w}_3)^{-1}
&=
\left(1-\vb{w}_2\vb{w}_3\over |1-\vb{w}_2\vb{w}_3|\right) 
\vb{w}_1
\left(1-\vb{w}_2\vb{w}_3\over |1-\vb{w}_2\vb{w}_3|\right)^{-1}
\notag\\
&
\vphantom{\Big|} = e^{-\vb{\Omega}_{2\oplus3}/2} \;\vb{w}_1\; e^{+\vb{\Omega}_{2\oplus3}/2}.
\end{align}
Thus,
\begin{multline}
\vb{w}_{1\oplus(2\oplus3) } 
= 
\{ 1-\vb{w}_2\vb{w}_3-  (e^{-\vb{\Omega}_{2\oplus3}/2} \vb{w}_1e^{+\vb{\Omega}_{2\oplus3}/2}) (\vb{w}_2 +\vb{w}_3)\}^{-1} \times \\
\{ \vb{w}_1 +  \vb{w}_2 +\vb{w}_3 - \vb{w}_2\vb{w}_3\vb{w}_1\}.
\label{E:1(23)}
\end{multline}
While structurally similar to the formulae  (\ref{E:(12)3a}) and (\ref{E:(12)3b}) for $\vb{w}_{(1\oplus2)\oplus3}$ the present result (\ref{E:1(23)}) for $\vb{w}_{1\oplus(2\oplus3)}$ is certainly different --- the Wigner angle $\vb{\Omega}_{2\oplus3}$ now makes an explicit appearance, also the form of the triple-product $\vb{w}_2\vb{w}_3\vb{w}_1$ is different.

\subsection{Combining 3 half-velocities: (Non)-associativity}
From  (\ref{E:(12)3a}) and (\ref{E:(12)3b}) for $\vb{w}_{(1\oplus2)\oplus3}$, and (\ref{E:1(23)}) for $\vb{w}_{1\oplus(2\oplus3)}$, it is clear that relativistic composition of velocities is in general not associative. (See for instance the discussion in references~\cite{sonego:2006, ungar:2006}, commenting on reference~\cite{sonego:2005}.) 

A sufficient condition for associativity,
 $\vb{w}_{(1\oplus2)\oplus3 } = \vb{w}_{1\oplus(2\oplus3) }$,   is to enforce
\begin{equation}
e^{-\vb{\Omega}_{2\oplus3}/2} \vb{w}_1e^{+\vb{\Omega}_{2\oplus3}/2} =  \vb{w}_1,
 \qquad\hbox{and} \qquad \vb{w}_1\vb{w}_2 \vb{w}_3 = \vb{w}_2\vb{w}_3\vb{w}_1.
\end{equation}
 That is, a sufficient condition for associativity is
 \begin{equation}
 [\vb{\Omega}_{2\oplus3}, \vb{w}_1]=0, \qquad\hbox{and} \qquad 
 [\vb{w}_1,\vb{w}_2 \vb{w}_3] = 0.
\end{equation}
But note $\vb{\Omega}_{2\oplus3}\propto [\vb{w}_2, \vb{w}_3]$ and $\vb{w}_2 \vb{w}_3 = {1\over2}
\{\vb{w}_2, \vb{w}_3\} + {1\over2} [\vb{w}_2, \vb{w}_3]$.
Since $\{\vb{w}_2, \vb{w}_3\} \in \mathbb{R}$, we then have $ [\vb{w}_1,\vb{w}_2 \vb{w}_3] =  {1\over2} [\vb{w}_1,[\vb{w}_2, \vb{w}_3]]$.
This now implies that these two sufficiency conditions are in fact identical; so a sufficient condition for associativity is
\begin{equation}
 [\vb{w}_1,[\vb{w}_2, \vb{w}_3]] = 0.
\end{equation}
 This sufficient condition for associativity can also be written as the vanishing of the vector triple product
\begin{equation}
\vec w_1 \times(\vec w_2 \times \vec w_3) = 0.
\end{equation}
Equivalently,
\begin{equation}
\vec v_1 \times(\vec v_2 \times \vec v_3) = 0.
\end{equation}

\subsection{Specific non-coplanar example}
As a final example of the power of the quaternion formalism, let us consider a specific  intrinsically  non-coplanar example. Let \( \w{1} = w_1 \ii \), \( \w{2} = w_2\, \jj \), and \( \w{3} = w_3 \kk \) be three mutually perpendicular half-velocities. (So this configuration does automatically satisfy the associativity condition discussed above.)

Then we have already seen that
\begin{equation}
\w{1} \oplus \w{2} = \frac{w_1(1-w_2^2)\ii + w_2(1+w_1^2)\jj}{1+w_1^2w_2^2},
\qq{and}
w_{1\oplus2}^2 = {w_1^2 + w_2^2\over 1+w_1^2 w_2^2}.
\end{equation}
Furthermore, since $\w{1} \oplus \w{2}$ is perpendicular to $\w{3}$, we have
\begin{equation}
(\w{1} \oplus \w{2})\oplus \w{3} 
 = \frac{w_{1\oplus2}(1-w_3^2)\vb{\hat n}_{1\oplus2} + w_3(1+w_{1\oplus2}^2)\kk}{1+w_{1\oplus2}^2w_3^2},
\end{equation}
and
\begin{equation}
w_{(1\oplus2)\oplus3}^2 = {w_{(1\oplus2)}^2+w_3^2\over 1 + w_{(1\oplus2)}^2 w_3^2}
=
{w_1^2+w_2^2+w_3^2+w_1^2w_2^2w_3^2
\over 1 + w_1^2 w_2^2 + w_2^2 w_3^2 + w_3^2 w_1^2}.
\end{equation}

A little algebra now yields the manifestly non-commutative result
\begin{multline}
(\w{1}\oplus\w{2})\oplus\w{3} =
\frac{1}{1+w_1^2w_2^2 +  w_2^2w_3^2 +w_3^2w_1^2} \bigg\{ (1-w_2^2)(1-w_3^2)\w{1} + \\
+ (1+w_1^2)(1-w_3^2)\w{2} + (1+w_1^2)(1+w_2^2)\w{3} \bigg\}.
\end{multline}
In this particular case we can also explicitly show that
\begin{equation}
(\w{1}\oplus\w{2})\oplus\w{3} = \w{1}\oplus(\w{2}\oplus\w{3}),
\end{equation}
though (as discussed above) associativity fails in general.

\section{Summary}

In this chapter, we analysed a simple, elegant, and novel algebraic method for combining special relativistic 3-velocities using quaternions:
\begin{equation}
\vb{w}_{1\oplus2} = \vb{w}_1 \oplus \vb{w}_2
=(1-\vb{w}_1\vb{w}_2)^{-1} (\vb{w}_1 +\vb{w}_2)
= (\vb{w}_1 +\vb{w}_2)(1-\vb{w}_2\vb{w}_1)^{-1}.
\end{equation}
The construction also leads to a simple, elegant, and novel formula for the Wigner angle:
\begin{equation}
e^{\vb{\Omega}} = e^{\Omega \; \vb{\hat\Omega} } = (1-\vb{w}_1\vb{w}_2)^{-1} (1-\vb{w}_2\vb{w}_1),
\end{equation}
in terms of which
\[
{\vb{\hat{w}}}_{1\oplus2} = \mathrm{e}^{\vb{\Omega}/2} \;\;
{\vb{w}_1+\vb{w}_2\over |\vb{w}_1+\vb{w}_2|}; 
\qquad\qquad
{\vb{\hat{w}}}_{2\oplus1} = \mathrm{e}^{-\vb{\Omega}/2} \;\;
{\vb{w}_1+\vb{w}_2\over |\vb{w}_1+\vb{w}_2|}.
\]
We saw how all of the non-commutativity associated with non-collinearity of 3-velocities is automatically and rather efficiently dealt with by the quaternion algebra. 

This concludes our discussion of the theory of special relativity, and we will now move onto the theory of general relativity in the following chapters.
\chapter{General relativity}\label{C:GR-maths-background}

We now move to the setting of general relativity. 
As the name suggests, general relativity is a generalisation of the theory of special relativity.
Indeed, it is a generalisation in the sense that it moves from the ``flat space'' of special relativity (Minkowski space) to more general ``curved spaces''.
These curved spaces are represented by manifolds, and the mathematical language used to describe them is differential geometry.
In this chapter, we will present some of the main mathematical tools from differential geometry used in general relativity and introduce the idea of a spacetime rigorously.
From there, we will discuss Einstein's equivalence principle and how it leads to the theory of general relativity.
This will then allow us to analyse specific spacetime models within the framework of general relativity in the following chapters.

The concept of a spacetime in general relativity is developed by adding additional mathematical structure to a four-dimensional manifold.
Rigorously, we may define a spacetime as  follows \cite{Hawking-Ellis, Visser:lorentzian-wormholes}.

\begin{definition}\label{def:spacetime}
A spacetime is a pair (\( \M, \g \)), where \( \M \) is a connected, four-dimensional \( C^\infty \) manifold and \( \g \) is a metric with Lorentzian signature.
\end{definition}

Although most of the terms in this definition have not yet been defined (they will be in the following sections), we can see a familiar theme with special relativity: spacetime is four-dimensional.
As in special relativity, three of these are spacial dimensions, whilst the fourth is a temporal dimension. 
Thus, intuitively, one may think of spacetime in general relativity as a four-dimensional object that encodes both spatial and temporal information. 
However, in order to understand all of the terms in definition \ref{def:spacetime}, we will need the mathematical framework presented in the following sections.

\section{Manifolds}

Intuitively, a manifold is a space which may be curved on a global scale (characterised by curvature tensors defined below), yet locally it must resemble Euclidean (flat) space.
We formalise the notion of `resembling flat space' with the following definition. 

\begin{definition}\label{def:Euclid-space}
A \( n \)-dimensional locally Euclidean space is a set \( \mathcal{E} \) together with a \( C^r \) atlas \( \{U_\alpha, \phi_\alpha\} \), i.e. a collection of charts \( (U_\alpha, \phi_\alpha) \) where the \( U_\alpha \) are subsets of \( \mathcal{E} \) and the \( \phi_\alpha \) are injective maps from the \( U_\alpha \) into open subsets of \( \RR^n \) (endowed with the standard topology) such that:
    \begin{enumerate}
    \item[(1)] the \( U_\alpha \) cover \( \mathcal{E} \), i.e. \( \mathcal{E} = \bigcup_\alpha U_\alpha \);
    \item[(2)] if \( U_\alpha \cap U_\beta \) is non-empty, then the map 
    \[ \phi_\alpha \circ \phi_\beta^{-1}: \phi_\beta(U_\alpha \cap U_\beta) \rightarrow \phi_\alpha(U_\alpha\cap U_\beta) \] 
    is a \( C^r \) map from an open subset of \( \RR^n \) to an open subset of \( \RR^n \).
    \end{enumerate}
\end{definition}

Here we define a \( C^r \) map to be one that is \( r \)-times continuously differentiable.
We can now define a manifold as follows \cite{Visser:464-notes}.

\begin{definition}\label{def:manifold}
A manifold \( \M \) is a locally Euclidean space which:
    \begin{enumerate}
    \item[(1)] has the same dimension everywhere,
    \item[(2)] is Hausdorff,
    \item[(3)] has at least one countable atlas.
    \end{enumerate}
\end{definition}

A space is said to be Hausdorff if whenever \( p \) and \( q \) are distinct points in \( \M \), there exists disjoint open subsets \( U \) and \( V \) of \( \M \) such that \( p\in U \) and \( q\in V \).
A topological space is said to be connected if it is \emph{not} the union of two disjoint non-empty open sets. 
Both properties of connectedness and Hausdorffness are simply ``physically reasonable'' constraints that we place on the manifold in order to ensure that the mathematical notion of a spacetime resembles as closely as possible the universe we observe. 
The canonical example of a two-dimensional manifold is the 2-sphere, \( S^2 \).
Globally it is a curved surface, yet a small patch on the surface of the sphere will look like a flat piece of \( \RR^2 \) if we `zoom in' close enough. 

\section{The metric tensor}

Perhaps the most important mathematical object in classical general relativity is the metric tensor which one endows on their manifold.
In general, we have the following definition \cite{Hawking-Ellis}.

\begin{definition}\label{def:metric}
A metric tensor \( \g \) (usually just called ``the metric'') at a point \( p\in\M \) is a symmetric tensor of type \( (0,2) \) at \( p \).
\end{definition}

If a coordinate basis \( \{\partial/\partial x^a\} \) is used, then one can express the metric in terms of its components as
\begin{equation}
\g = g_{ab}\dd{x^a}\otimes\dd{x^b}.
\label{eq:metric-compts-def}
\end{equation}
The metric is extremely useful as it allows us to define the notion of a path length between two points along a curve in a manifold.
Indeed, suppose that for two points \( p,q\in\M \) there is a smooth curve \( \gamma \) parametrised by some parameter \( \lambda\in\RR \) such that \( p=\gamma(a) \) and \( q=\gamma(b) \).
Then the path length between the points \( p \) and \( q \) is given by
\begin{equation}
L = \int_a^b \sqrt{ g_{ab} \dv{x^a}{\lambda} \dv{x^b}{\lambda} } \dd{\lambda}.
\label{eq:path-length-def}
\end{equation}
We may express equations \eqref{eq:metric-compts-def} and \eqref{eq:path-length-def} via one relation as
\begin{equation}
\dd s^2 = g_{ij}\dd x^i \dd x^j,
\label{eq:line-element-def}
\end{equation}
which represents the infinitesimal arc determined by the coordinate displacement \( x^i \rightarrow x^i + \dd x^i \).
In the context of general relativity, the quantity \( \dd s^2 \) as defined by equation \eqref{eq:line-element-def} is commonly called the ``line element'' of the spacetime, and is in direct one-to-one correspondence with the metric endowed on the spacetime manifold.
As such, one may see the terms `metric' and `line-element' used interchangeably within the context of a spacetime.

One of the powerful properties of the metric is that it allows us to interchange between contravariant and covariant tensor quantities. 
We call a metric non-degenerate at a point \( p\in\M \) if the matrix \( (g_{ab}) \) of components of \( \g \) is non-singular at \( p \) (i.e. the matrix is invertible at \( p \)).
For a non-degenerate metric \( \g \), we can always define a unique non-degenerate symmetric tensor of type \( (2,0) \) (sometimes called the ``contravariant-metric'') by the relation
\begin{equation}
g^{ab}g_{bc} = \delta^a_c.
\end{equation}
That is, \( (g^{ab}) \) is the matrix-inverse of \( (g_{ab}) \).
Thus, if \( X^a \) are the components of a contravariant vector, then we can uniquely define a covariant vector with components given by \( X_a = g_{ab}X^b \).
Hence, one may also write \( X^a = g^{ab}X_b \).
Similarly, for a type \( (0,2) \) tensor \( X_{ab} \) we may write \( X\indices{^a_{\,b}} = g^{ac}X_{cb} \), \( X\indices{_a^b} = g^{bc}X_{ac} \), and \( X^{ab} = g^{ac}g^{bd}X_{cd} \).

Using the metric tensor, we can classify non-zero vectors into three distinct classes:
\begin{enumerate}
    \item[(i)] Timelike: \( g_{ab}X^aX^b < 0 \),
    \item[(ii)] Null: \( g_{ab}X^aX^b = 0 \),
    \item[(iii)] Spacelike: \( g_{ab}X^aX^b > 0 \).
\end{enumerate}
Analogous to special relativity, this corresponds to (i) massive particles, (ii) massless particles, and (iii) tachyonic particles.
 
The last term to unpack in our definition of a spacetime (definition \ref{def:spacetime}) is the concept of the signature of a metric.
We have the following definition \cite{Hawking-Ellis}.

\begin{definition}\label{def:metric-sig}
The signature of a metric \( \g \) at a point \( p\in\M \) is defined as the number of positive eigenvalues of the matrix \( (g_{ab}) \) at \( p \), minus the number of negative ones.
\end{definition}

Furthermore, if \( \g \) is non-degenerate, the signature of \( \g \) will be constant on \( \M \) \cite{Hawking-Ellis}.
Commonly, one may see the signature of a metric presented as a \( n \)-tuple of \( `+' \) and \( `-' \) signs, representing the positive (\( + \)) and negative (\( - \)) eigenvalues of the metric.
For example, \( (-,+,+) \) or \( (+,+,+,+,+) \).
This encodes the signature of the metric, as in definition \ref{def:metric-sig}, but is not so easily adapted to the description of metrics on high-dimensional manifolds.
However, as a spacetime is restricted to four dimensions, this representation of the metric signature is commonly seen in classical general relativity.

\begin{definition}\label{def:R/L-metric}
A positive-definite metric (sometimes called a Riemannian metric) on a \( n \)-dimensional manifold is a metric with signature \( n \) (i.e. all positive eigenvalues).
Contrastingly, a Lorentzian metric on a \( n \)-dimensional manifold is a metric with signature \( n-2 \) (i.e. one negative eigenvalue).
\end{definition}

Now that we have all of the framework necessary to understand the definition of a spacetime (c.f. definition \ref{def:spacetime}), we can begin to investigate some of the more geometrical properties of manifolds and the associated spacetimes.

\section{Covariant differentiation and connections}

In general, the partial derivative of a tensor is not a tensor in of itself.
There are three standard, distinct ways to define a notion of the derivative of a general tensor quantity which invoke additional complications in order to bypass this problem. 
The first of these notions (the exterior derivative) restricts the class of tensors it operates on so as to ensure that it produces tensor quantities, whilst the second (the Lie derivative) restricts the `directions' that you can differentiate tensors in to again ensure that it produces tensor quantities. 
Both of these notions of a derivative have a multitude of uses in classical general relativity, but will not be needed for any of the analysis presented in this thesis.
As such, details of exterior- and Lie derivatives will not be discussed here for purposes of keeping this chapter compendious.
Thus, we will focus our analysis on the third way of defining a derivative of a tensor quantity, the covariant derivative. 
The covariant derivative solves the issue of differentiating a tensor by adding additional mathematical structure by way of an ``affine connection''\footnote{Sometimes simply called a ``connection'', or the ``Christoffel symbols''. However, as we will soon see, the Christoffel symbols are in fact a very specific, special type of connection.}.
Various ways of defining or constructing the covariant derivative can be found in the literature \cite{Hawking-Ellis, MTW:gravitation, Visser:464-notes, Wald:general-relativity}.
We define the covariant derivative for vectors axiomatically below, as in \cite{Visser:464-notes}. 
The definition adopted here greatly simplifies the discussion with regards to covariant differentiation, but does not provide much insight into the construction of the affine connection.
As such, many authors prefer to start by constructing the affine connection and then go on to use this in their definition of the covariant derivative. 
However, as covariant differentiation will be used explicitly in the following sections, I have used the following construction to simplify the discussion around it.
If the reader would like additional insight into the affine connection and its relationship to the parallel transport of vectors, these details may be found in any of the standard textbooks on general relativity \cite{Carroll:spacetime, Hartle:gravity, Hawking-Ellis, MTW:gravitation, Wald:general-relativity}.

\begin{definition}\label{def:cov-diff}
Suppose we have two contravariant vector fields \( u^a \) and \( v^a \), and define the vectors \( u \) and \( v \) to be the directional derivatives \( u = u^a\partial_a \) and \( v = v^a\partial_a \).
We now define our covariant derivative operator to be the linear operator \( \nabla_u = u^a\nabla_a \) which satisfies 
    \begin{enumerate}
    \item[(1)] \( \nabla_u f = u^a\partial_a f \),
    \item[(2)] \( \nabla_{fu} v = f\nabla_u v \),
    \item[(3)] \( \nabla_u (fv) = (\nabla_u f)v + f\nabla_u v \),
    \end{enumerate}
where \( f \) is a scalar function.
\end{definition}

Now, note that \( \partial_a \) is a vector field, and so \( \nabla_{\partial_a}\partial_b \) is also a vector field and can hence be expanded as a linear combination of \( \partial_c \)'s.
That is, in terms of the expansion coefficients \( \connec{a}{bc} \), we have \( \nabla_{\partial_a}\partial_b = \connec{a}{bc}\partial_c \).
We can now use this result to derive expressions for the covariant derivative of tensor quantities of general rank.
Full details of this process van be found in any of \cite{Hawking-Ellis, Wald:general-relativity, Visser:464-notes}, but here I will just present the main results:
\begin{subequations}\label{eq:cov-diff-expressions}
    \begin{itemize}
    \item \( (1,0) \) tensor: 
        \begin{equation} 
            \nabla_a X^b = \partial_a X^b + \connec{b}{ca}X^c, 
            \label{eq:cov-diff-contravec}
        \end{equation}
    \item \( (0,1) \) tensor: 
        \begin{equation} 
            \nabla_a X_b = \partial_aX_b - \connec{c}{ba}g_c, 
            \label{eq:cov-diff-covec}
        \end{equation}
    \item general \( (r,s) \) tensor: 
        \begin{multline}
            \hspace{-1.5em} \nabla_a X\indices{^{b_1b_2\dots b_r}_{c_1c_2\dots c_s}} = \partial_a X\indices{^{b_1b_2\dots b_r}_{c_1c_2\dots c_s}} + \sum_{i=1}^r \connec{b_i}{ma}X\indices{^{b_1b_2\dots b_{i-1}mb_{i+1}\dots b_r}_{c_1c_2\dots c_s}} \\
            - \sum_{j=1}^s \connec{m}{c_j a} X\indices{^{b_1b_2\dots b_r}_{c_1c_2\dots c_{j-1}mc_{j+1}\dots c_s}}.
        \label{eq:cov-diff-general}
        \end{multline}
    \end{itemize}
\end{subequations}
The \( \connec{a}{bc} \) form the components of the affine connection, which plays an extremely important role in general relativity. 
In fact, classical general relativity imposes a geometrical constraint on the connection.
This extra condition is that it is torsion-free.
That is, \( \connec{a}{[bc]} \equiv \frac{1}{2} [ \connec{a}{bc} - \connec{a}{cb} ] = 0 \).
This extra condition has been experimentally checked, and so far all evidence seems to suggest that we do indeed live in a torsion-free universe.
Furthermore, it has the added benefit that it allows one to use an extremely natural choice of connection for their spacetime, the Christoffel connection.
This is typically denoted by \( \Gamma \), and has components defined in terms of partial derivatives of the metric: 
\begin{equation}
\chris{a}{bc} = \frac{1}{2}g^{ad}\big( \partial_b g_{cd} + \partial_c g_{bd} - \partial_d g_{bc} \big).
\label{eq:chris-connec}
\end{equation}
In fact, the Christoffel connection is the unique metric-compatible\footnote{Metric compatible simply means \( \nabla_a g_{bc} = 0 \).}, torsion-free affine connection one can construct.
It should be noted that there are alternative theories of gravity that do not impose zero torsion on their connection (for example, certain string-inspired models, teleparallel gravity, \textit{etcetera}).
However, this thesis is primarily concerned with classical general relativity, and so we will we use the torsion-free Christoffel connection throughout the rest of the thesis.

We can now start with a metric for a given manifold, use this to construct the Christoffel connection via equation \eqref{eq:chris-connec}, and hence define a notion of covariant differentiation for general tensors in our spacetime via equations \eqref{eq:cov-diff-expressions}.
In the following sections, we will use our covariant derivative to obtain various notions for the `curvature' of a spacetime, and then use this to define the concept of the `straightest possible path' in a curved space by way of the geodesic equation.
Finally, we will use these concepts to construct the theory of general relativity. 

\hfill

\section{Curvature}

In general, covariant derivatives do not commute.
As a result of this, if one starts at a point \( p\in\M \) and parallel transports a vector \( v_1 \) along a curve \( \gamma_1 \) that also ends at \( p \), you will obtain a vector \( v_2 \) which is in general different to \( v_1 \).
Furthermore, if one parallelly transports along a different curve \( \gamma_2 \) that also ends at \( p \), you will in general obtain yet another vector \( v_3 \neq v_2 \neq v_1 \).
This `non-integrability' of parallel transport directly corresponds to the non-commutativity of the covariant derivative.
Thus, we can define a tensor \( R \) which measures the extent of this non-commutativity when acting on vector fields \( X, Y, \) and \( Z \):
\begin{equation}
R(X,Y)Z = \nabla_X (\nabla_Y Z) - \nabla_Y (\nabla_X Z) - \nabla_{[X, Y]} Z.
\label{eq:def-Riemann}
\end{equation}
This tensor is called the Riemann (curvature) tensor, and has components \cite{MTW:gravitation, Wald:general-relativity}
\begin{equation}
\R{a}{bcd} = \partial_c \chris{a}{bd} - \partial_d \chris{a}{bc} + \chris{a}{mc}\chris{m}{bd} - \chris{a}{md}\chris{m}{bc}.
\label{eq:Riemann-compts}
\end{equation}
As we will soon see, the Riemann tensor plays a central role in the mathematical formulation of general relativity.

Using the Riemann tensor, we can construct a whole host of other tensor quantities.
Below, I list some of the more commonly used tensors relevant to general relativity.

Contracting on the first and second indices of the Riemann tensor we yield another tensor, the Ricci tensor:
\begin{equation}
R_{ab} \equiv \R{c}{acb} = \partial_c \chris{c}{ab} - \partial_b \chris{c}{ac} + \chris{c}{dc}\chris{d}{ab} - \chris{c}{db}\chris{d}{ac}.
\label{eq:def-Ricci}
\end{equation}
We may contract once more to obtain a scalar, the Ricci scalar\footnote{Sometimes called the ``scalar curvature''.}:
\begin{equation}
R \equiv g^{ab}R_{ab} = g^{ab} \big( \partial_c \chris{c}{ab} - \partial_b \chris{c}{ac} + \chris{c}{dc}\chris{d}{ab} - \chris{c}{db} \big).
\label{eq:def-Ricci-scalar}
\end{equation}
Note that as this is a scalar quantity it is invariant under changes of coordinates.
As the name suggests, another tensor quantity of central importance in general relativity is the Einstein tensor:
\begin{equation}
G_{ab} \equiv R_{ab} - \frac{1}{2}R g_{ab}.
\label{eq:def-Einstein}
\end{equation}
Contacting all indices on the Riemann tensor we obtain the Kretschmann scalar:
\begin{equation}
K \equiv R^{abcd}R_{abcd}.
\label{eq:def-Kretcsh}
\end{equation}
By defining the Weyl (conformal) tensor (for a manifold of dimension \( n\geq4 \)) as
\begin{multline}
C_{abcd} \equiv R_{abcd} + \frac{1}{n-2}\big( g_{ad}R_{bc} + g_{bc}R_{ad} - g_{ac}R_{bd} - g_{bd}R_{ac} \big) \\
+ \frac{R}{(n-1)(n-2)}\big( g_{ac}g_{bd} - g_{ad}g_{bc} \big),
\label{eq:def-Weyl}
\end{multline}
we can re-write the Kretschmann scalar as
\begin{equation}
K = C^{abcd}C_{abcd} + \frac{4}{n-2}R^{ab}R_{ab} - \frac{2}{(n-1)(n-2)} R^2.
\label{eq:Kretsch-weyl}
\end{equation}

All of the tensors presented above give slightly different notions of the geometrical curvature of a spacetime.
As a spacetime is intrinsically four-dimensional, it is often hard to obtain a mental picture of what this physically means for a spacetime (how can time be curved?).
In section \ref{sec:field-eqs-equiv-principle}, we will provide a way of connecting this mathematical framework back to physical reality. 
Before we get to this, however, we still need one more piece of mathematical machinery: geodesics.

\section{Geodesics}

Intuitively, we want to define a geodesic to be a curve that is ``as straight as possible'' in a curved manifold.
Building on this intuition, we wish to require that the tangent vector \( T^a \) to the curve points in the same direction as itself when parallel propagated.
That is, we require
\begin{equation}
T^a\nabla_a T^b \propto T^b.
\end{equation}
With a suitable parameterisation of our curve, we can turn this equation in to a much simpler condition:
\begin{equation}
T^a\nabla_a T^b = 0.
\label{eq:def-geodesic}
\end{equation}
Such a parameterisation is called an ``affine'' parameterisation, and we will assume throughout the rest of this thesis that we are using such a parameterisation unless otherwise stated. 
Choosing some affine parameter \( \lambda \), our tangent vector is simply \( T^a = \dd x^a/\dd \lambda \), and so using equation \eqref{eq:cov-diff-contravec} we can re-write equation \eqref{eq:def-geodesic} as 
\begin{equation}
\dv[2]{x^a}{\lambda} + \chris{a}{bc}\dv{x^b}{\lambda}\dv{x^c}{\lambda} = 0.
\label{eq:geodesic-eq}
\end{equation}
This equation is commonly referred to as the geodesic equation, and any curves \( x^a(\lambda) \) which satisfy it are called geodesics.

\sectionmark{Equivalence principle and field equations}
\section[Einstein's equivalence principle and field equations]{Einstein's equivalence principle and field\\ equations}\label{sec:field-eqs-equiv-principle}
\sectionmark{Equivalence principle and field equations}

We now have constructed the mathematical framework necessary to understand Einstein's general theory of relativity.
However, as with all physical theories, we have to start with a set of fundamental axioms, or ``laws'' of physics, on top of which we can apply our mathematical framework to obtain the fundamental equations of motion.
It is worth noting again that, as with all ``laws'' of physics, these axioms are just a summary of very good experimental evidence. 
Using the equations of motion, we can then make predictions about how the universe behaves. 
In Newtonian mechanics, we have Newton's three laws; in special relativity, we have Einstein's postulates; in general relativity, we have the equivalence principle.
In all cases, laws of physics are just statements which encode and summarise a lot of rigorous experimental evidence.
The situation is no different in general relativity.

\subsection{The equivalence principle}

The equivalence principle asserts that all freely falling particles follow the same trajectories independent of their internal composition \cite{Visser:lorentzian-wormholes}. 
In the language of Newtonian mechanics, this is equivalent to asserting that an objects gravitational mass (how the mass responds to-/generates a gravitational field) is identical to its inertial mass (how the mass `resists' changes to its velocity).
The assertion that gravitational mass and inertial mass are the same quantity is now an extremely accurately tested principle: In 1999 Baessler et al. experimentally verified this to around one part in \( 10^{13} \) \cite{Baessler:1999}. 
Although this was not so well experimentally verified at the time Einstein developed his theory, it was still commonly believed to be true by the physicists of the time.
Thus, Einstein decided to fundamentally encode the equivalence principle into his theory with the following postulates, commonly called the Einstein equivalence principle.

\begin{postulate}[Equivalence principle]
Given a spacetime \( (\M, \g) \), the gravitational field is represented by the Christoffel connection and free fall corresponds to geodesic motion. 
Furthermore, in the flat spacetime limit, the metric must reduce to the Minkowski metric \( \eta \) of special relativity.
\end{postulate}

The last of these two statements implies, that in suitably small local coordinate patches, we must be able to reproduce the laws of special relativity from the more general theory. 
Hence, we have the following hierarchy 
\begin{equation*}
\text{General relativity} \underset{\text{flat space}}{\longrightarrow} \text{Special relativity} \underset{\text{low energy}}{\longrightarrow} \text{Newtonian mechanics}.
\end{equation*}

\subsection[Einstein's field equations and the stress-energy tensor]{Einstein's field equations and the stress-energy\\ tensor}

In order for general relativity to be a complete physical theory, we need a set of equations of motion.
Like all field theories, it is desirable to derive the equations of motion via an action principle.
Therefore, consider the action
\begin{equation}
S = -\frac{1}{16\pi} \int_\M \sqrt{-g} \, (R - 2\Lambda) \, \dd[4]x + \int_\M \sqrt{-g} \, \LL_M \, \dd[4]x,
\label{eq:EH-action-full}
\end{equation}
where \( \Lambda \) is a constant (usually called the ``cosmological'' constant), \( g = \det(g_{ab}) \), and \( \LL_M \) is a Lagrangian which is related to the matter-content (or stress-energy) of the spacetime.
Generally, one may write
\begin{equation}
S = S_{gr} + S_M,
\end{equation}
highlighting the gravitational and matter-content contributions of a given spacetime to the general action.
The action \eqref{eq:EH-action-full} has been accredited to David Hilbert, who published it less than a week after Einstein published his own formulation of the field equations \cite{Hilbert:1915}.
As such, it is typically referred to as the Einstein-Hilbert action\footnote{Traditionally, the Einstein-Hilbert action is only defined to be the \( (\sqrt{-g} \, R) \) term in the action presented above. However, including the other terms allows for the derivation of the most general form of the Einstein equations.}.
Varying the first integral in the action yields
\begin{equation}
\delta S = -\frac{1}{16\pi} \int_\M \bigg[ \delta\big(\sqrt{-g}\,\big) \, (R - 2\Lambda) + \sqrt{-g}\,\delta R \bigg] \dd[4]x .
\label{eq:vary-EHa1}
\end{equation}
The variation of \( \sqrt{-g} \) is a standard result \cite{Hawking-Ellis, Wald:general-relativity}:
\begin{equation}
\int_\M \delta(\sqrt{-g}) \dd[4]x = \int_M -\frac{1}{2}\sqrt{-g} \, g_{ab} \, \delta g^{ab} \, \dd[4]x.
\label{eq:vary-g}
\end{equation}
However, the variation of the Ricci scalar is a little more subtle:
\begin{align}
\int_\M \delta R \dd[4]x &= \int_\M \delta(g^{ab}R_{ab}) \dd[4]x = \int_\M \bigg[ \delta(g^{ab})R_{ab} + g^{ab} \delta(R_{ab}) \bigg] \dd[4]x 	\notag \\
&= \int_\M \bigg[ \delta(g^{ab})R_{ab} + \nabla^a v_a \bigg] \dd[4]x,
\label{eq:vary-R}
\end{align}
where \( v_a = \nabla^b \delta(g_{ab}) - g^{cd} \nabla_a \delta(g_{cd}) \) \cite{Wald:general-relativity}.
Thus, equation \eqref{eq:vary-EHa1} can be written as
\begin{multline}
\delta S = -\frac{1}{16\pi} \int_\M \sqrt{-g} \, \bigg[ R_{ab} - \frac{1}{2} R \, g_{ab} + g_{ab} \, \Lambda \bigg] \delta(g^{ab}) \, \dd[4]x \\
-\frac{1}{16\pi} \int_\M \nabla^a v_a \sqrt{-g}\,\dd[4]x.
\label{eq:vary-EHa2}
\end{multline}
But the second integral is just the integral of the divergence \( \nabla^a v_a \) with respect to the natural volume element \( (\sqrt{-g}\,\dd[4]x) \), and so by Stokes' theorem contributes only a boundary term -- the Gibbons--Hawking surface term \cite{Visser:lorentzian-wormholes}:
\begin{equation}
-\frac{1}{16\pi} \int_\M \nabla^a v_a \sqrt{-g}\,\dd[4]x = \delta\left( \frac{1}{8\pi} \int_{\partial\M} K \sqrt{^3g}\,\dd[3]x \right),
\label{eq:Gibbons-Hawking}
\end{equation}
where \( K = K\indices{^a_a} \) is the trace of the extrinsic-curvature of the spacetime, and \( ^3g \) is the determinant of the induced three-metric.
It is standard to ignore this term in the derivation of the equations of motion, as one can just simply subtract it from the original action presented in equation \eqref{eq:EH-action-full}.
Furthermore, it will be identically zero for variations where \( g^{ab} \) and its derivatives are held fixed on the boundary \cite{Wald:general-relativity}. 
Hence, moving forward, we will ignore this term in the derivation of the equations of motion. 

We have not yet considered the contribution of the matter-content action \( S_M \).
To do this, we simply define a \( 4\times4 \) type \( (0,2) \) tensor \( T_{ab} \) by
\begin{equation}
T_{ab} \equiv - \frac{1}{8\pi} \frac{1}{\sqrt{-g}} \frac{\delta S_M}{\delta g^{ab}},
\label{eq:def-Tab}
\end{equation}
where, as above, the action \( S_M \) represents the matter-content of the spacetime:
\begin{equation}
S_M = \int_\M \sqrt{-g} \, \LL_M \, \dd[4]x.
\end{equation}
As such, the tensor \( T_{ab} \) is referred to as the stress-energy tensor of the spacetime (sometimes called the `stress-energy-momentum' tensor).
Its components are related to the matter in the spacetime by
\begin{equation}
T_{ab} = \left(
    \begin{array}{c | c}
        \rho & S_j \\
        \hline
        S_i & \pi_{ij}
    \end{array}
\right),
\label{eq:Tab-compts}
\end{equation}
where \( \rho \) is the energy density, \( S_i \) is the energy-flux, and \( \pi_{ij} \) is the stress (\( i,j\in\{1,2,3\} \)).
Typically, \( S_i \) is considered a generalisation of the Poynting vector and \( \pi_{ij} \) is considered a generalisation of the notion of pressure \cite{MTW:gravitation, Visser:lorentzian-wormholes}.
In an orthonormal frame, all of the components have dimension [energy/volume].

Finally, requiring that the action \eqref{eq:EH-action-full} be stationary under variations with respect to the metric, \( \delta S/\delta g^{ab} = 0 \), we obtain the equations of motion for a general spacetime in general relativity (the Einstein equations):
\begin{equation}
R_{ab} - \frac{1}{2}Rg_{ab} + g_{ab}\Lambda = 8\pi T_{ab}.
\label{eq:Einstein-eqs-full}
\end{equation}
In most instances, the cosmological constant is set to zero as the latest (2021) experimental evidence seems to suggest that it is a very small number (roughly \( 10^{-52} \, \text{m}^{-2} \) \cite{Planck:2020, particle-data}).
In such instances, the Einstein equations are commonly written in terms of the Einstein tensor \( G_{ab} \) as\footnote{In SI units: \( G_{ab} = (8\pi G_N/c^2) T_{ab} \)}.
\begin{equation}
G_{ab} = 8\pi T_{ab}.
\label{eq:E-eqs}
\end{equation}
Even in cases where the cosmological constant is not set to zero, one can re-define the stress energy tensor as \( (T_{ab})_{\text{new}} = (T_{ab})_{\text{old}} - \Lambda g_{ab}/8\pi \), thereby writing the full Einstein equations \eqref{eq:Einstein-eqs-full} in the form of equation \eqref{eq:E-eqs}.

Although it may look simple, equation \eqref{eq:E-eqs} actually constitutes a system of ten, coupled, non-linear, second order, partial differential equations. 
As such, finding exact solutions to the Einstein equations is, in general, a very hard task.
Historically, the Einstein equations have only been solved in situations involving high degrees of mathematical symmetry \cite{Kramer:1981}.
For example, the Schwarzschild solution is the unique, non-rotating (static), time-independent (stationary), spherically symmetric solution; whilst the Kerr solution is a rotating (non-static), time-independent (stationary), solution with azimuthal symmetry.
The Schwarzschild solution was discovered very shortly after Einstein published his theory\footnote{Although, it took another 40 odd years before it was realised that the solution represented a black hole spacetime (to be defined in chapter \ref{C:BH-mimickers}).} \cite{Schwarzschild:1916}, whilst the Kerr solution was not discovered until around 50 years later \cite{Kerr:1963}.

\subsection{Energy conditions}\label{subsubsec:energy}

In the case of a spherically symmetric spacetime, the stress-energy tensor of equation \eqref{eq:Tab-compts} takes the form \cite{Visser:lorentzian-wormholes}
\begin{equation}
T_{\hat{a}\hat{b}} =
    \begin{pmatrix}
    \rho & 0 & 0 & 0 \\
    0 & p_r & 0 & 0 \\
    0 & 0 & p_t & 0 \\
    0 & 0 & 0 & p_t 
    \end{pmatrix}_{\hat{a}\hat{b}},
\label{eq:Tab-spherical}
\end{equation}
where \( \rho \) is the energy density, \( p_r \) is radial pressure, \( p_t \) denotes the transverse pressure, and hats on the indices denote components in an orthonormal basis.

It is commonly believed all physically reasonable classical matter will satisfy a set of seven equations, known as the energy conditions.
These conditions are: the null, weak, strong, and dominant energy conditions (NEC, WEC, SEC, DEC); as well as the averaged null, weak, and strong energy conditions.
In this thesis, we will primarily be concerned with the four non-averaged energy conditions. 

\subsubsection{Null energy condition}
The null energy condition is the statement that
\begin{equation}
T_{ab}k^ak^b\geq0,
\label{eq:nec-def}
\end{equation}
for any \emph{null} vector \( k^a \).
In terms of the components of the stress-energy tensor \eqref{eq:Tab-spherical}, this is the statement that
\begin{equation}
\text{NEC} \iff \rho+p_r\geq0 \qq{and} \rho+p_t\geq0.
\label{eq:nec-compts}
\end{equation}

\subsubsection{Weak energy condition}
The weak energy condition asserts that
\begin{equation}
T_{ab}V^aV^b\geq0,
\label{eq:wec-def}
\end{equation}
for any \emph{timelike} vector \( V^a \).
In terms of the components of the stress-energy tensor \eqref{eq:Tab-spherical}:
\begin{equation}
\text{WEC} \iff \rho\geq0 \qq{and} \text{NEC}.
\label{eq:wec-compts}
\end{equation}
Thus, satisfaction of the WEC will imply that the NEC holds also.
Equivalently, if the NEC fails to hold, then so will the WEC.
Physically, the WEC is the statement that the local energy density as measured by a timelike observer must be non-negative. 

\subsubsection{Strong energy condition}
The strong energy condition is the statement that
\begin{equation}
\left(T_{ab} - \frac{T}{2}g_{ab}\right) V^aV^b \geq 0,
\label{eq:sec-def}
\end{equation}
for any \emph{timelike} vector \( V^a \), where \( T = T\indices{^a_a} = -\rho+p_r+2p_t \) is the trace of the stress-energy tensor.
In terms of the components of the stress-energy tensor \eqref{eq:Tab-spherical}:
\begin{equation}
\text{SEC} \iff \rho+p_r+2p_t\geq0 \qq{and} \text{NEC}.
\label{eq:sec-compts}
\end{equation}
Thus, satisfaction of the SEC will automatically imply satisfaction of the NEC, but not necessarily the WEC.
Equivalently, if the NEC fails to hold, then so will the SEC.

\subsubsection{Dominant energy condition}
The dominant energy condition asserts that for any \emph{timelike} vector \( V^a \),
\begin{equation}
T_{ab}V^aV^b\geq0, \qq{and \( T_{ab}V^a \) is a non-spacelike vector.}
\label{eq:dec-def}
\end{equation}
In terms of the components of the stress-energy tensor \eqref{eq:Tab-spherical}:
\begin{equation}
\text{DEC} \iff \rho\geq0 \qq{and} p_r,p_t \in [-\rho,+\rho].
\label{eq:dec-compts}
\end{equation}
Physically, this is the statement that the local energy density appears non-negative \emph{and} that the energy flux is timelike or null.
Note that satisfaction of the DEC automatically implies satisfaction of the WEC and the NEC, but not necessarily the SEC.
Equivalently, if the NEC \emph{or} the WEC fails to hold, then so will the DEC.

The interested reader can find a discussion of how the various energy conditions are related to the singularity theorems in Ref. \cite[pp. 118--119]{Visser:lorentzian-wormholes}.
A somewhat non-technical discussion regarding the validity of the energy conditions outside of general relativity is given in Ref. \cite{Barcelo:2002bv}.
\chapter[Black hole mimickers: extensions to Einstein's theory]{Black hole mimickers: extensions to Einstein's theory}\label{C:BH-mimickers}
\chaptermark{Black hole mimickers}

In this chapter, we discuss what is considered to be one of the main issues with classical general relativity -- the prediction of spacetime singularities.
We then discuss how black hole mimickers may provide a classical resolution in the specific case of black hole singularities, and then give some examples of the different types of black hole mimickers commonly studied in the literature and in this thesis.

\section{Singularities}

As we saw in chapter \ref{C:GR-maths-background}, if we have a spacetime \( (\M,\g) \), we can calculate various curvature tensors which provide insight into the geometrical nature of the manifold.
For example, consider the line element for the Schwarzschild spacetime in curvature coordinates
\begin{equation}
\dd s^2 = -\left(1-\frac{2m}{r}\right)\dd t^2 + \frac{\dd r^2}{1-\frac{2m}{r}} + r^2 \big( \dd\theta^2 + \sin^2\theta \dd\phi^2 \big).
\label{eq:sch-metric}
\end{equation}
The Kretschmann scalar for this spacetime is 
\begin{equation}
K \equiv R^{abcd}R_{abcd} = \frac{48m^2}{r^6},
\label{eq:R-sch}
\end{equation}
and so \( K\rightarrow\infty \) as \( r\rightarrow0 \).
That is, the Kretschmann scalar is singular (in the mathematical sense) at \( r=0 \).
In this instance, one would say that there is a curvature singularity at \( r=0 \) in the Schwarzschild spacetime. 

Note, however, that the although the metric presented in equation \eqref{eq:sch-metric} is singular at \( r=2m \), this is purely a result of the coordinate system we are using.
That is, a suitable change of coordinate system would remove the singular nature of the metric components at \( r=2m \)\footnote{For example, the Painlev\'e--Gullstrand coordinates (details may be found in references \cite{Lense-Thirring, MTW:gravitation}. See also reference \cite{Wald:general-relativity}).}.
Thus, this type of singular nature of the metric is typically referred to as a ``coordinate artefact'', and is not considered to be a true spacetime singularity. 
That being said, defining precisely what one means by a `true' spacetime singularity is in fact a very difficult task.

Naively, one may wish to define a singularity as a point in the spacetime where at least one curvature invariant diverges to infinity. 
However, one could simply remove such a point from the spacetime manifold and then claim that the resulting spacetime is non-singular.
Thus, the issue defining whether or not a spacetime has a singularity has now become a problem of determining whether or not any singular points have been removed.
The notion of whether or not a space has any `holes' in it is easily formalised for Riemannian manifolds.

Consider a manifold \( \M \) and take a curve \( \gamma:[a,b]\subset\RR \rightarrow \M \).
Intuitively, one can see how the point \( p = \gamma(b) \) could be considered an endpoint of the curve \( \gamma \), and how in a `space without holes' we would want any endpoints of a curve to be contained in the space itself.
We can formalise the notion of the endpoint of a curve \( \gamma:I\rightarrow\M \) for metric spaces (i.e. Riemannian manifolds) with the following definition \cite{Hawking-Ellis}.

\begin{definition}\label{def:endpoint}
A point \( p \) in a manifold \( \M \) is said to be an endpoint of a curve \( \gamma:I\rightarrow\M \) if for every neighbourhood \( U \) of \( p \), there is a \( \lambda\in I \) such that \( \gamma(\lambda_1)\in U \) for every \( \lambda_1\in I \) with \( \lambda_1\geq\lambda \).
\end{definition}

This allows us to formalise what we mean by a `space without holes'.

\begin{definition}\label{def:complete-space}
A Riemannian manifold \( \M \) is said to be metrically complete if every \( C^1 \) curve of finite length has an endpoint.
Furthermore, it is said to be geodesically complete if every geodesic can be extended to arbitrary values in its affine parameter. 
\end{definition}

For the case of Lorentzian metrics, one cannot define a metric space, and so neither a notion of metric completeness. 
However, as Lorentzian metrics admit the construction of geodesics, geodesic completeness can still be defined. 
Lorentzian metrics further allow the classification of timelike, null, or spacelike geodesic completeness depending on the sign of the norm of its tangent vector.
If a timelike geodesic is incomplete, this would imply the possibility of a physical observer whose history does not exist after, or before, a finite amount of proper time, which is certainly a physically objectionable property of a (classical) spacetime.
Thus, we will say that \textit{if a spacetime is either timelike or null geodesically incomplete, it contains a singularity}.
Note that we do not require the converse statement to be true, as there are examples of spacetimes, such as the one constructed by Geroch \cite{Geroch:1968}, which exhibit singularity-like properties but are still geodesically complete.
As such, some authors now require a more general condition for a spacetime to be considered singularity-free \cite{Hawking-Ellis}.

\begin{definition}\label{def:b-complete}
A spacetime \( (\M,\g) \) is said to be bundle complete if every affinely-parameterised \( C^1 \) curve of finite length has an endpoint.
\end{definition}

\begin{definition}\label{def:singularity}
A spacetime is said to be singularity free, or contain no singularities, if it is bundle complete.
\end{definition}

We will adopt definition \ref{def:singularity} as our formal definition of a singularity.
However, most of the time geodesic completeness is enough to classify singularities, and bundle completeness is only needed for very technical reasons.
The interested reader may find more details in any of \cite{Geroch:1968, Hawking-Ellis, Wald:general-relativity}.

Defining singular spacetimes by the presence of incomplete curves of certain classes is necessary for proving the singularity theorems (discussed below), but does not provide us with any information about the location or type of singularity present in the spacetime. 
However, all spacetimes which contain curvature singularities are definitely bundle (and hence geodesic) incomplete.
Thus, in practice, one would usually look for singular points in the curvature invariants of a particular spacetime of interest.

In the following chapters, we consider a type of singular spacetime known as a black hole spacetime.
In these spacetimes it will be clear where the singularity is located by analysing the curvature invariants.
In fact, if we know we have a black hole spacetime, or the weaker condition of a trapped surface, then in classical general relativity we can prove that such a spacetime must necessarily contain a singularity. 
This is known as the Penrose singularity theorem.
We only present a simplified version of the theorem below, as the full details are not directly relevant to the main work in this thesis.

\begin{theorem}\label{theorem:Penrose-sing}
A spacetime \( (\M,\g) \) cannot be null geodesically complete if 
    \begin{enumerate}
    \item[(1)] \( R_{ab}k^ak^b \geq0 \) for any null vector \( k^a \);
    \item[(2)] there is a Cauchy surface \( \HH \) in \( \M \);
    \item[(3)] \( \M \) contains a closed trapped surface \( \T \).
    \end{enumerate}
\end{theorem}

Condition (1) is known as the \textit{null convergence condition}, and is implied by the weak energy condition.
Thus, a sufficient condition for it to hold is that the energy density of the spacetime is positive for any observer (a physically reasonable assumption for macroscopic spacetimes). 

A Cauchy surface \( \HH \) is a subset of the manifold \( \M \) which is intersected \emph{exactly once} by every differentiable timelike curve in \( \M \).
There are spacetimes which satisfy conditions (1) and (3) which do not contain singularities, and as such condition (2) must be included.
The technical reasons for as to why the spacetime must contain a Cauchy surface are not germane to the theme of this thesis, and are best understood by completing the proof of the theorem.
Details of the proof may be found in \cite[pp. 263--265]{Hawking-Ellis}. 

Formally, a closed trapped surface is a closed, compact, spacelike two-surface without boundary, such that the two families of null geodesics orthogonal to \( \T \) are converging at \( \T \) \cite{Hawking-Ellis, MTW:gravitation, Wald:general-relativity}.
Intuitively, one may view \( \T \) as being the surface at which even outgoing null geodesics are dragged back inwards and forced to converge.
As nothing physical can travel faster than light, any matter inside \( \T \) will be trapped inside a succession of two-surfaces of ever decreasing area.
Eventually, we will get to a situation where we a forcing a massive object into a region so small that it can no longer be considered physically reasonable. 
This apparent breakdown of the theory is neatly captured by Penrose's singularity theorem.
Note that theorem \ref{theorem:Penrose-sing} implicitly assumes that the spacetime metric \( \g \) is an exact solution to the Einstein equations \eqref{eq:Einstein-eqs-full}, and so is strictly a theorem within the framework of classical general relativity. 

Perhaps the most important class of spacetimes containing trapped surfaces is the black hole spacetimes.
The formal definition of a black hole spacetime is very technical, and will not be necessary for the work in this thesis.
As such, we will use a simplified definition of a black hole spacetime which contains all of the ingredients necessary for the work in the following chapters.
The definition used here follows that of \cite{Visser:lorentzian-wormholes}.

\begin{definition}\label{def:BH}
For each asymptotically flat region, the associated future/past event horizon is defined as the boundary of the region from which causal curves can reach asymptotic future/past null infinity.
Furthermore, we define a black hole to be the region contained inside an event horizon, and say that a spacetime containing a black hole is a black hole spacetime.
\end{definition}

Note the important fact: \emph{an event horizon is a trapped surface}.
As such, any black hole spacetime satisfying conditions (1) and (2) in theorem \ref{theorem:Penrose-sing} will necessarily contain a singularity. 

\begin{figure}[t]
\centering
    \includegraphics[scale=0.45]{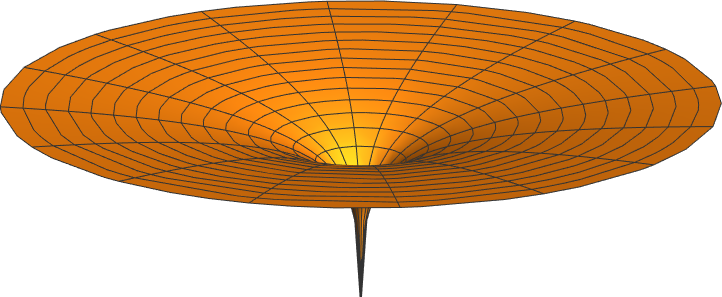}
    \caption[Black hole.]{A black hole spacetime embedded as a two-surface in four-dimensional spacetime. The black hole contains a curvature singularity at its core.}
\end{figure}

From a purely mathematical point of view, the fact the black holes contain singularities causes no issues.
However, as black holes are valid solutions to the Einstein equations, and furthermore they can be shown to form by real, physical stellar collapse models, this presents a real issue from a physical standpoint.
In spite of this, most of the time singularities are not an issue for every-day physics as they are `hidden' behind an event horizon, and so physics outside of the black hole is unaffected. 
However, there are cases where singularities do become an issue, such as during the very last stages of a black hole's evaporation -- a time that plays an important role in determining the compatibility of quantum mechanics and general relativity.
Many researchers now believe that a consistent theory of quantum gravity will invoke sufficient changes to black holes at the Planck scale, which will stop singularities forming \cite{Carballo-Rubio:2018pmi, Carballo-Rubio:2018jzw, Carballo-Rubio:2019fnb, Carballo-Rubio:2019nel}.
However, it is unlikely that such a theory will be achieved any time in the near future, and as such it is useful to consider classical modifications to black hole spacetimes which remove any singularities which may be present. 
If such a modified spacetime mimics the observable qualities of the singular black holes of classical general relativity, we will call it a black hole mimicker.
Note that as black hole mimickers are non-singular geometries, by Penrose's theorem they can not be exact solutions to the vacuum Einstein equations.
However, we can always assume that they satisfy the non-vacuum equations, and then calculate what the required stress-energy tensor will be for the non-singular spacetime.

We will now move on to discuss various types of black hole mimickers, specific models of which will be analysed in subsequent chapters. 
Note that this is not intended to be an exhaustive list; there are types of black hole mimickers that we will not discuss in this thesis (for example, the Mazur--Mottola gravastars \cite{gravastar-original, gravastars-bh-alternative} and subsequent refinements \cite{anisotropic, lobo-gravastars, Lobo:2015, MartinMoruno:2011, visser-gravastars}).
We will, however, discuss in some detail regular black holes and traversable wormholes.

\section{Regular black holes}

The first class of black hole mimickers we will discuss is the so-called regular black holes.
A regular black hole is a black hole (in that it has a well-defined event horizon) that does not contain a singularity at its core.
In this sense, one may say that the singularity has been ``regularised''.

Regular black holes were first suggested as alternatives to black holes by Bardeen in 1968, where he advocated the metric (now known as the Bardeen regular black hole) \cite{Bardeen:1968}
\begin{equation}
\dd s^2 = -f(r)\dd{t^2} + \frac{\dd{r^2}}{f(r)} + r^2\dd{\Omega}^2, \quad f(r) = 1-\frac{2mr^2}{\big(r^2+q^2\big)^{3/2}}.
\label{eq:bardeen-rbh}
\end{equation}
Here \( \dd\Omega^2 = \dd\theta^2+\sin^2\theta\,\dd\phi^2 \) is the infinitesimal solid angle in spherical symmetry, and \( q = (2ml^2)^{1/3} \) where \( l \) is a length scale typically associated with the Planck length.
As \( l \) is such a small number, \( l \approx 1.616\times10^{-35}m \) \cite{particle-data}, the Bardeen spacetime was devised to be a regular spacetime which differs only minimally from the Schwarzschild spacetime (c.f. equation \eqref{eq:sch-metric}).
A simple (but tedious) calculation verifies that this is indeed a black hole spacetime in the sense of definition \ref{def:BH}, with a horizon characterised by the location at which \( g^{ab}\nabla_ar\nabla_br = g^{rr} = f(r) \) changes sign.

One can easily check that all curvature invariants remain finite everywhere in the spacetime\footnote{This is best done with computer algebra software packages. Commonly used packages include \text{Maple}, \text{Mathematica}, and \text{Sage Math}.}, but we will only present the calculation of the Ricci scalar for purposes of brevity:
\begin{equation}
R = \frac{6mq^2(4q^2-r^2)}{(r^2+q^2)^{7/2}}.
\label{eq:RicciS-Bardeen}
\end{equation}
Note that as \( r\rightarrow0 \) we have \( R\rightarrow24m/q^3 = 12/l^2 \), and so the spacetime is indeed regular.
Also worthy of note is that in the \( r\rightarrow0 \) limit we have \( R \rightarrow \O(l^{-2}) \), that is, the regular corrections are on order of magnitude of the Planck scale.

Since this is a black hole spacetime without a singularity, it must have a non-zero stress-energy tensor (this can easily be seen by that fact that \( R\neq0 \)).
Indeed, the stress-energy tensor is found by the non-vacuum Einstein equations \eqref{eq:E-eqs} to have components
\begin{subequations}
    \begin{equation}
    T\indices{^t_t} = T\indices{^r_r} =  -\frac{6mq^2}{8\pi(r^2+q^2)^{5/2}},
    \label{eq:Ttt-Bardeen}
    \end{equation}
    \begin{equation}
    T\indices{^\theta_\theta} = T\indices{^\phi_\phi} = \frac{3mq^2(3r^2-2q^2)}{8\pi(r^2+q^2)^{7/2}}.
    \label{eq:Tthth-Bardeen}
    \end{equation}
\label{eq:Tab-Bardeen}
\end{subequations}

As we are working in spherical symmetry, the stress-energy tensor has the form of equation \eqref{eq:Tab-spherical}, and we can write \( T\indices{^a_b} = \diag(-\rho, \, p_r, \, p_t, \, p_t)\indices{^a_b} \).
Note that \emph{inside} the horizon, \( \rho \) and \( p_r \) switch places, and so the stress-energy tensor has the form \( T\indices{^a_b} = \diag(p_r, \, -\rho, \, p_t, \, p_t)\indices{^a_b} \).
Thus, outside the horizon, we have
\begin{subequations}
    \begin{equation}
    \rho  = -p_r = \frac{6mq^2}{8\pi(r^2+q^2)^{5/2}},
    \label{eq:rho-pr-Bardeen}
    \end{equation}
whilst
    \begin{equation}
    p_t = \frac{3mq^2(3r^2-2q^2)}{8\pi(r^2+q^2)^{7/2}}.
    \label{eq:pt-Bardeen}
    \end{equation}
\label{eq:Tab-compts-Bardeen}
\end{subequations}
This then implies that \( \rho + p_r = 0 \), whilst 
\begin{equation}
\rho+p_t = \frac{6mq^2}{8\pi(r^2+q^2)^{5/2}} + \frac{3mq^2(3r^2-2q^2)}{8\pi(r^2+q^2)^{7/2}} = \frac{15mq^2r^2}{8\pi(r^2+q^2)^{7/2}} \geq 0.
\label{eq:NEC-Bardeen}
\end{equation}
Thus, we can conclude from equation \eqref{eq:nec-compts} that the NEC is satisfied for the Bardeen regular black hole.
Similarly, since \( \rho\geq0 \), we can conclude that the WEC is globally satisfied (i.e. the local energy density as measured by a timelike observer will be non-negative).
The SEC, however, is a little more subtle.
We have that
\begin{equation}
\rho+p_r+2p_t = \frac{3mq^2(3r^2-2q^2)}{4\pi(r^2+q^2)^{7/2}},
\label{eq:sec-Bardeen}
\end{equation}
and so from equation \eqref{eq:sec-compts}, we can conclude that the SEC will only be satisfied in the region
\begin{equation}
r\geq\frac{2}{3}q = \frac{2}{3}(2ml^2)^{1/3}.
\label{eq:sec-Bardeen-satisfy}
\end{equation}

This is a common theme that we will see with black hole mimickers: \textit{we can force a regular black hole metric to satisfy the Einstein equations by constructing the relevant stress-energy tensor, however said stress-energy tensor does not globally satisfy all of the classical energy conditions.}
This has lead to much discussion regarding the physical viability of black hole mimickers \cite{Berry:2020tky, Berry:2020ntz, Carballo-Rubio:2018pmi, Carballo-Rubio:2018jzw, Carballo-Rubio:2019fnb, Carballo-Rubio:2019nel}.
All black hole mimickers studied in this thesis will fail to globally satisfy at least one of the classical energy conditions.
This is not necessarily an issue, however, as there are a number of physical systems from a wide range of areas of physics which violate one or more of the energy conditions. 
Perhaps the best well-known instance of this is the Casimir effect, in which a physical system is shown to have negative energy density, thereby violating the WEC and DEC.
That being said, the scales at which the Casimir effect is detectable induces a negative energy density orders of magnitude smaller than what would be necessary to form, or stop the collapse of, most black hole mimickers.
The interested reader can find a discussion of the Casimir effect and its relation to the energy conditions in Ref. \cite[pp. 121--126]{Visser:lorentzian-wormholes} (see also the original paper by Casimir \cite{Casimir:1948dh}, and the technical references \cite{Birrell-Davies-1982, Blau:1988kv}).

\section{Wormholes}

The next class of black hole mimicker we will introduce is the (traversable) wormhole. 
Wormholes have a long history in general relativity: within only a year of Einstein finalising his formulation of the field equations, Flamm published an article hinting of possible spacetime ``shortcuts'' \cite{Flamm:1916}.
From here, wormholes were then investigated as solutions to the Einstein equations by Weyl in the 1920s \cite{Weyl:1949}, Einstein and Rosen in the 1930s \cite{Einstein:1935}, and by Wheeler in the 1950s \cite{Wheeler:1962}.
After this, though, the field lay relatively dormant for close to 30 yers until the wormhole `renaissance' in the late 1980s with the seminal papers by Morris, Thorne, and Yurtsever \cite{Morris:1988cz, Morris:1988tu}.
This lead to rapid investigation into wormhole physics with many metrics of interest being investigated \cite{visser-wormhole-examples, baby}, and new ways of `constructing' wormholes being discovered \cite{poisson-visser, visser-surgical}.
Since then, the amount of research into wormholes has slowed, but there is still a consistent and significant flow of wormhole physics papers being published \cite{Berry:2020tky, expmetric, Dadhich:2002, Lobo:2020, blackbounce, visser-kar-dadhich}.

A particularly lucid definition of a traversable wormhole spacetime is provided by Morris and Thorne\footnote{Morris and Thorne actually provide many more conditions than just the ones presented in definition \ref{def:wormhole}. However, the extra conditions they provide are primarily concerned with making the wormhole physically traversable for a real, living human being -- something we will not worry about in this thesis.} \cite{Morris:1988cz}.

\begin{definition}\label{def:wormhole}
A spacetime is said to contain a traversable wormhole if it satisfies the following four conditions:
    \begin{enumerate}
    \item[(1)] The metric must satisfy the Einstein field equations \eqref{eq:Einstein-eqs-full} for some non-zero stress-energy tensor.
    \item[(2)] The spacetime must contain a region known as the `throat'. The throat region must satisfy two conditions:
        \begin{enumerate}
        \item[(i)] it must connect two asymptotically flat regions of spacetime;
        \item[(ii)] it must satisfy the `flare-out' condition: the areas of the induced spatial hypersurfaces on either side of the throat must be strictly increasing functions of the distance from the throat.
        \end{enumerate}
    \item[(3)] The spacetime cannot contain any horizons.
    \item[(4)] The spacetime cannot contain any singularities.
    \end{enumerate}
\end{definition}

\begin{figure}[t]
\centering
    \includegraphics[scale=0.45]{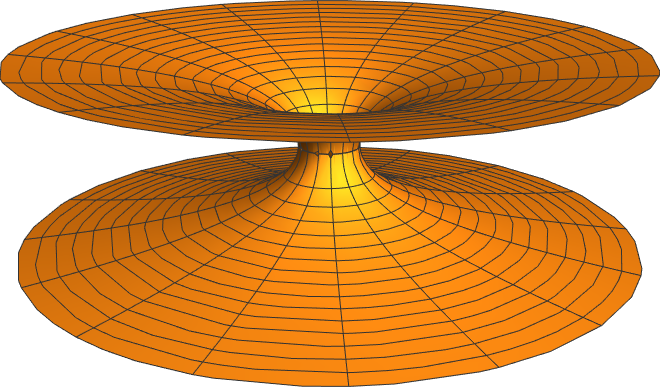}
    \caption[Inter-universe wormhole.]{An inter-universe wormhole embedded as a two-surface in four-dimensional spacetime.}
\label{fig:wormhole}
\end{figure}

Condition (1) simply ensures that we stick to the confines of classical general relativity.
As with regular black holes, in practice one will typically specify a metric that satisfies conditions (2)--(4) then use the Einstein equations \eqref{eq:Einstein-eqs-full} or \eqref{eq:E-eqs} to calculate the necessary stress-energy tensor required to support such a geometry. 
Condition (2) specifies the geometry of the wormhole spacetime. 
This agrees with our intuitive notion of a wormhole as a ``tunnel'' between two regions of spacetime (c.f. figure \ref{fig:wormhole}).
Conditions (3) and (4) enforce the `traversability' of the spacetime. 
If the spacetime contained a horizon, only one-way propagation of timelike observer would be allowed (anytime timelike or null geodesic would be dragged back inwards towards the wormhole throat).
If the spacetime contained a singularity, the immense tidal forces it would produce would be sure to destroy anything physical.
Furthermore, the singularity would cause the throat to be highly unstable and collapse in on itself in a very short amount of time \cite{Morris:1988cz}.
Note that in spherical symmetry, conditions (2) and (3) imply condition (4).

Although traversable wormholes do not contain horizons, we will still consider them to be black hole mimickers. 
The reason for this is two-fold:
\begin{enumerate}
    \item[(1)] By definition, wormhole geometries do not contain singularities. As a result, they rectify the physical concerns that result from singular black hole spacetimes.
    \item[(2)] We can construct the spacetime so that the wormhole throat lies arbitrarily close to any would-be horizons in a corresponding black hole geometry (see chapter \blue{\ref{C:thin-shell-wormhole}}). Thus, it would be hard for any astronomical observations to distinguish between the two objects (the wormhole is ``mimicking'' the black hole in an observational sense).
\end{enumerate}

The canonical example of a wormhole spacetime is the Morris-Thorne traversable wormhole \cite{Morris:1988cz, Morris:1988tu}.
In curvature coordinates, this spacetime has line element 
\begin{equation}
\dd s^2 = -\dd t^2 + \dd r^2 + (r^2+b^2)(\dd\theta^2+\sin^2\theta\dd\phi^2),
\label{eq:M-T-wormhole}
\end{equation}
where \( b \) is a constant with dimensions of length.
One can easily check that this spacetime satisfies conditions (2)--(4) in definition \ref{def:wormhole}, and represents a static, spherically symmetric spacetime.
Requiring that it be a traversable wormhole solution to the Einstein field equations with \( \Lambda=0 \) (i.e. condition (1)), forces the stress-energy tensor to have components 
\begin{equation}
\rho = p_r = -p_t = -\frac{b^2}{8\pi(r^2+b^2)^2}.
\label{eq:Tab-M-T-wormhole}
\end{equation}
Thus, 
\begin{equation}
\rho+p_r = -\frac{b^2}{4\pi(r^2+b^2)^2} < 0,
\label{eq:NEC-M-T-wormhole}
\end{equation}
and so the NEC is violated globally in the Morris-Thorne wormhole spacetime.
This guarantees that the WEC, SEC, and the DEC will also be similarly violated globally.
Violation of the energy conditions is a common theme amongst wormhole geometries, and is one of the reasons why physicists question their physical validity \cite{Kar:2004, Visser:lorentzian-wormholes, visser-kar-dadhich}.

\enlargethispage{10pt}
Wormholes also bring to light many questions regarding causality and time travel \cite{Hawking:1992, Morris:1988tu, baby, time-machine}.
If one were to construct (or discover) a traversable intra-universe wormhole (i.e. connecting two regions of the same universe), they could travel through the throat, along a specially chosen timelike geodesic, and arrive back before they ever left.
Many other paradoxes arise when one considers using wormholes as time travel machines \cite{Hawking:1992, Lobo:2003, Lobo:2010sz, Morris:1988cz, Visser:lorentzian-wormholes}, which is yet another reason why some physicists believe that they cannot possibly exist in nature.
This is perhaps best summarised by Hawking's chronology protection conjecture, which asserts that the laws of physics will conspire in such a way as to prevent using wormholes for time travel, except for perhaps on microscopic or quantum scales.
This is elegantly encapsulated by Hawking's famous quote \cite{Hawking:1992}: \textit{``It seems there is a chronology protection agency, which prevents the appearance of closed timelike curves and so makes the universe safe for historians.''}

Of course, there are other avenues of thought on the physical reality of wormholes/time travel.
One such example is Novikov's self-consistency conjecture.
This states that if time travel via wormholes is possible, then the laws of physics describing it will be such that the overall result is always consistent.
In their 1990 paper, Novikov et al. state \cite{Novikov:1990}: \textit{``...the only solutions to the laws of physics that can occur locally in the real Universe are those which are globally self-consistent.''}
This is particularly nice, as it does not conjecture any new law of physics that we have no experimental evidence for.
Instead, it simply claims that the universe will continue to behave in the same way that we have observed it to so far; we have no evidence that the universe is in any way internally inconsistent.
Novikov's conjecture also agrees with Hawking's (pseudo)-experiment \cite{Hawking:1992}: \textit{``we have not been invaded by hordes of tourists from the future.''}
\chapter{Regular black hole with asymptotically Minkowski core}\label{C:Mink-core}
\chaptermark{RBH with asymptotically Minkowski core}

In this chapter, we analyse a specific example of a regular black hole spacetime with an asymptotically Minkowski core.
We calculate the radius of the photon sphere and the extremal stable timelike circular orbit (ESCO), which are (at least, in theory) physically observable quantities. 
The manner in which the photon sphere and ESCO relate to the presence (or absence) of horizons is much more complex than for the Schwarzschild black hole. 
We find situations in which photon spheres can approach arbitrarily close to (near extremal) horizons, situations in which some photon spheres become stable, 
and situations in which the locations of both photon spheres and ESCOs become multi-valued, with both ISCOs (innermost stable circular orbits) and OSCOs (outermost stable circular orbits).  This provides an extremely rich phenomenology of potential astrophysical interest.

\section{Introducing the spacetime}

Without any loss of generality, any static spherically symmetric spacetime can be described by a metric of the form
\begin{equation}
\dd{s}^2 = - \e^{-2\Phi(r)}\left(1-\frac{2m(r)}{r}\right)\dd{t^2}+\frac{\dd{r^2}}{1-\frac{2m(r)}{r}}+r^2\left(\dd{\theta^2}+\sin^2\theta\dd{\phi^2}\right).
\end{equation}
For the standard Schwarzschild metric (equation \eqref{eq:sch-metric}) one sets \( \Phi(r)=0 \) and \( m(r)=m_0 \).
Over the past century, a vast host of black hole spacetimes, qualitatively distinct from that of Schwarzschild, have been investigated by multiple researchers~\cite{ Lense-Thirring, Baines:2020, kerr-original, kerr-schild, kerr-newmann, nordstrom-original, reissner-original, vaidya-original1, vaidya-original2, vaidya-original3, kerr-intro, weyl-original, kerr-book}.

Furthermore, the field has now grown to not only include classical black holes, but also quantum-modified black holes~\cite{quantum-bh3, quantum-aspects-of-bhs_book, quantum-bh1, quantum-bh2}, regular black holes~\cite{rbhs-review, Bardeen:1968, Carballo-Rubio:2018pmi, frolov-rbh, hayward-rbh}, and various other exotic spherically symmetric spacetimes that are fundamentally different from black holes but mimic many of their observable phenomena (\emph{e.g.} traversable wormholes~\cite{expmetric, natural, Dadhich:2002, Kar:2004, Lobo:2020, Morris:1988cz, Morris:1988tu, poisson-visser, blackbounce2, blackbounce, visser-surgical, visser-wormhole-examples, baby, time-machine, Visser:lorentzian-wormholes, visser-kar-dadhich}, gravastars~\cite{anisotropic, lobo-gravastars, Lobo:2015, MartinMoruno:2011, gravastars-bh-alternative, gravastar-original, visser-gravastars}, ultracompact objects~\cite{Cunha1,Cunha2}, \textit{etcetera}~\cite{Carballo-Rubio:2019nel, observability, small-dark-heavy}; see~\cite{Carballo-Rubio:2018jzw} for an in-depth discussion).

The model spacetime investigated in this work is a regular black hole with an asymptotically Minkowski core, as discussed in~\cite{Berry:2020tky, asymptot-mink-core}.
This is an example of a metric with an exponential mass suppression, and is described by the line element
\begin{equation}
\dd s^2 = -\left(1-\frac{2m\,\e^{-a/r}}{r}\right)\dd{t}^2 + \frac{\dd{r}^2}{1-\frac{2m\,\e^{-a/r}}{r}} + r^2\left(\dd{\theta}^2 + \sin^2\theta \dd{\phi}^2\right).
\label{metric}
\end{equation}
A rather different (extremal) version of this model spacetime, based on nonlinear electrodynamics, has been previously discussed by Culetu~\cite{Culetu:2013}, with follow-up on some aspects of the non-extremal case in references~\cite{Culetu:2015a, Culetu:2014, Culetu:2015b}. See also~\cite{Junior:2015, Rodrigues:2015}.

Most regular black holes have a core that is asymptotically de~Sitter (with constant positive curvature)~\cite{rbhs-review, Bardeen:1968, frolov-rbh, hayward-rbh}.
However, the regular black hole described by the metric \eqref{metric} has an asymptotically Minkowski core (in the sense that the stress-energy tensor asymptotes to zero).
This model has some attractive features compared to the more common de~Sitter core regular black holes: the stress-energy tensor vanishes at the core, greatly simplifying the physics in this region; and many messy algebraic expressions are replaced by simpler expressions involving the exponential and Lambert $W$ functions, whilst still allowing for explicit closed form expressions for quantities of physical interest~\cite{asymptot-mink-core}.
Additionally, the results obtained in this work reproduce the standard results for the Schwarzschild metric by letting the parameter \( a \rightarrow 0 \).
Thus, the value of the parameter \( a \) determines the extent of the ``deviation'' from the Schwarzschild spacetime.

If $0<a<2m/\e$ then the spacetime described by the metric \eqref{metric} has two horizons located at
\begin{equation}
r_{H^-} = 2m\;\e^{W_{-1}\left(-\frac{a}{2m}\right)}, 
\qquad\hbox{and} \qquad
r_{H^+} = 2m\;\e^{W_{0}\left(-\frac{a}{2m}\right)}.
\label{horizon}
\end{equation}
Here \( W_{-1}(x) \) and   \( W_{0}(x) \) are the real-valued branches of Lambert $W$ function.
We could also write
\begin{equation}
r_{H^-} = {a\over |W_{-1}\left(-\frac{a}{2m}\right)|}, 
\qquad\hbox{and} \qquad
r_{H^+} = {a\over |W_{0}\left(-\frac{a}{2m}\right)|}.
\label{horizon2}
\end{equation}

Perturbatively, for small $a$ we have
\begin{equation}
r_{H^+} = 2m -a +\O(a^2),
\label{E:outer-pert}
\end{equation}
nicely reproducing Schwarzschild in the $a\to0$ limit.  For the inner horizon, since $r_{H_-} < 2m$ then
\begin{equation}
r_{H^-} = {a\over \ln(2m/r_{H_-})}
\end{equation}
implies $r_{H^-} < a$, whence we have a strict upper bound given by the simple analytic expression:
\begin{equation}
r_{H^-} < {a\over \ln(2m/a)}.
\label{E:inner-bound}
\end{equation}
Certainly $\lim_{a\to0} r_{H^-}(m,a)=0$ as we would expect to recover Schwarz\-schild; 
but the form of $r_{H^-}(m,a)$ is not analytic. This bound can also be viewed as the first term in an asymptotic expansion~\cite{Corless:1996} based on (as $x\to 0^+$) 
\begin{equation}
W_{-1}(-x) = \ln(x) + \O(\ln(-\ln(x))) = -\ln(1/x) + \O(\ln(\ln(1/x))). 
\end{equation}
This leads to
\begin{equation}
r_{H^-} = {a\over \ln(2m/a) +\O(\ln(\ln(2m/a)))} 
= {a\over \ln(2m/a)} +
\O\left(a\ln(\ln(2m/a))\over (\ln(2m/a))^2 \right).
\label{E:inner-asymp}
\end{equation}
More specifically (as $a/m\to 0$ or $m/a\to\infty$) 
\begin{equation}
{r_{H^-} \over a} = {1\over \ln(2m/a)} +
\O\left(\ln(\ln(2m/a))\over (\ln(2m/a))^2 \right).
\label{E:inner-asymp2}
\end{equation}

If $a=2m/\e$ then the two horizons merge at $r_H=2m/\e=a$ and one has an extremal black hole. 
If $a >2m/\e$ then there are no horizons, and one is dealing with a regular horizonless, extended but highly localised object, (the energy density peaks at $r=a/4$).

This object could either be extended all the way down to $r=0$, or alternatively be truncated at some finite value of  $r$,  to be used as the exterior geometry for some static and spherically symmetric mass source that \emph{isn't} a black hole. This is potentially useful as a model for planets, stars, \emph{etc.} Consequently, we will also incorporate aspects of the analysis for $a>2m/\e$ as and when required to generate astrophysical observables in the case when equation~(\ref{metric}) is modelling a compact object other than a black hole.

\section{Geodesics and the effective potential}

Continuing the analysis of~\cite{asymptot-mink-core}, we will now calculate the location of the photon sphere and extremal stable circular orbit (ESCO) for the regular black hole with line element given by equation \eqref{metric}.
Photon spheres, (or more precisely the closely related black hole silhouettes),  have been recently observed for the massive objects M87 and  Sgr A*~\cite{eht-1,eht-2,eht-3,eht-4,eht-5,eht-6}.  As such they are, along with the closely related ESCOs, practical and useful quantities to calculate for black hole mimickers.

We begin by considering the affinely parameterised tangent vector to the worldline of a massive or massless particle in our spacetime \eqref{metric}:
\begin{multline}
g_{\mu\nu}\dv{x^\mu}{\lambda}\dv{x^\nu}{\lambda} = -\left(1-\frac{2m\,\e^{-a/r}}{r}\right)\left(\dv{t}{\lambda}\right)^2 + \left(\frac{1}{1-\frac{2m\,\e^{-a/r}}{r}}\right)\left(\dv{r}{\lambda}\right)^2  \\
+ r^2 \left[ \left(\dv{\theta}{\lambda}\right)^2 + \sin^2\theta \left(\dv{\phi}{\lambda}\right)^2 \right] = \epsilon,
\label{tangent}
\end{multline}
where \( \epsilon \in \{-1,0\} \); with $-1$ corresponding to a massive (timelike) particle and 0 corresponding to a massless (null) particle. (The case $\epsilon=+1$ would correspond to tachyonic particles following spacelike geodesics, a situation of no known physical applicability.) 
Since we are working with a spherically symmetric spacetime, we can set \( \theta = \pi/2 \) without any loss of generality and reduce equation \eqref{tangent} to 
\begin{equation}
-\left(1-\frac{2m\,\e^{-a/r}}{r}\right)\left(\dv{t}{\lambda}\right)^2 + \left(\frac{1}{1-\frac{2m\,\e^{-a/r}}{r}}\right)\left(\dv{r}{\lambda}\right)^2 + r^2 \left(\dv{\phi}{\lambda}\right)^2 = \epsilon.
\label{tangentreduced}
\end{equation}
Due to the presence of time-translation and angular Killing vectors, we can now define the conserved quantities
\begin{equation}
E = \left(1-\frac{2m\,\e^{-a/r}}{r}\right)\left(\dv{t}{\lambda}\right) \qq{and} L = r^2\left(\dv{\phi}{\lambda}\right),
\end{equation}
corresponding to the energy and angular momentum of the particle, respectively. 
Thus, equation \eqref{tangentreduced} implies
\begin{equation}
E^2 = \left(\dv{r}{\lambda}\right)^2  + \left(1-\frac{2m\,\e^{-a/r}}{r}\right)\left(\frac{L^2}{r^2}-\epsilon\right).
\end{equation}
This defines an ``effective potential'' for geodesic orbits
\begin{equation}
V_\epsilon (r) = \left(1-\frac{2m\,\e^{-a/r}}{r}\right)\left(\frac{L^2}{r^2}-\epsilon\right),
\end{equation}
with the circular orbits corresponding to extrema of this potential.

\section{Photon spheres}

We subdivide the discussion into two topics: First the \emph{existence} of circular photon orbits (photon spheres) and then the \emph{stability} of circular photon orbits.
The discussion is considerably more complex than for the Schwarz\-schild spacetime, where there is only one circular photon orbit, at $r=3m$, and that circular photon orbit is unstable. Once the extra parameter $a$ is nonzero, and in particular sufficiently large, 
the set of photon orbits exhibits more diversity. 

\subsection{Existence of photon spheres}

For null trajectories we have
\begin{equation}
V_0(r) = \left(1-\frac{2m\,\e^{-a/r}}{r}\right)\frac{L^2}{r^2}.
\end{equation}
So for circular photon orbits
\begin{equation}
V_0'(r_c) = \frac{2L^2}{r_c^5} \left[ m\,\e^{-a/r_c}(3r_c-a)-r_c^2 \right] = 0.
\end{equation}

\begin{figure}[t]
\centering
\begin{subfigure}{.5\textwidth}
  \centering
  \includegraphics[width=.95\linewidth]{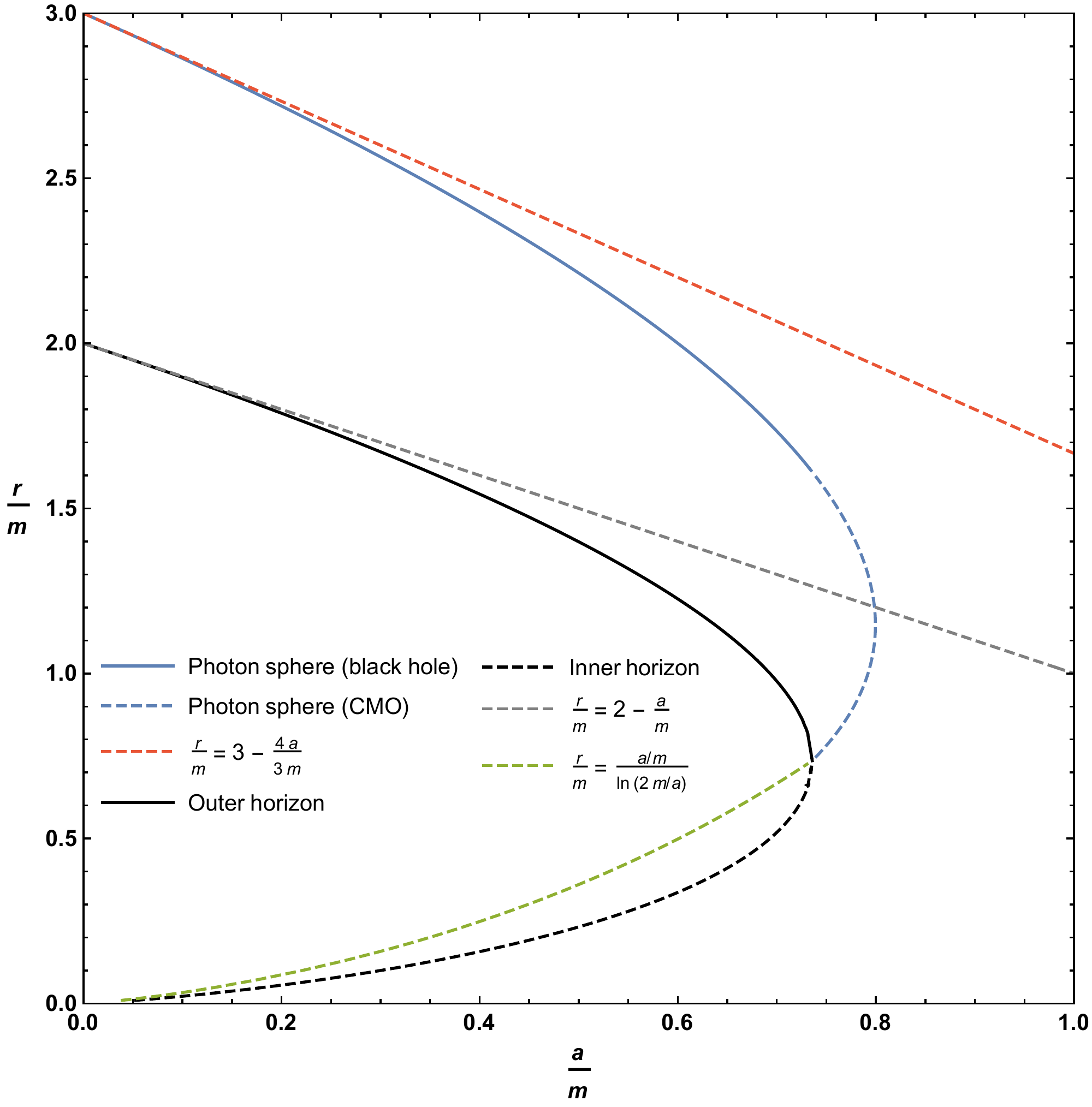}
  \caption{}
  \label{fig:iscom}
\end{subfigure}%
\begin{subfigure}{.5\textwidth}
  \centering
  \includegraphics[width=.95\linewidth]{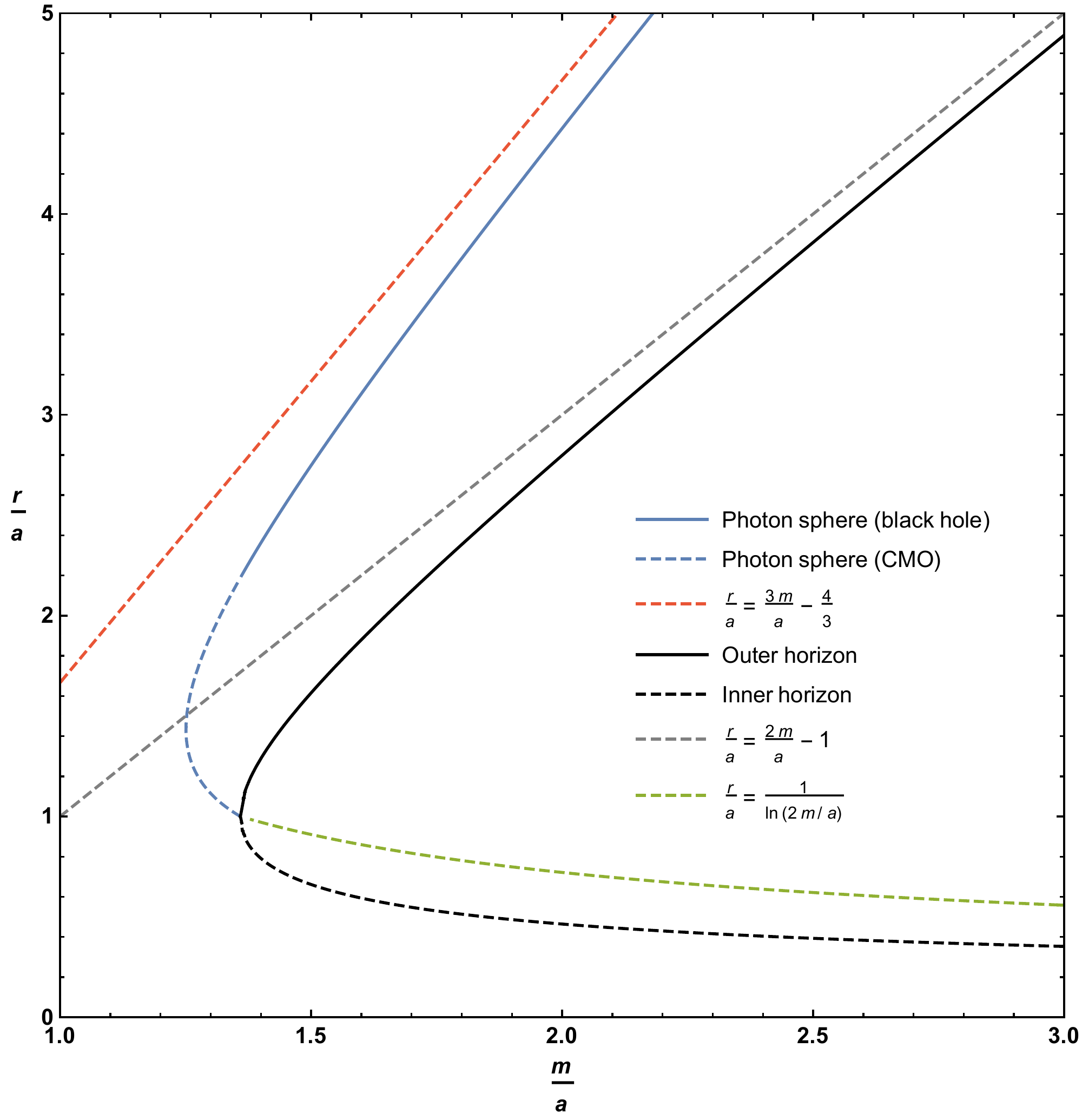}
  \caption{}
  \label{fig:iscoa}
\end{subfigure}
\caption[Asymptotically Minkowski core regular black hole:\break Photon sphere.]{Location of the photon sphere, inner horizon,  and outer horizon as a function of the parameters \( a \) and \( m \). The dashed blue line represents the extension of the photon sphere to horizonless compact massive objects (CMOs), whilst the dashed red line is the asymptotic solution for small values of the parameter \( a \). (Equation \eqref{E:small-a}.) The dashed grey line is the asymptotic solution to the outer horizon for small values of \( a \). (Equation \eqref{E:outer-pert}.)  The dashed green line is the simple analytic bound and asymptotic estimate for the location of the inner horizon. (Equations \eqref{E:inner-bound} and \eqref{E:inner-asymp2}.) }
\label{fig:photonsphere}
\end{figure}

To be explicit about this, the location of a circular photon orbit, \( r_c \), is given implicitly by the equation
\begin{equation}
r_c^2 = m\,\e^{-a/r_c}(3r_c-a),
\label{eq;photonsphere}
\end{equation}
where \( a \) and \( m \) are fixed by the geometry of the spacetime.\footnote{As $a\to 0$ we have $r_c\to 3m$, as expected for Schwarzschild spacetime.}
The curve described by the loci of these circular photon orbits has been plotted in two distinct ways in figure~\ref{fig:photonsphere}.

For clarity, defining  \( w = r_c/a \) and \( z = m/a \), we can re-write the condition for circular photon orbits as
\begin{equation}
w^2 = z \, \e^{-1/w} (3w-1);
\qquad \implies \qquad 
 z = {w^2 \; \e^{1/w}\over3w-1}.
 \label{E:z-for-photon}
\end{equation}
In figure \ref{fig:photonsphere} we also plot the locations of both inner and outer horizons. 

The inner and outer horizons merge at $a/m=2/\e = 0.7357588824...$;
that is, at $m/a = \e/2=1.359140914...$.   For $a/m>2/\e$; that is for $m/a < \e/2$; one is dealing with a horizonless compact object and we see that there is a region where there are \emph{two} circular photon orbits. 
Note that the curve described by the loci of circular photon orbits terminates once one hits a horizon, that is, at $w=1$. Sub-horizon curves of constant $r$ are spacelike (tachyonic), and \emph{cannot} be  lightlike, so they are explicitly excluded. 
That is, photon spheres can only \emph{exist} in the region $w\in(1,\infty)$. 

Can we be more explicit about the key qualitative and quantitative features of this plot? 
Specifically, let us now analyze stability \emph{versus} instability, and find the exact location of the various turning points.

\subsection{Stability \emph{versus} instability for circular photon orbits} 

To check the \emph{stability} of these circular photon orbits we now need to investigate
\begin{equation}
V_0''(r_c) = \frac{2L^2}{r_c^7} \left[ 3r_c^3 - m\,\e^{-a/r_c}(6r_c-a)(2r_c-a) \right].
\label{E:V''}
\end{equation}

\subsubsection{Perturbative analysis (small $a$)} 

We note that determining $r_c(m,a)$ from equation (\ref{eq;photonsphere}) is  not analytically feasible, but $r_c(m,a)$ can certainly be estimated perturbatively for small $a$. 
We have
\begin{equation}
r_c(m,a) = 3m - \frac{4ma}{r_c} + \mathcal{O}(a^2) 
\quad\implies\quad 
r_c(m,a) = 3m - {4\over3} a + \mathcal{O}(a^2).
\label{E:small-a}
\end{equation}
So, for small values of \( a \), we recover the standard result for the location of the photon sphere in  Schwarzschild spacetime.

Estimating $V_0''(r_c)$ by  now substituting the approximate location of the photon sphere as \( r_c(m,a) = 3m - 4a/3 +\mathcal{O}(a^2),\) we find
\begin{equation}
V_0''(r_c(m,a)) =  -{2L^2\over81 m^4}\left(1 +  {4\over3}\, {a\over m} +\mathcal{O}(a^2)\right) .
\label{E:V''-small-a}
\end{equation}
This quantity is manifestly negative for small $a$. 
That is, (within the limits of the current small-$a$ approximation), photons are in an unstable orbit at the small-$a$ photon sphere.

\subsubsection{Non-perturbative analysis} 

However, if we rephrase the problem then we can make some much more explicit exact statements that are no longer perturbative in small $a$: Whereas determining $r_c(m,a)$ is analytically infeasible it should be noted that in contrast both $a(m,r_c) $ and $m(r_c,a)$ are easily determined analytically:
\begin{equation}
a(m,r_c) = r_c (3 - W(r_c \e^3/m)); \qquad m(r_c,a) = {r_c^2 \; \e^{a/r_c} \over(3r_c-a)}.
\label{E:non-pert}
\end{equation}
Consequently, at the peak we can write
\begin{equation}
V_0(r_c,m) = {L^2\over r_c^2} \left( 1- {2\over W(r_c \e^3/m)}\right);
\qquad
V_0(r_c,a) = {L^2\over r_c^2} \; {r_c-a\over 3r_c-a}. 
\end{equation}
Regarding stability, in the first case, substituting~(\ref{E:non-pert}\;a) into \eqref{E:V''}, we have
\begin{equation}
V_0''(r_c,m) = -{2L^2
\left( W(r_c \e^3/m)^2- W(r_c \e^3/m)-3\right)\over r_c^4  W(r_c \e^3/m)} .
\end{equation}
Using properties of the Lambert $W$ function, we quickly see that this is negative for $r_c/m >  {1\over 2} (1+\sqrt{13}) \; \e^{-5/2 +\sqrt{13}/2} = 1.146702958...$, implying instability of the circular photon orbits in this region, (and stability outside this region).

That is, on the curve of circular photon orbits, $V''(r_c)=0$ at the point
\begin{equation}
(r_c/m, a/m)_* = ( 1.146702958..., 0.7995092385...).
\end{equation}

In the second case, substituting~(\ref{E:non-pert}\;b) into \eqref{E:V''}, we have
\begin{equation}
V_0''(r_c,a) = -{2L^2\over r_c^5} \; {3r_c^2-5ar_c+a^2\over 3r_c-a}. 
\end{equation}
This will certainly be negative for $r_c/a > (5+\sqrt{13})/6 = 1.434258546...$, implying instability of the circular photon orbits in this region, (and stability outside this region).

That is, on the curve of circular photon orbits, $V''(r_c)=0$ at the point
\begin{equation}
(r_c/a,m/a)_* = ( 1.434258546..., 1.250767286...).
\end{equation}
Consequently, on the curve of circular photon orbits we have \emph{existence} and \emph{stability} in the region $w\in(1,1.434258546...)$; and \emph{existence} and \emph{instability} in the region $w\in(1.434258546...,\infty)$. 
Precisely at the point $w=1.434258546...$ the photon sphere exhibits neutral stability.

\subsection{Turning points} 

To evaluate the exact location of the turning points on the curve described by the loci of circular photon orbits, recall that using  \( w = r_c/a \) and \( z = m/a \) we can write this curve as 
\begin{equation}
w^2 = z \, \e^{-1/w} (3w-1) \qquad \implies \qquad z = {w^2 \e^{1/w}\over (3w-1)}.
\label{E:z-for-photon2}
\end{equation}
This allows us to calculate
\begin{equation}
\dv{z}{w} = \e^{1/w} \, \frac{3w^2 - 5w +1}{(3w-1)^2},
\end{equation}
which has a zero located at \( w = (5+\sqrt{13})/6 \), where we have already seen that $V_0''(r_c,a)=V_0''(w)=0$.

At this point $z$ takes on its maximum value
\begin{equation}
z = \e^{6/(5+\sqrt{13})} \frac{(5+\sqrt{13})^2}{18(3+\sqrt{13})}  =
\e^{(5-\sqrt{13})/2}\;{(2+\sqrt{13})\over 9}.
\end{equation}
Consequently, no photon sphere can exist if 
\begin{equation}
\frac{a}{m} > \e^{-(5-\sqrt{13})/2}\;(\sqrt{13}-2) = 0.7995092385...;
\label{eq;photonlimit1}
\end{equation}
or equivalently
\begin{equation}
\frac{m}{a} < \e^{(5-\sqrt{13})/2}\;{(2+\sqrt{13})\over 9} =1.250767286....
\label{eq;photonlimit2}
\end{equation}
Note that this happens when
\begin{equation}
{r_c\over m} > {1\over2} (1+\sqrt{13}) \e^{-(5-\sqrt{13})/2};
\qquad\qquad
{r_c\over a} > {5+\sqrt{13}\over 6},
\end{equation}
which was where, as we have already seen, $V_0''(r_c,m)=0$.

\begin{figure}[t]
\centering
\begin{subfigure}{.5\textwidth}
  \centering
  \includegraphics[width=.95\linewidth]{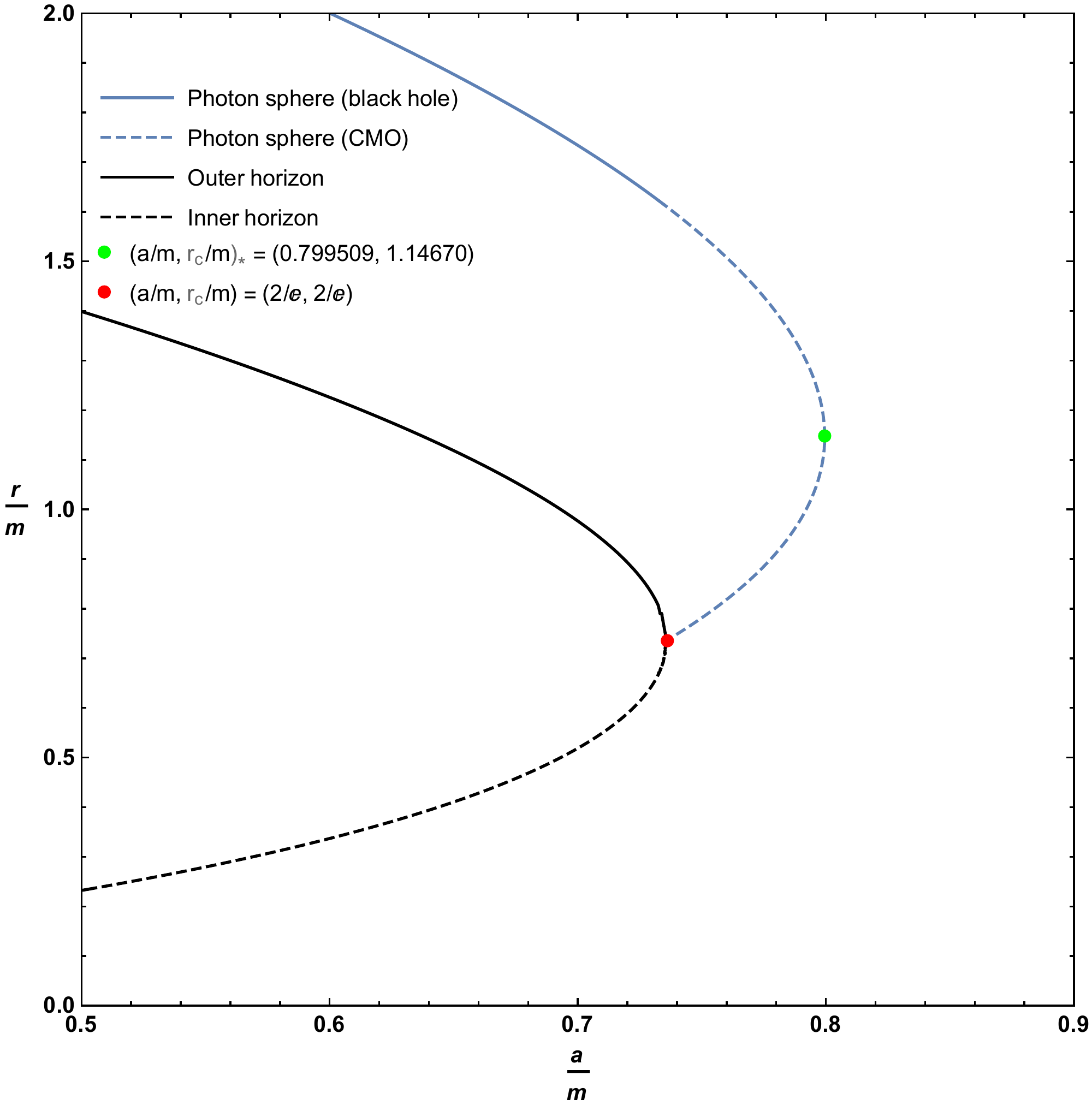}
  \caption{}
  \label{fig:iscom}
\end{subfigure}%
\begin{subfigure}{.5\textwidth}
  \centering
  \includegraphics[width=.95\linewidth]{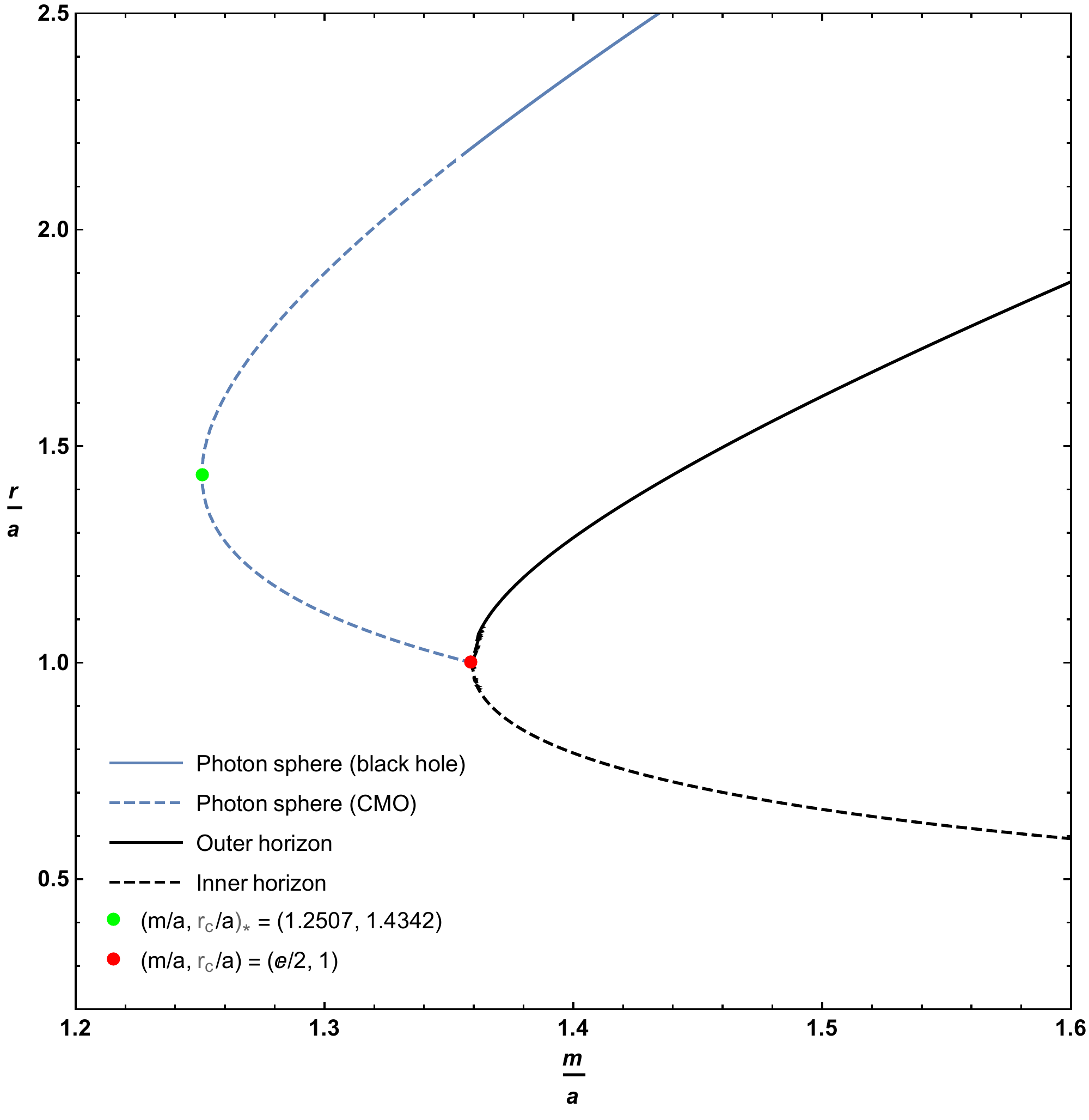}
  \caption{}
  \label{fig:iscoa}
\end{subfigure}
\caption[Asymptotically Minkowski core regular black hole:\break Photon sphere zoomed in.]{Zoomed in plots of the location of the photon sphere, inner horizon,  and outer horizon as a function of the parameters \( a \) and \( m \), focussing on the extremal and merger regions. The dashed blue line represents the extension of the photon sphere to horizonless compact massive objects (CMOs). Whenever the location of the photon sphere is double-valued the upper branch corresponds to an unstable photon orbit while the lower branch corresponds to a stable photon orbit.}
\label{fig:photonsphere2}
\end{figure}

As can be seen, originally from figure~\ref{fig:photonsphere}, and now in more detail in the zoomed-in plot in figure \ref{fig:photonsphere2}, for horizonless compact massive objects there is a region where there are two possible locations for the photon sphere for fixed values of \( m \) and \( a \). Furthermore when this happens it is the upper branch that corresponds to an unstable photon orbit, while the lower branch is a stable photon orbit.

\vfill

\section{Timelike circular orbits}
\def\ISCO{{\hbox{\tiny ISCO}}}
\def\ESCO{{\hbox{\tiny ESCO}}}

Let us first check the \emph{existence}, and then the \emph{stability}, of timelike circular orbits.
Even in Schwarzschild spacetime ($a\to0$) this is not entirely trivial: Timelike circular orbits \emph{exist} for all $r_c\in (3m,\infty)$; they are unstable for $r_c\in (3m,6m)$,
exhibit neutral stability at $r_c=6m$, and are stable for $r_c\in (6m,\infty)$. 
Once the parameter $a$ is non-zero the situation is much more complex.

\subsection{Existence of circular timelike orbits} 

For timelike trajectories, the effective potential is given by
\begin{equation}
V_{-1}(r) = \left(1-\frac{2m\,\e^{-a/r}}{r}\right)\left(1+\frac{L^2}{r^2}\right),
\end{equation}
and so the locations of the circular orbits can be found from
\begin{equation}
V_{-1}'(r_c) = -\frac{2}{r_c^5}\left\{ L^2r_c^2+m\,\e^{-a/r_c} [ a(L^2+r_c^2)-r_c(3L^2+r_c^2) ] \right\} = 0.
\end{equation}
That is, all timelike circular orbits (there will be infinitely many of them) must satisfy
\begin{equation}
 L^2r_c^2+m\,\e^{-a/r_c} [ a(L^2+r_c^2)-r_c(3L^2+r_c^2) ]  = 0.
\end{equation}

This is not analytically solvable for \( r_c(L,m,a) \), however we \emph{can} solve for the required angular momentum $L_c(r_c,m,a)$ of these circular orbits:
\begin{equation}
L_c(r_c,m,a)^2 = {\frac{r_c^2 \, m(r_c-a)}{ma-3mr_c+r_c^2\,\e^{a/r_c}}}.
\label{E:Lsq}
\end{equation}
Physically we must demand $0\leq L_c^2 <\infty$, so the boundaries for the \emph{existence} region of circular orbits (whether stable or unstable) are given by
\begin{equation}
r_c = a; \qquad\qquad {ma-3mr_c+r_c^2\,\e^{a/r_c}}=0.
\end{equation}
The first of these conditions $r_c=a$, comes from the fact that in this spacetime gravity is effectively repulsive for $r<a$. 
Remember that $g_{tt} = -(1-2m\e^{-a/r}/r)$, and that the pseudo-force due to gravity depends on $\partial_r g_{tt}$. Specifically
\begin{equation}
\partial_r g_{tt} = - {2m\over r^2} \; \e^{-a/r} \; \left(1-{a\over r}\right),
\end{equation}
and this changes sign at $r=a$. 
So for $r>a$ gravity attracts you to the centre, but for $r<a$ gravity repels you from the centre.

And if gravity repels you, there is no way to counter-balance it with a centrifugal pseudo-force, and so there is simply no way to get a circular orbit, regardless of whether it be stable or unstable.
Precisely at $r=a$ there are stable ``orbits'' where the test particle just sits there, with zero angular momentum, no sideways motion required.
Since by construction $r_c>r_{H^+} \geq a$, this constraint is relevant only for horizonless CMOs. 

 The second of these conditions is exactly the location of the photon orbits considered in the previous sub-section. (Physically what is going on is this: At large distances it is easy to put a massive particle into a circular orbit with $L_c\propto\sqrt{mr_c}$. As one moves inwards and approaches the photon orbit, the massive particle must move more and more rapidly, and the angular momentum per unit mass must diverge when a particle with nonzero invariant mass tries to orbit at the photon orbit.)
 
\begin{figure}[t]
\centering
\begin{subfigure}{.5\textwidth}
  \centering
  \includegraphics[width=.95\linewidth]{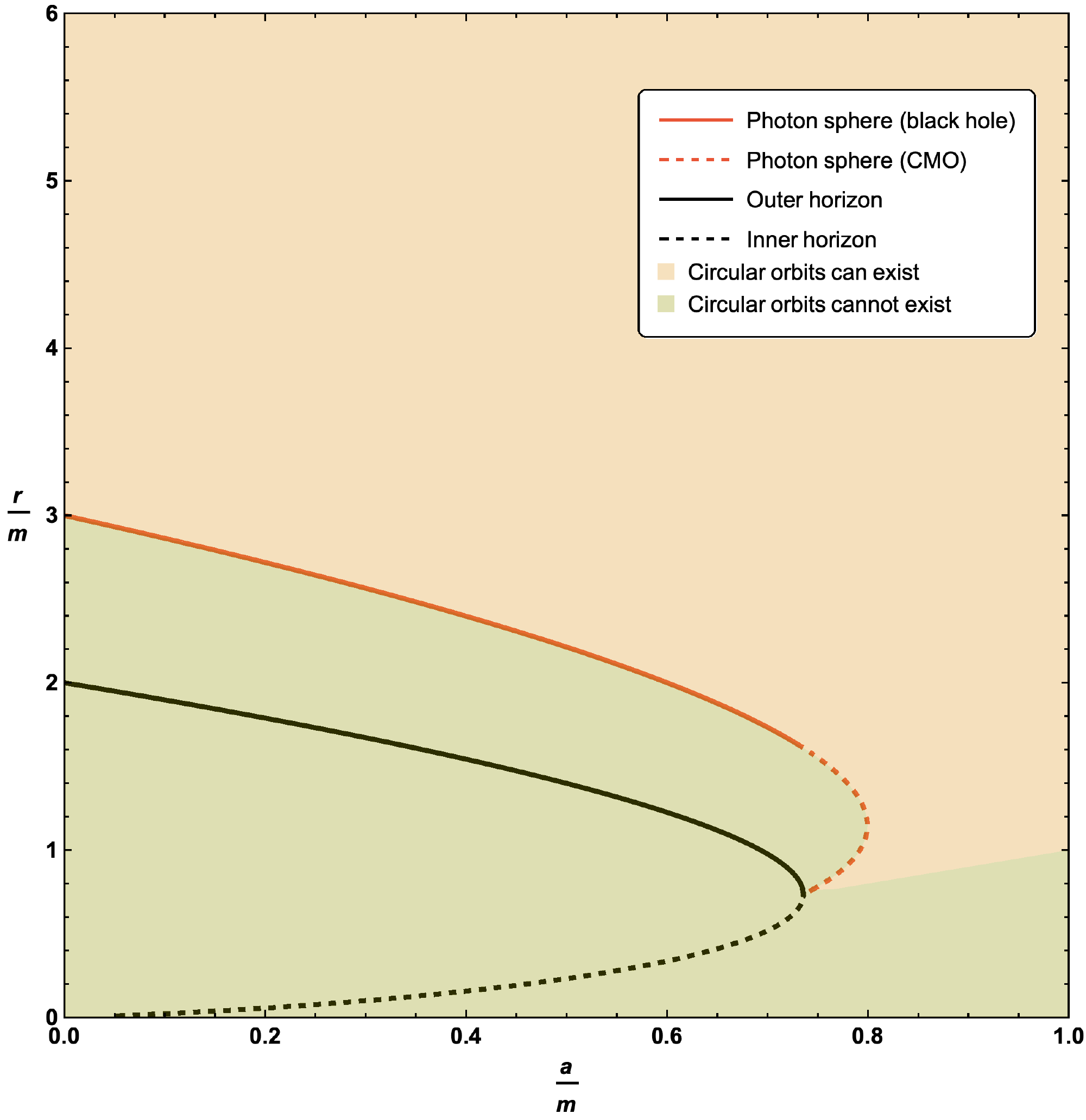}
  \caption{}
  \label{fig:iscom}
\end{subfigure}%
\begin{subfigure}{.5\textwidth}
  \centering
  \includegraphics[width=.95\linewidth]{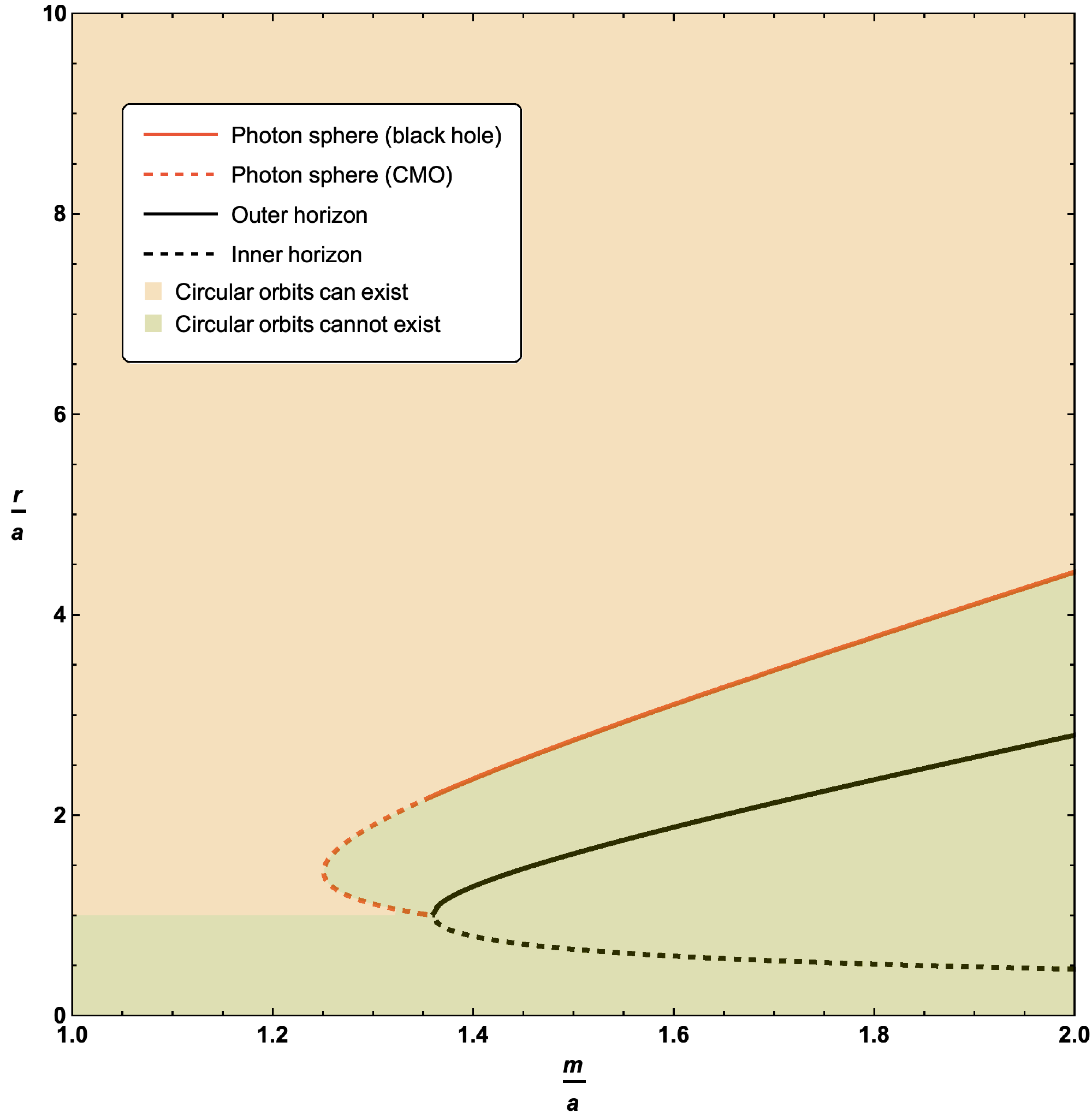}
  \caption{}
  \label{fig:iscoa}
\end{subfigure}
\caption[Asymptotically Minkowski core regular black hole:\break Existence of circular orbits.]{Locations of the \emph{existence} region for timelike circular orbits in terms of the circular null geodesics, outer horizon, and inner horizon for various values of the parameters \( a \) and \( m \). }
\label{fig:isco-existence}
\end{figure}

 Thus the existence region (rather than just its boundary) for timelike circular orbits is therefore:
 \begin{equation}
r_c > a; \qquad\qquad {ma-3mr_c+r_c^2\,\e^{a/r_c}}>0.
\end{equation}
See figure~\ref{fig:isco-existence}.

\subsection{Stability \emph{versus} instability for  circular timelike orbits} 

Now consider the general expression
\begin{equation}
V_{-1}''(r) = {6L^2 r^3- 2m(2r^4-4ar^3+(12L^2+a^2)r^2-8L^2ar+L^2a^2)\e^{-a/r}\over r^7},
\end{equation}
and substitute the known value of $L\to L_c(r_c)$ for circular orbits, see \eqref{E:Lsq}. Then
\begin{equation}
V_{-1}''(r_c) = -{2m e^{-a/r_c}(2m(3r_c^2-3ar_c+a^2) \e^{-a/r_c} -r_c(r_c^2+ar_c-a^2))\over
(r_c^2 - m(3r_c-a) \e^{-a/r_c})r^4}.
\end{equation}
Note that $V_{-1}''(r_c) \to\infty$ at the photon orbit, (where the denominator has a zero).

To locate the \emph{boundary} of the region of \emph{stable} circular orbits, the ESCO (extremal stable circular orbit),  we now need to set $V_{-1}''(r_c)=0$, leading to the equation
\begin{equation}
2m(3r_c^2-3ar_c+a^2) \e^{-a/r_c} = r_c(r_c^2+ar_c-a^2).
\label{E:stable}
\end{equation}
We note that locating this boundary is equivalent to extremizing $L_c(r_c)$. 
To see this, consider the quantity $[V_{-1}'(L(r),r)]=0$ and differentiate:
\begin{equation}
{\dd{} [V_{-1}'(L(r),r)]\over \dd r }  = 
\left.{\partial V_{-1}'(L,r)\over\partial L}\right|_{L=L(r)} \times {\dd L(r)\over \dd r}
+ \left.V''_{-1}(L,r)\right|_{L=L(r)}.
\end{equation}
This implies
\begin{equation}
0  = 
\left.{\partial V_{-1}'(L,r)\over\partial L}\right|_{L=L(r)} \times {\dd L(r)\over \dd r}
\;+\; \left.V_{-1}''(L,r)\right|_{L=L(r)},
\end{equation}
and so
\begin{equation}
 \left.V_{-1}''(L,r)\right|_{L=L(r)}  = - \left.{\partial V_{-1}'(L,r)\over\partial L}\right|_{L=L(r)} 
 \times {\dd L(r)\over \dd r}.
 \end{equation}
But it is easily checked that ${\partial V_{-1}'(L,r)/\partial L}$ is non-zero outside the photon sphere, (that is, in the existence region for circular timelike geodesics).
Thus,
\begin{equation}
 \left.V_{-1}''(L,r)\right|_{L=L(r)}  = 0
 \qquad \Longleftrightarrow  \qquad
 {\dd L(r)\over \dd r}=0.
 \end{equation}
So one might a well extremize $L^2_c(r_c)$, as in equation (\ref{E:Lsq}), and one again finds equation (\ref{E:stable}). 

Defining  \( w = r_c/a \) and \( z = m/a \) the curve describing the boundary of the region of stable timelike circular orbits can be rewritten as
\begin{equation}
\label{E:dd}
2z (3w^2-3w+1)\e^{-1/w} = w(w^2+w-1).
\end{equation}

\begin{figure}[t]
\centering
\begin{subfigure}{.5\textwidth}
  \centering
  \includegraphics[width=.95\linewidth]{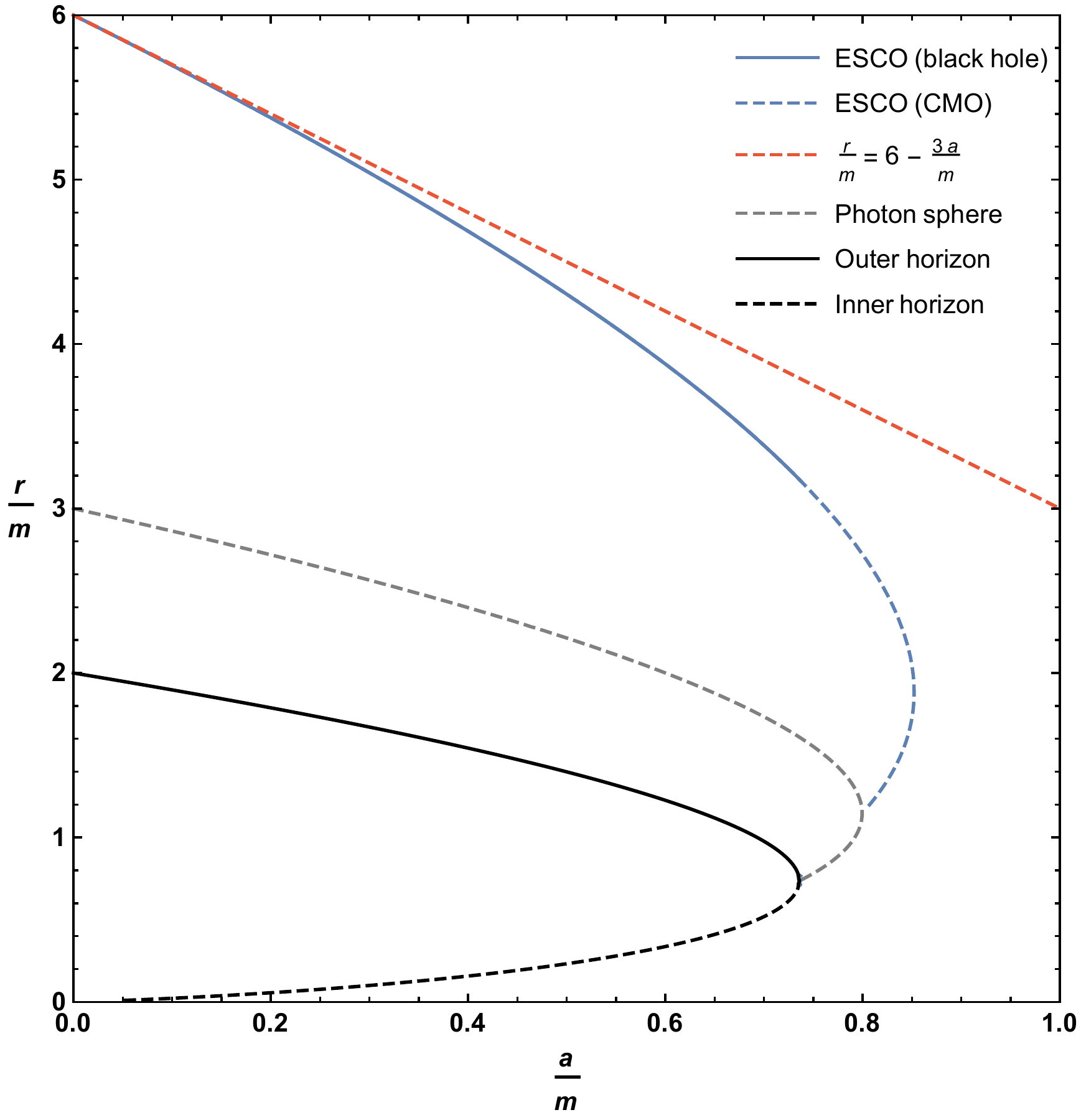}
  \caption{}
  \label{fig:iscom}
\end{subfigure}%
\begin{subfigure}{.5\textwidth}
  \centering
  \includegraphics[width=.95\linewidth]{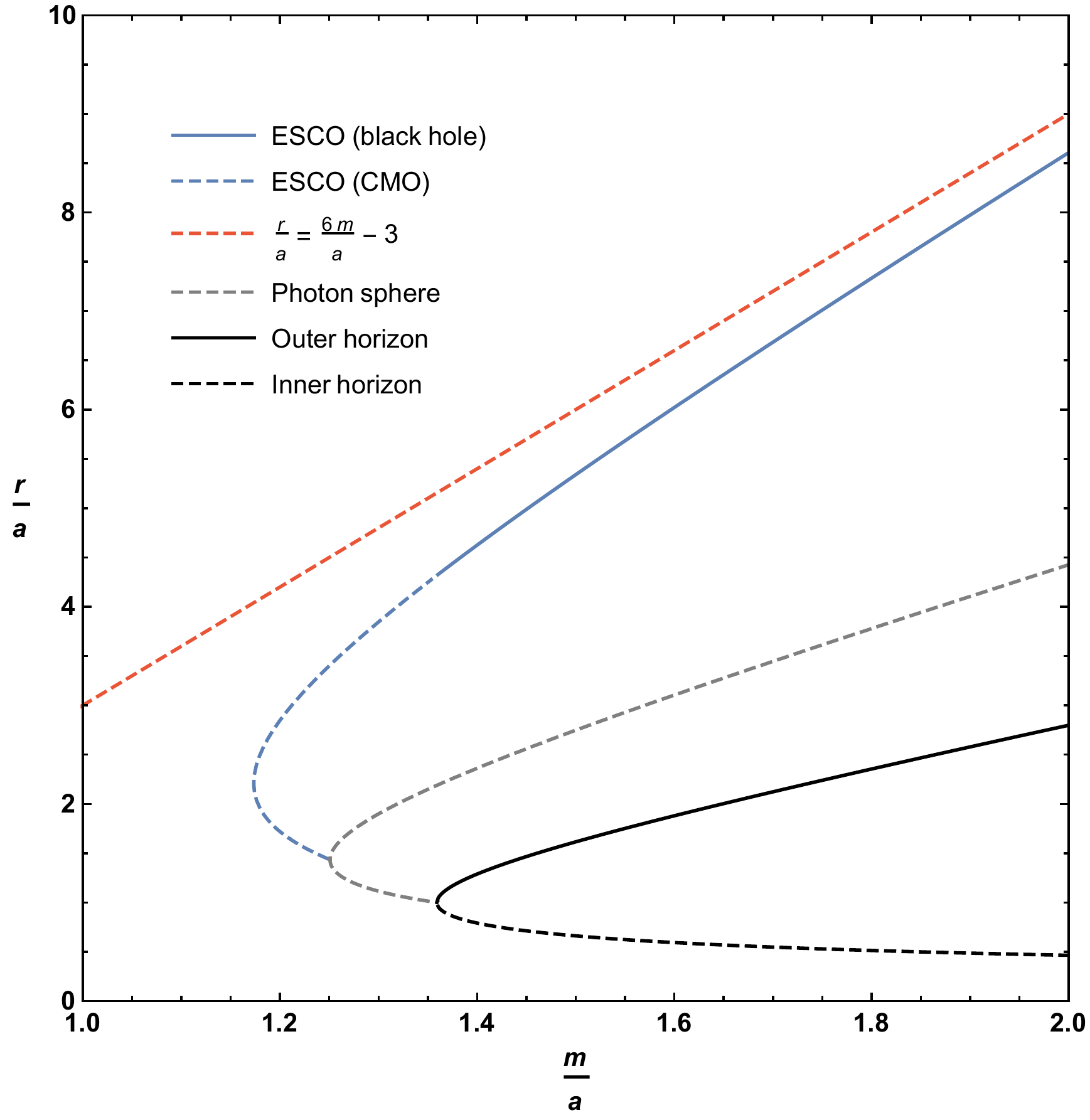}
  \caption{}
  \label{fig:iscoa}
\end{subfigure}
\caption[Asymptotically Minkowski core regular black hole:\break Photon sphere and ESCO.]{Locations of the ESCO, photon sphere, outer horizon, and inner horizon for various values of the parameters \( a \) and \( m \).
The dashed blue line represents the extension of the ESCO to CMOs. The dashed red curves in sub-figure (a) and (b) is the asymptotic location of the ISCO for small values of \( a \) (approaching the Schwarzschild solution).}
\label{fig:esco}
\end{figure}

\begin{figure}[t]
\centering
\begin{subfigure}{.5\textwidth}
  \centering
  \includegraphics[width=.95\linewidth]{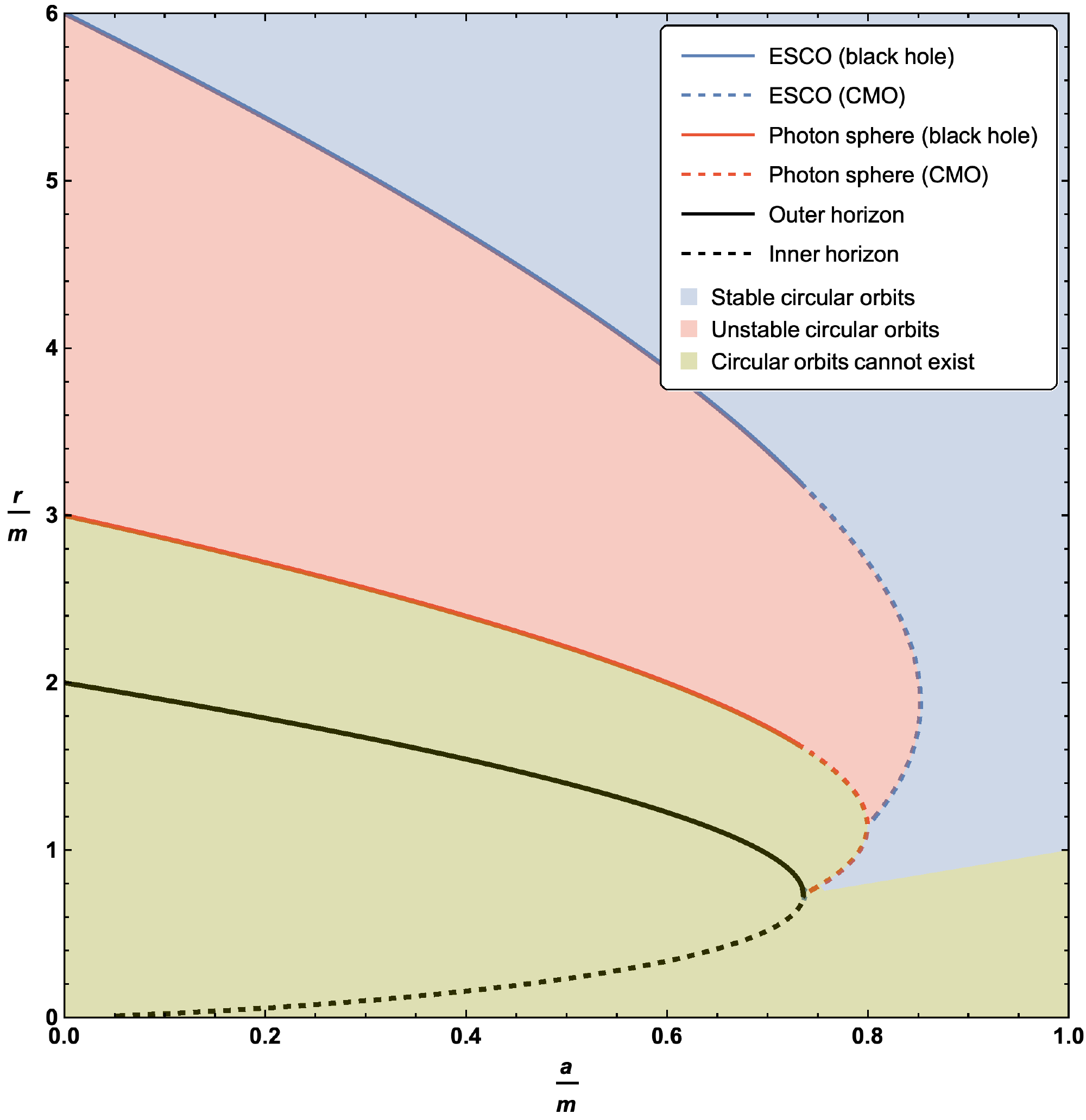}
  \caption{}
  \label{fig:iscom}
\end{subfigure}%
\begin{subfigure}{.5\textwidth}
  \centering
  \includegraphics[width=.95\linewidth]{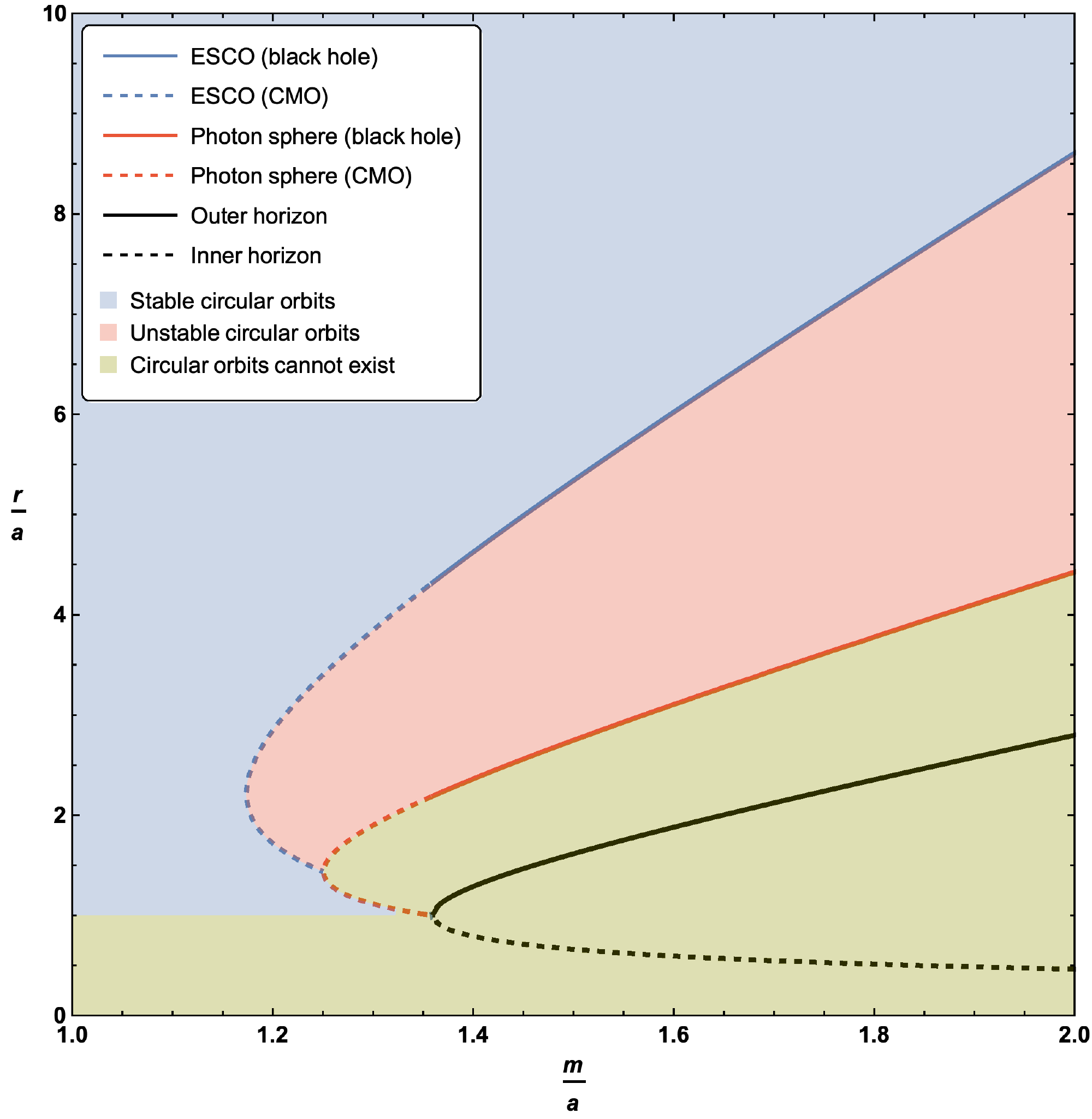}
  \caption{}
  \label{fig:iscoa}
\end{subfigure}
\caption[Asymptotically Minkowski core regular black hole:\break Photon sphere and ESCO stability regions.]{Locations of the ESCO, photon sphere, outer horizon, and inner horizon for various values of the parameters \( a \) and \( m \).
The dashed blue line represents the extension of the ESCO to CMOs. 
The dashed red line represents the extension of the photon sphere to CMOs.
The blue region denotes stable timelike circular orbits, while the red region denotes unstable timelike circular orbits, and the green region denotes the non-existence of timelike circular orbits. 
Where the ESCO is single-valued it is an ISCO. Where the ESCO is double-valued the upper branch is an ISCO and the lower branch is an OSCO (outermost stable circular orbit). 
}
\label{fig:esco-2}
\end{figure}

Plots of the boundary implied by equation \eqref{E:stable}, or equivalently \eqref{E:dd}, can be seen in figure~\ref{fig:esco}.
As for the photon sphere, we have the interesting result that the extension of the ESCO to horizonless compact massive objects results in up to two possible ESCO locations for fixed values of \( a \) and \( m \). 
Perhaps unexpectedly, the curve of ESCOs does not terminate at the horizon --- it terminates once it hits the curve of circular photon orbits at a very special point.
Let us now turn to the detailed analysis of both the qualitative behaviour and the various turning points presented in figures~\ref{fig:esco} and~\ref{fig:esco-2}.
Note that where the ESCO is single-valued it is an ISCO (innermost stable circular orbit). Where the ESCO is double-valued the upper branch is an ISCO and the lower branch is an OSCO (outermost stable circular orbit)~\cite{OSCO}.


\subsubsection{Perturbative analysis (small $a$)} 

Let us first investigate the existence region perturbatively for small $a$. We have
\begin{equation}
L_c(r_c,m,a)^2 = {mr_c^2\over r_c-3m} -{2mr_c(r_c-m)\over(r_c-3m)^2} \; a + \mathcal{O}(a^2).
\end{equation}
Note that this approximation diverges at the Schwarzschild photon sphere $r=3m$.
So for small $a$ the boundary for the region of \emph{existence} of timelike circular orbits is still  $r=3m$.

Now investigate the \emph{stability} region perturbatively for small $a$. Rearranging  
equation (\ref{E:stable}) we see
\begin{equation}
r_c = {6m(r_c^2-ar_c+a^2/3) \e^{-a/r_c}\over r_c^2+ar_c-a^2} 
= 6m\left( 1-{3a\over r_c} +\O(a^2) \right),
\end{equation}
and so
\begin{equation}
r_c = 6m -3a +\O(a^2). 
\end{equation}
Which sensibly reproduces the Schwarzschild ISCO to lowest order in $a$, and explains the asymptote in figure~\ref{fig:esco} (b).

Furthermore, for small $a$, substituting $L_c(r_c)$ into $V''_{-1}(L,r_c)$ and expanding
\begin{equation}
V_{-1}''(r_c) = {2m(r_c-6m)\over r_c^3(r_c-3m)} 
+  {4m^2(7r_c-15m)\over r^4(r_c-3m)^2} \; a  +\O(a^2)
\end{equation}
Demanding that this quantity be zero self-consistently yields $r_c = 6m -3a +\O(a^2)$. 

\subsubsection{Non-perturbative analysis} 

\begin{figure}[t]
\centering
\begin{subfigure}{.5\textwidth}
  \centering
  \includegraphics[width=.95\linewidth]{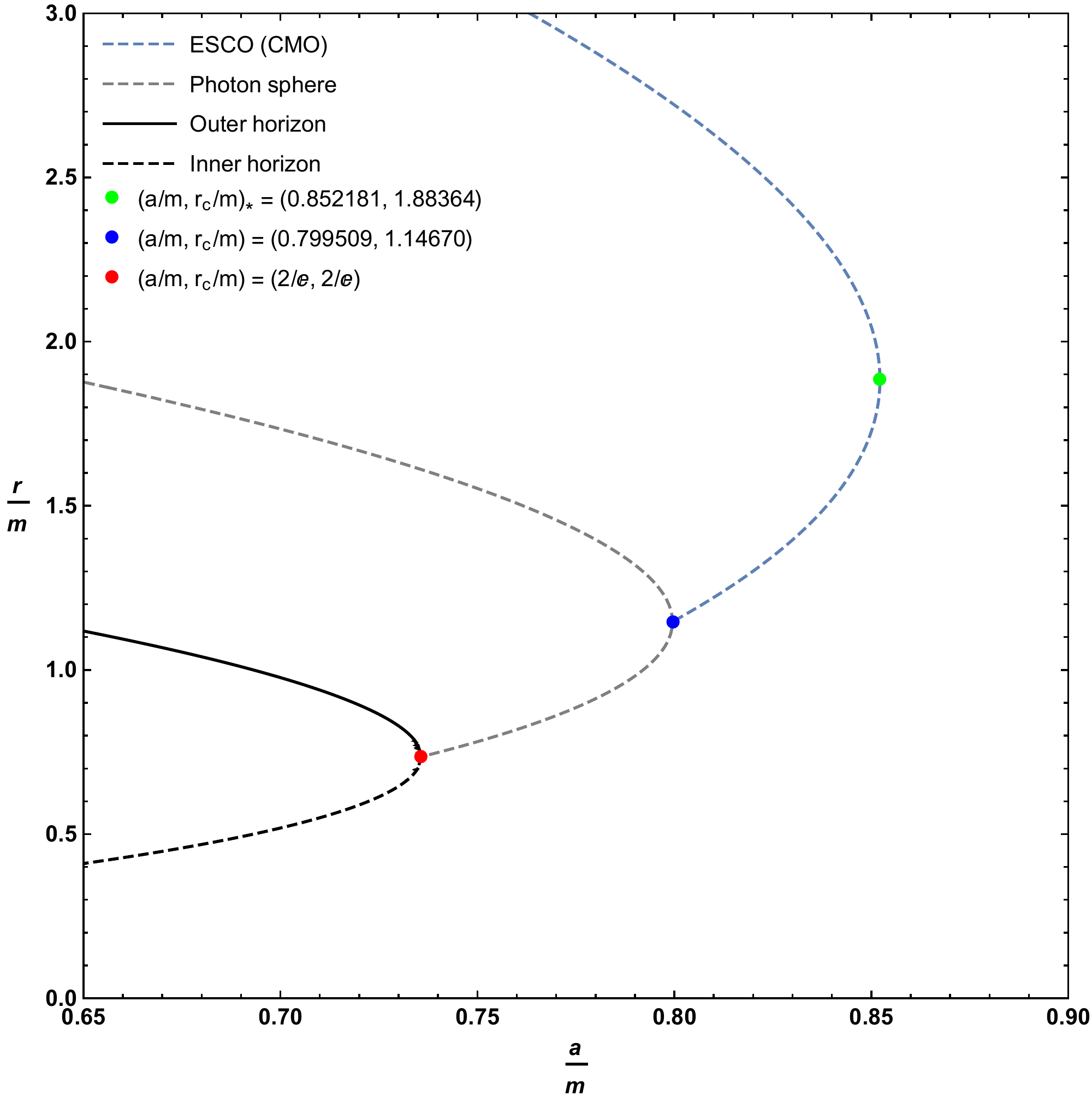}
  \caption{}
  \label{fig:iscom}
\end{subfigure}%
\begin{subfigure}{.5\textwidth}
  \centering
  \includegraphics[width=.95\linewidth]{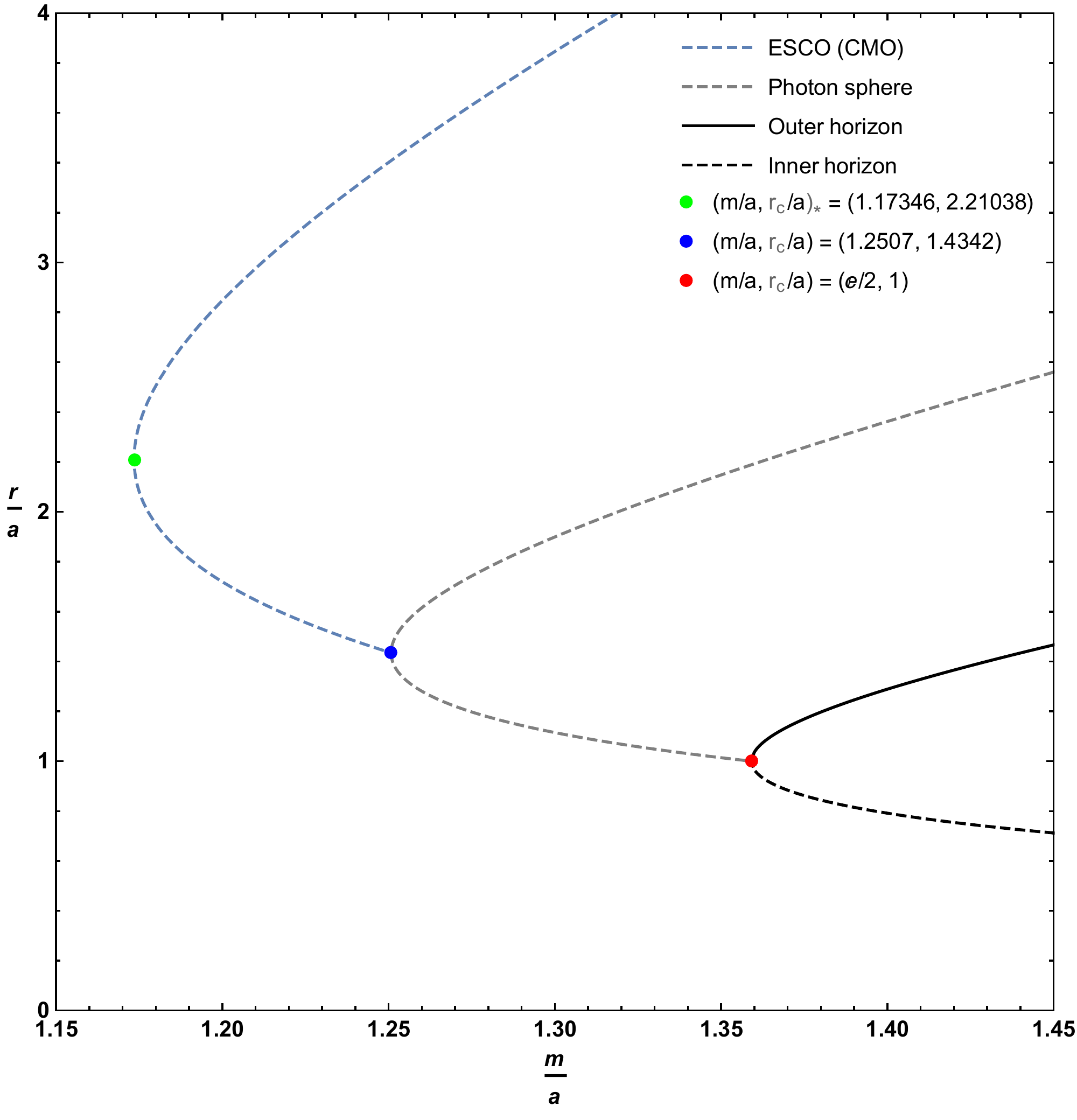}
  \caption{}
  \label{fig:iscoa}
\end{subfigure}
\caption[Asymptotically Minkowski core regular black hole:\break ESCO zoomed in.]{Zoomed in plot of the locations of the ESCO, outer horizon, and inner horizon for various values of the parameters \( a \) and \( m \), focussing on the turning points.  The dashed blue line represents the extension of the ESCO to CMOs. Where the ESCO is single-valued it is an ISCO. Where the ESCO is double-valued the upper branch is an ISCO and the lower branch is an OSCO.}
\label{fig:esco-3}
\end{figure}

We have already seen that, defining  \( w = r_c/a \) and \( z = m/a \), the curve describing the boundary of the region of stable timelike circular orbits can be rewritten as
\begin{equation}
\label{E:dd3}
2z (3w^2-3w+1)\e^{-1/w} = w(w^2+w-1).
\end{equation}
That is,
\begin{equation}
    z = {w(w^2+w-1)\e^{1/w}\over 2(3w^2-3w+1)}.
    \label{E:z-for-esco}
\end{equation}
Let us look for the turning points of $z(w)$. The derivative is
\begin{equation}
{\dd z\over\dd w} = {(w-1)(3w^4-6w^3-3w^2+4w-1)\e^{1/w} \over 2w(3w^2-3w+1)^2}.
\end{equation}
There is one obvious local extremum at $w=1$, corresponding to $z=\e/2$. Physically this corresponds to the point where inner and outer horizon merge and become extremal --- but from inspection of figure~\ref{fig:esco}, the descriptive plots of figure~\ref{fig:esco-2}, and the zoomed-in plots of figure~\ref{fig:esco-3}, we see that the curve of ESCOs hits the photon orbit (and becomes unphysical) before getting to this point.
In terms of the variables used when plotting figures~\ref{fig:esco}--\ref{fig:esco-3} this unphysical (from the point of view of ESCOs) point corresponds to
\begin{equation}
(r_c/a, m/a)_* = (1, \e/2) \qquad (r_c/m, a/m)_* = (2/\e, 2/\e). 
\end{equation}

The other local extremum is located at the only physical root of the quartic polynomial
\begin{equation}
3w^4-6w^3-3w^2+4w-1 =0.
\end{equation}
While this can be solved analytically, the results are too messy to be enlightening and so we resort to numerics. 
Two roots are complex, one is negative, the only physical root is $w= 2.210375896...$, corresponding to $z=1.173459017...$. 
Physically this implies that the ESCO curve should exhibit a non-trivial local extremum --- and from inspection of figure~\ref{fig:esco} we see that the curve of ESCOs does indeed have a local extremum at this point. In terms of the variables used when plotting figure~\ref{fig:esco} this extremal point corresponds to
\begin{equation}
(r_c/a, m/a)_* = (2.210375896, 1.173459017), 
\end{equation}
and
\begin{equation}
(r_c/m, a/m)_* = (1.883641323, 0.8521814444). 
\end{equation}

\subsection{Intersection of ESCO and photon sphere } 

We can rewrite the curve for the loci of the photon spheres \eqref{E:z-for-photon} as 
\begin{equation}
\e^{-1/w} z = {w^2\over (3w-1)}.
\end{equation}
Similarly, for the loci of ESCOs rewrite \eqref{E:z-for-esco} as
\begin{equation}
\e^{-1/w} z = {w(w^2+w-1)\over 2 (3w^2-3w+1)}.
\end{equation}
These curves cross at
\begin{equation}
{w\over (3w-1)} = {(w^2+w-1)\over 2 (3w^2-3w+1)}.
\end{equation}
That is, at
\begin{equation}
(w-1)(3w^2-5w+1)=0,
\end{equation}
with explicit roots at
\begin{equation}
1, \quad {5\pm\sqrt{13}\over6}.
\end{equation}
The physically relevant root is  $w = {5+\sqrt{13}\over6}= 1.434258546...$, which was where we previously determined that  the photon sphere became stable, and at the point where the curve of photon spheres maximised the value of $z=m/a$.

\subsection{Explicit result for the angular momentum} 

We can rewrite the curve for the angular momentum \eqref{E:Lsq} as 
\begin{equation}
L_c^2 = 
a^2  \left( \e^{-1/w} z \; w^2 (w-1) \over  w^2 - \e^{-1/w} z (3w-1)\right) .
\end{equation}
Similarly, for the loci of ESCOs we can rewrite \eqref{E:z-for-esco} as
\begin{equation}
\e^{-1/w} z = {w(w^2+w-1)\over 2 (3w^2-3w+1)}.
\end{equation}
We then substitute this into back into $L_c$:
\begin{equation}
L_c^2 = a^2\; {w^2(w^2+w-1)\over 3w^2-5w+1}.
\end{equation}
This has a pole at $w = {5+\sqrt{13}\over6}= 1.434258546...$, 
and is then positive and finite for all $w >{5+\sqrt{13}\over6}$.  
(Of course the point $w = {5+\sqrt{13}\over6}$ on the ESCO curve is exactly where the ESCO curve hits the photon curve, so we would expect the angular momentum to go to infinity there.)
Asymptotically for large $r$ (large $w=r_c/a$) we have $L_c^2 \sim a^2 w^2/3$ and $m/a = z \sim w/6$, so $L_c^2 \sim 2 m r_c$ as expected from the large-distance Newtonian limit.

\section{Regge--Wheeler and quasi-normal modes}\label{subsection:RW-mink-core}

Up until this point in our analysis, we have allowed for the possibility of horizonless objects where \( a \in (2m/\e, \infty) \).
However, in the following sections, we must strictly enforce the existence of the outer horizon (or at the very least the extremal horizon) in order to ensure that our problem has the correct radiative boundary conditions when solving the Regge--Wheeler equation.
Consequently, in the following section we are explicitly assuming \( a \in [0,2m/\e] \).

\subsection{Regge--Wheeler potential}

We will now calculate the Regge--Wheeler potential for spin 0 and spin 1 perturbations in our spacetime.
The spin 2 perturbations are somewhat messier, and hence do not lend themselves nicely to the WKB approximation and subsequent computation of the quasi-normal modes.
First, we implicitly define the tortoise coordinate by
\begin{equation}
\dd{r^*} = \frac{\dd{r}}{1-\frac{2m\,\e^{-a/r}}{r}}.
\label{tortoise}
\end{equation}
Although this equation is not analytically integrable, we can still conduct an analysis of the Regge--Wheeler potential through this \emph{implicit} definition. 
The coordinate transformation \eqref{tortoise} allows us to write the spacetime metric \eqref{metric} in the following ``isothermal'' form\footnote{Coordinates of this form are also commonly called Buchdahl coordinates.}:
\begin{equation}
\dd{s}^2 = \left( 1- \frac{2m\,\e^{-a/r}}{r} \right) \bigg\{ -\dd{t}^2 + \dd{r_*}^2 \bigg\} + r^2 \left( \dd{\theta}^2 + \sin^2\theta \dd{\phi}^2 \right),
\end{equation}
which we may re-write as
\begin{equation}
\dd{s}^2 = A(r_*)^2 \big\{ -\dd{t}^2 + \dd{r}_*^2 \big\} + B(r_*)^2 \left( \dd{\theta}^2 + \sin^2\theta \dd{\phi}^2 \right).
\end{equation}

In Regge and Wheeler's original work \cite{RW-original}, they show for perturbations in a black hole spacetime that assuming a separable wave form of the type
\begin{equation}
\Psi(t,r,\theta,\phi) = \e^{i\omega t} \psi(r) Y(\theta,\phi),
\label{sepwaveform}
\end{equation}
results in the differential equation (now called the Regge--Wheeler equation):
\begin{equation}
\pdv[2]{\psi(r)}{r_*} + \big\{ \omega^2 - \V_S \big\} \psi(r) = 0.
\label{RWeq}
\end{equation}
Here \( Y(\theta,\phi) \) represents the spherical harmonic functions, while \( \psi(r) \) is a propagating field in our spacetime, \( \V \) is a spin-dependent potential (the ``Regge--Wheeler potential''), and \( \omega \) is some temporal frequency component in the Fourier domain \cite{AneeshBoseKarRW, expmetric, VisserRW, FizievRW, RW-original, RW-QNMs-Schwarzschild, blackbounce}.

The method for solving equation \eqref{RWeq} is dependent on the spin of the perturbations and the background spacetime.
For scalar perturbations (\(S=0\)), one must solve the massless Klein--Gordon equation, \( \square\psi(r) = 0 \); whilst for electromagnetic (\(S=1\)) perturbations, one must analyse the the electromagnetic four-potential subject to Maxwell's equations.
Further details can be found in references \cite{RW-QNMs-dS-AdS-Schwarzschild, VisserRW, RW-original, RW-QNMs-Schwarzschild}.
For spin 0 and spin 1 perturbations, this yields the result \cite{expmetric, blackbounce}: 
\begin{equation}
\V_{S} = \left\{\frac{A^2}{B^2}\right\} \ell(\ell+1) + (1-S) \frac{\partial^2_{r_*}B}{B}.
\end{equation}

For our spacetime, we have \( \partial_{r_*} = \left(1-\frac{2m\,\e^{-a/r}}{r}\right) \partial_r \) and \( B(r_*) = r \).
Hence,
\begin{align}
\frac{\partial^2_{r_*}B}{B} 
&= \frac{\left(1-\frac{2m\,\e^{-a/r}}{r}\right)\partial_r\left[1-\frac{2m\,\e^{-a/r}}{r}\right]}{r} \notag \\
&= \left( \frac{r - m\,\e^{-a/r}}{r^3} \right) \left( \frac{2m\,\e^{-a/r}(r-a)}{r^2} \right),
\end{align}
and so we have the exact result:
\begin{equation}
\V_{S} = \left( \frac{r - 2m\,\e^{-a/r}}{r^3} \right) \left\{ \ell(\ell+1) + (1-S) \frac{2m\,\e^{-a/r}(r-a)}{r^2} \right\}.
\end{equation}
That is,
\begin{equation}
\V_{S} = \left( 1- \frac{2m\,\e^{-a/r}}{r} \right) \left\{ {\ell(\ell+1)\over r^2}  + (1-S) \frac{2m\,\e^{-a/r}(1-a/r)}{r^3} \right\}.
\label{eq:RW-Mink-core}
\end{equation}
Thus, at the horizon, \( r_H = 2m \,\e^{W\left(-\frac{a}{2m}\right)} \), the Regge--Wheeler potential vanishes.
By taking the limit \( a\rightarrow0 \) we recover the known Regge--Wheeler potentials for spin 0 and spin 1 perturbations in the Schwarzschild spacetime:
\begin{equation}
\V_{\mathrm{Sch};\,S} = \lim_{a\rightarrow0} \V_{0,1} = \left(1-\frac{2m}{r}\right) \left\{ \frac{\ell(\ell+1)}{r^2} + (1-S) \frac{2m}{r^3} \right\}.
\end{equation}

Note that as we are only analysing spin 0 and spin 1 perturbations, this is a different result to the spin 2 case analysed in Regge and Wheeler's original work (reference \cite{RW-original}).
However, it agrees well with later results extending the work of Regge and Wheeler to spin 0 and spin 1 perturbations, for example reference \cite{RW-QNMs-Schwarzschild}.

\subsection{Quasi-normal modes}

We now wish to calculate the quasi-normal modes for our spacetime.
We define the quasi-normal modes in the standard way: they are the \( \omega \) which are solutions to equation \eqref{sepwaveform}, and satisfy the ``radiation'' boundary conditions that \( \Psi \) is purely outgoing at spatial infinity and purely ingoing at the horizon \cite{QNMs-Mashhoon, RW-QNMs-Schwarzschild}.
Due to the inherent difficulty of analytically solving the Regge--Wheeler equation for the quasi-normal modes, a standard approach in the literature is to use the WKB approximation.
Although the WKB method was originally constructed to solve Schr\"odinger-type equations in quantum mechanics, the close resemblance between the Regge--Wheeler equation \eqref{RWeq} and the Schr\"odinger equation allows for it to be relatively easily adapted to the general relativistic setting. 
The WKB approximation was first applied to the calculation of quasi-normal modes by Iyer, Schutz, and Will \cite{Iyer:1987, Schutz:1985}.
See also the earlier work on quasi-normal modes by Blome and Mashhoon which does not use the WKB approximation \cite{QNMs-Mashhoon}. 

Computing a WKB approximation to first-order yields a simple and tractable approximation to the spin-dependent quasi-normal modes for a black hole spacetime \cite{ QNMs-Mashhoon, QNMs-WKB-expansion, RW-QNMs-Schwarzschild}:
\begin{equation}
\omega_S^2 \approx \bigg[ \V_S(r) + i \big(n+\textstyle{1\over2}\big) \sqrt{-2 \, \partial_{r_*}^2 \V_S(r)} \bigg]_{r=r_0}.
\end{equation}
Here \( S\in\{0,1\} \), while  \( n \in \mathbb{N} \) is the overtone number, and \( r = r_0 \) is the coordinate location which maximises the Regge--Wheeler potential.
In-depth calculations of the WKB approximation up to higher orders in a general setting can be found in references \cite{ QNMs-Mashhoon, QNMs-WKB-expansion, RW-QNMs-Schwarzschild}.

\subsubsection{Spin 1}

For spin 1 particles the $\V_1(r)$ Regge--Wheeler potential is proportional to the $V_0(r)$ effective potential used for determining the photon sphere for massless particles. Specifically, we find
\begin{equation}
\pdv{\, \V_1(r)}{r} = \frac{2\ell(\ell+1)}{r^5}\bigg\{ m \, \e^{-a/r}(3r-a) - r^2 \bigg\},
\end{equation}
and so by comparison with equation \eqref{eq;photonsphere}, we see that the Regge--Wheeler potential is maximised at the location of the photon sphere. 
Thus, we can immediately obtain the spin 1 first-order WKB approximation for the real part of our quasi-normal modes in terms of the approximate photon sphere location \eqref{E:small-a}:
\begin{equation}
\Re(\omega_1^2) \approx \V_1\left(3m - \tfrac{4}{3}a\right) 
=  9 \, \ell(\ell+1) \left( \frac{9m-4a-6m\,\e^{3a\over4a-9m}}{(9m-4a)^3} \right).
\end{equation}
Equivalently, using equations \eqref{E:non-pert}, we can eliminate \( a \) or \( m \) and express this analytically (although \emph{implicitly}) in terms of the exact location of the photon sphere:
\begin{equation}
\Re(\omega_1^2) 
\approx \frac{\ell(\ell+1)}{r_c^2} \left( 1 - \frac{2}{3r_c - a} \right) 
= \frac{\ell(\ell+1)}{r_c^2} \left( 1- \frac{2m}{\e^3 r_c^2} \e^{\e^3r_c\over m} \right).
\end{equation}
We now wish to calculate the imaginary part of the quasi-normal modes.
First, we note that in general
\begin{multline}
\eval{ \pdv[2]{\V_S}{r_*} }_{r=r_0} = \left(1-\frac{2m\,\e^{-a/r}}{r}\right) \times \\
\eval{ \left[ \left(1-\frac{2m\,\e^{-a/r}}{r}\right)\pdv[2]{\V_S}{r} + \frac{2m\,\e^{-a/r}(r-a)}{r^3} \pdv{\V_S}{r} \right] }_{r=r_0}.
\end{multline}
But \( \partial\V_S/\partial r \eval_{r=r_0} = 0 \), and so this reduces to
\begin{equation}
\eval{ \pdv[2]{\V_S}{r_*} }_{r=r_0} = \eval{ \left(1-\frac{2m\,\e^{-a/r}}{r}\right)^2 \pdv[2]{\V_S}{r} }_{r=r_0}.
\end{equation}
Thus, for our spin 1 particle we find
\begin{equation}
\eval{ \pdv[2]{\V_1}{r_*} }_{r=r_0} 
= \ell(\ell+1) \left(1-\frac{2m\,\e^{-a/r_c}}{r_c}\right)^2 \left\{ \frac{6}{r_c^4} - (a-2r_c)(a-6r_c)\frac{2m\,\e^{-a/r_c}}{r_c^7} \right\}.
\end{equation}
In terms of \( m(r_c,a) \) and \( a(r_c,m) \) from equations \eqref{E:non-pert}, we have:
\begin{align}
\Im(\omega_1^2) &\approx 
\frac{(2n+1)\sqrt{\ell(\ell+1)(r_c-a)(3r_c^2-5r_ca+a^2)}}{(3r_c-a) r_c^{5/2}} \notag \\
&= \frac{(2n+1)}{r_c^2 W(\e^3 r_c/m)} \bigg\{\ell(\ell+1)\big[2-W(\e^3 r_c/m)\big] \times \notag \\
&\hspace{10em}\big[W(\e^3 r_c/m)^2-W(\e^3 r_c/m)-3\big]\bigg\}^{1/2}.
\end{align}
Alternatively, one could use the approximate photon sphere location \eqref{E:small-a} and obtain the approximate (although \emph{explicit}) result:
\begin{multline}
\Im(\omega_1^2) \approx 9\sqrt{3} \, (2n+1) \left(1 - \frac{6m\,\e^{3a\over4a-9m}}{9m-4a} \right) \times \\
\sqrt{\ell(\ell+1) \left( \frac{27m\,\e^{3a\over4a-9m} (18m-11a)(2m-a) - (9m-4a)^3}{(9m-4a)^7} \right)}.
\end{multline}

In summary, our first-order WKB approximation of the spin 1 quasi-normal modes can be written analytically (but \emph{implicitly}) as
\begin{align}
\omega_1^2 
&\approx \frac{\ell(\ell+1)}{r_c^2(3r_c-a)} \left\{ 3r_c-a+2+i(2n+1)\sqrt{\frac{(r_c-a)(3r_c^2-5r_ca+a^2)}{l(l+1)r_c}} \right\}	\notag \\
&= \frac{\ell(\ell+1)m\,\e^{\e^3r_c/m}}{\e^3r_c^3} \left\{ \frac{m\,\e^{\e^3r_c/m}}{\e^3r_c} - {2\over r_c} \, + \right. \notag \\
&\hspace{2em} \left. i(2n+1)\sqrt{ \frac{\big[2-W(\e^3 r_c/m)\big]\big[W(\e^3 r_c/m)^2-W(\e^3 r_c/m)-3\big]} {\ell(\ell+1)} } \right\}.
\end{align}
Or approximately (but \emph{explicitly}) as
\begin{multline}
\omega_1^2 \approx 9\ell(\ell+1) \left( \frac{9m-4a-6m\,\e^{3a\over4a-9m}}{(9m-4a)^3} \right) \times	\\
\left\{ 1 + (2n+1) \sqrt{3} \, i \, \sqrt{\frac{27m\,\e^{3a\over4a-9m}(18m-11a)(2m-a) - (9m-4a)^3}{\ell(\ell+1)(9m-4a)^3}} \right\}.
\end{multline}
In either case, the Schwarzschild \( a\rightarrow0 \) limit yields
\begin{equation}
\omega^2_{Sch.,1} \approx \frac{\ell(\ell+1)}{27m^2}\left( 1 + \frac{i(2n+1)}{\sqrt{\ell(\ell+1)}} \right),
\end{equation}
which agrees with existing work in the literature \cite{QNMs-Mashhoon, RW-QNMs-Schwarzschild}.

Note that, although spin 1 perturbations have a direct physical relevance due to their representation of photons and classical electro-magnetic fields, spin 0 perturbations are not so physically relevant. 
Additionally, spin 0 calculations are considerably more algebraically tedious due the extra term present in the Regge--Wheeler potential \eqref{eq:RW-Mink-core}.
As such, we will not endeavour to calculate them in this thesis.

\section{Summary} 

Overall, we see that the boundary of the stability region for timelike circular orbits 
is rather complicated. In terms of the variable $w=r_c/a$:
\begin{itemize}
\item For $w\in ({5+\sqrt{13}\over6}, \infty)$ we have an ESCO. \\[3pt]
This ESCO then subdivides as follows:
\begin{itemize}
\item For $w \in( 2.210375896, \infty)$ we have an ISCO.
\item For $w \in ({5+\sqrt{13}\over6}, 2.210375896)$ we have an OSCO.
\end{itemize}
\item For $w\in (1, {5+\sqrt{13}\over6})$ the stability region is bounded by a stable photon orbit.
\item The line $w=1$ bounds the stability and existence region for timelike circular orbits from below.
\end{itemize}
This is considerably more complicated than might reasonably have been expected.

The spin-1 quasi-normal modes were found to be fairly tractable in the first-order WKB approximation.
However, higher order calculations would be algebraically non-trivial.
Similarly, the spin 0 quasi-normal-modes were not calculated due to the intractability of the results and their lack of physical relevance.
\chapter{From regular black hole to thin-shell wormhole}\label{C:thin-shell-wormhole}
\chaptermark{\small{Regular black hole to thin-shell wormhole}}

Using the regular black hole with asymptotically Minkowski core from chapter \ref{C:Mink-core} as a template, we will construct a spherically symmetric thin-shell traversable wormhole using the ``cut-and-paste'' technique, thereby constructing yet another black hole mimicker.
We calculate the surface stress-energy at the wormhole throat, and the stability of the wormhole is analysed.
An important result is that, (as compared to their Schwarzschild thin-shell counterparts),  increasing the exponential suppression of the Mis\-ner--Sharp quasi-local mass by increasing the suppression parameter $a$, also considerably increases  the stability regions for these thin-shell wormholes, and furthermore minimises the amount of energy condition violating exotic matter required to keep the wormhole throat open.

\section{Thin--shell wormhole framework}

\subsection{Background} 

After the renaissance of wormhole physics in the late 1980s~\cite{Morris:1988cz, Morris:1988tu}, there was very rapid progress of investigations into thin-shell wormholes. 
See, for instance, references \cite{visser-surgical, visser-wormhole-examples} and \cite{Visser:lorentzian-wormholes}.  
A relatively recent general analysis and summary can be found in reference \cite{Garcia:2011aa}, whilst a very recent brief and cogent literature survey can be found in~\cite{Lobo:2020}.

The central idea behind thin-shell wormholes is to take two bulk spacetimes, excise two regions with isometric boundaries, and then identify the boundaries \cite{visser-surgical, visser-wormhole-examples}. 
This is effectively a modification of the abstract mathematical notion of the ``connected sum'' of manifolds, wherein one uses metrical information, not just topological information (further details can be found in references \cite{Adams:2008, Massey:1991}). 
Key ingredients of the analysis are the two bulk metrics, the (isometric) induced metrics (intrinsic 3-metrics) on the boundaries (the first fundamental forms), and the extrinsic curvatures of these boundaries in the two bulk spacetimes (the second fundamental forms). 
On the boundary itself, there is a delta-function distribution of stress-energy that is related to the discontinuity in the extrinsic curvatures \cite{visser-surgical, visser-wormhole-examples} in a very precise and specific manner \cite{Visser:lorentzian-wormholes}. 

We shall now apply this very general and flexible formalism in the specific case of spherical symmetry, choosing the bulk spacetimes to be the regular black hole with asymptotically Minkowski core studied in the last chapter (i.e. metric \eqref{metric}).

\subsection{Construction}

We start with the spacetime of the regular black hole with an asymptotically Minkowski core, which we will reproduce in this chapter for ease of reference:
\begin{equation}
\dd s^2 = -\left(1-\frac{2m\,\e^{-a/r}}{r}\right)\dd{t}^2 + \frac{\dd{r}^2}{1-\frac{2m\,\e^{-a/r}}{r}} + r^2\left(\dd{\theta}^2 + \sin^2\theta \dd{\phi}^2\right).
\label{metricworm}
\end{equation}
Recall that this spacetime possesses horizons located at
\begin{equation}
r_H = 2m\;\e^{W\left(-\frac{a}{2m}\right)} = {a\over |W\left(-\frac{a}{2m}\right)|},
\label{horizon}
\end{equation}
where $W(x)$ is the real-valued Lambert $W$ function, which is negative for those negative arguments where it is defined.
Equation \eqref{horizon} implies that an outer horizon and an inner horizon exist, which are obtained by either taking the $W_0$ or the $W_{-1}$ branch of the Lambert $W$ function, respectively.
Recall also, that in order for horizons to be present, equation \eqref{horizon} forces the parameter $a$ to lie in the interval $a \in \left(0, {2m}/{\e}\right]$, and in particular $a \leq 2m/\e$.
For the specific case of $a=2m/\e$, one has $W\left(-{a}/{2m}\right)\to 
W\left(-1/\e\right)= -1$. 
Then the two horizons merge at $r_{H^\pm }= a$ and the regular black hole is extremal.
If $a>2m/\e$, the horizon locations are undefined and we are dealing with a horizonless compact object.

For the purposes of thin--shell construction, if horizons are present, then we shall perform spacetime surgery \emph{outside} the outer horizons, where we have good control over the physics, and hence we shall have a thin-shell located at some $r>r_{H^+}>a$.
If horizons are not present, $a>2m/\e$, then we could in principle perform spacetime surgery at any nonzero value of $r$.

In the following, we will consider two copies of the regular black hole spacetime given by the line element \eqref{metricworm}, and subsequently analyse the manifold formed by surgically removing the regions 
$r \in (0,R(\tau))$, with the surface $R(\tau)$ lying \emph{outside} both \emph{outer} horizons (if present) of each spacetime, and ``gluing'' them together along this new boundary.

\subsection{Energy conditions in the bulk spacetime} 

The bulk spacetime has the following stress-energy tensor profile:
\begin{subequations}
\begin{align}
\rho &= -p_r = \frac{ma \, \e^{-a/r}}{4\pi r^4}, 	\label{rhopr} \\
p_t &= - \frac{ma(a-2r) \e^{-a/r}}{8\pi r^5},
\end{align}
\end{subequations}
where $\rho$ is the energy density, $p_r$ and $p_t$ are the pressures in the radial and tangential directions, respectively (c.f. equation \eqref{eq:Tab-spherical}).

Recall that in order to satisfy the null energy condition (NEC), we require \( \rho + p_r \geq 0 \) and \( \rho + p_t \geq 0 \). 
Indeed, we have \( \rho + p_r = 0 \) globally, however 
\begin{equation}
\rho + p_t = \frac{r}{2}\rho' = \frac{ma \, \e^{-a/r}}{8\pi r^5}(a-4r),
\end{equation}
and so the NEC is only satisfied in the region \( r \leq a/4 \). 
In view of the fact that the outer horizon (if it exists) is located at $r_{H^+} = 2me^{W_{0}\left(-\frac{a}{2m}\right)}>a$, corresponding to possible locations $r_{H^+}\in\left(a, +\infty\right)$, and we `chop' the spacetime outside any horizons that are present, we may conclude that the transverse NEC is manifestly violated in the bulk regions of the constructed spacetime.\footnote{If horizons are not present, then one might be able to satisfy the NEC for small enough $r$.}

Similarly, we find
\begin{equation}
\rho + p_r + 2p_t = \frac{ma(2r-a) \e^{-a/r}}{4\pi r^5},
\end{equation}
and it can be clearly seen that this is only non-negative in the region \( r \geq a/2 \), and so, (regardless of whether or not horizons are present), there is no region in which both the {NEC} and the {SEC} are simultaneously satisfied. However, in the
presence of horizons, this aspect of the SEC will be globally satisfied in the bulk regions.\footnote{If horizons are not present, then one might be able to violate the SEC for small enough $r$.}
This violation of the energy conditions is in-keeping with every example of a black hole mimicker we have encountered so far in this thesis.

\subsection{Four-velocity, unit normal, and extrinsic curvature of the throat}

We now allow the boundary surface \( \Sigma \) to be dynamic. 
For tractability, we consider dynamic perturbations to the radial location of the wormhole throat \emph{only}. 
It follows that the intrinsic metric on \( \Sigma \) is given by:
\begin{equation}
\dd s_\Sigma^2 = -\dd \tau^2 + R(\tau)^2\; (\dd \theta^2 + \sin^2\theta \dd \phi^2),
\end{equation}
with coordinate chart \( x^\mu(\tau,\theta,\phi) = (t(\tau), R(\tau), \theta, \phi) \), where $\tau$ is the proper time of an observer comoving with $\Sigma$.
The implied form for the four-velocity of an observer (or a  piece of stress--energy) located on the junction surface is thus:
\begin{equation}
U^\mu_{\pm} = \left( \dv{t}{\tau}, \, \dv{R}{\tau}, \, 0, \, 0 \right),
\end{equation}
and takes the following explicit form
\begin{equation}
U^\mu_\pm = \left( \frac{\sqrt{\gpm + \dot{R}^2}}{\gpm}, \, \dot{R}, \, 0, \, 0 \right).
\end{equation}

The hyper-surface $\Sigma$ is defined by the function $f(x^\mu(\xi^i))=r-R(\tau)=0$, and so the unit normals to this surface are defined by
\begin{equation}
n_\mu = \pm \abs{ g^{\alpha\beta}\frac{\partial f}{\partial x^{\alpha}} \frac{\partial f}{\partial x^{\beta}} }^{-\frac{1}{2}} \pdv{f}{x^\mu}.
\label{normalSigma}
\end{equation}
A trivial but quite lengthy calculation yields the following unit normal vector to~$\Sigma$:
\begin{equation}
n^{\mu} = \pm\left( \frac{\dot{R}}{\gpm}, \, \sqrt{\gpm +\dot{R}^2}, \, 0, \, 0 \right).
\end{equation}

An essential ingredient in the thin-shell formalism is the extrinsic curvature, or second fundamental form, which is defined as $K_{ij}=n_{{(}\mu;\nu{)}}e^{\mu}_{(i)}e^{\nu}_{(j)}$,
where $n_{\mu}$ is the unit normal 4-vector (\ref{normalSigma}) to the surface $\Sigma$, and $e^{\mu}_{(i)}$ are the components of the holonomic basis of vectors tangent to $\Sigma$. Thus,  in terms of the above quantities, the extrinsic curvature can be expressed in the more tractable form:
\begin{equation}
    K_{ij}^{\pm} = -n_{\mu}\left(\frac{\partial^{2}x^{\mu}}{\partial\xi^{i}\partial\xi^{j}} + \Gamma^{\mu\pm}_{\ \alpha\beta}\frac{\partial x^{\alpha}}{\partial\xi^{i}}\frac{\partial x^{\beta}}{\partial\xi^{j}}\right).
\end{equation}

A quick calculation yields the $K_{\theta\theta}^{\pm}$ component, where the mixed tensor is given~by:
\begin{equation}
K^{\theta\pm}_{\ \theta} = g^{\theta\theta}\,K_{\theta\theta}^{\pm} = \pm\frac{1}{R}\sqrt{\gpm +\dot{R}^{2}}.
\end{equation}

A lengthy calculation yields the $K^{\tau\pm}_{\ \tau}$ component, but we make use of the formalism discussed in~\cite{Visser:lorentzian-wormholes}, which is rather pedagogical. 
To this effect, note that we have
\begin{align}
    K^{\pm}_{\tau\tau} &= K^{\pm}_{\mu\nu}U^{\mu}U^{\nu} = \nabla^{\pm}{}_{(\mu} \; n_{\nu)} U^{\mu}U^{\nu} 	\notag  \\
    &=  \left[\frac{1}{2}\left(\nabla^{\pm}_{\mu}n_{\nu}+\nabla^{\pm}_{\nu}n_{\mu}\right)\right]U^{\mu}U^{\nu} = \nabla^{\pm}_{\mu}n_{\nu}U^{\mu}U^{\nu}.
\end{align}
Taking into account $K^{\tau\pm}_{\ \tau} = -K^{\pm}_{\tau\tau}$, we have the following:
\begin{equation}
K^{\tau\pm}_{\ \tau} = -\left(\nabla^{\pm}_{\mu}n_{\nu}\right)U^{\mu}U^{\nu} 
= +U^{\mu}n_{\nu}\left(\nabla^{\pm}_{\mu}U^{\nu}\right)
= n_{\nu}\left(U^{\mu}\nabla^{\pm}_{\mu}U^{\nu}\right) 
= n_{\nu}A^{\nu}_{\pm}, 
\end{equation}
where $A^{\nu}_{\pm}$ is the $4$--acceleration of the throat. 
Spherical symmetry implies that $A^{\nu}_{\pm}\propto n^{\nu}$, i.e. $A^{\nu}_{\pm} = \abs{A_{\pm}} n^{\nu}$. 
Therefore:
\begin{equation}
K^{\tau\pm}_{\ \tau} = \left( n_{\nu} \abs{A_{\pm}} \right) n^{\nu} = \abs{A_{\pm}}.
\end{equation}
That is, $K^{\tau\pm}_{\ \tau}$ is simply equal to the magnitude of the $4$--acceleration of the throat.

The underlying bulk geometry possesses a Killing vector $k^{\mu} = \left(\partial_{t}\right)^{\mu} = \left(1, 0, 0, 0\right)^\mu$. 
Lowering the index on this Killing vector, we obtain (calculating \emph{at the throat} where $r=R(\tau)$)
\begin{equation} 
k_{\mu} = \left(-\left[\ggg\right], 0, 0, 0\right).
\end{equation}
We now examine the quantity $\frac{\d}{\d\tau}\left(k_{\mu}U^{\mu}\right)$, which we can compute in two different ways to obtain the magnitude of the $4$--acceleration as a function of $R$, its first and second derivatives, $a$ and $m$:

\begin{itemize}
\item First calculation (employing Killing's equation):   
    \begin{align}
        \dv{\tau}\left(k_{\mu}U^{\mu}\right) 
        &= U^{\nu}\nabla_{\nu}\left(k_{\mu}U^{\mu}\right)
           = \left(\nabla^{\pm}_{\nu}k_{\mu}\right)U^{\mu}U^{\nu}+k_{\mu}\dv{U^{\mu}}{\tau} 
        \nonumber \\
        &= k_{\mu}\dv{U^{\mu}}{\tau} = k_{\mu}A^{\mu}_{\pm} 
        = k_{\mu}\vert A_{\pm}\vert n^{\mu} = \vert A_{\pm}\vert\left(k_{\mu}n^{\mu}\right) 
        \nonumber \\
        &= \mp \vert A_{\pm}\;  \vert \dot{R}. \label{ktt1}
    \end{align}
\item Second calculation:
    \begin{align}
        \dv{\tau}\left(k_{\mu}U^{\mu}\right) &= \dv{\tau}\left(k_{t}U^{t}\right) 
        = -\dv{\tau}\left[\sqrt{\ggg +\dot{R}^{2}}\,\right] \nonumber \\
        &= -\frac{\dot{R}\left[\ddot{R}+\frac{m\,\e^{-\frac{a}{R}}}{R^{2}}\left(1-\frac{a}{R}\right)\right]}{\sqrt{\ggg +\dot{R}^{2}}}. \label{ktt2}
    \end{align}
\end{itemize}

\noindent Comparing equations \eqref{ktt1} and \eqref{ktt2}, we obtain:
\begin{equation}
\mp\vert A_{\pm}\vert\dot{R} = -\frac{\dot{R}\left[\ddot{R}+\frac{m\,\e^{-\frac{a}{R}}}{R^{2}}\left(1-\frac{a}{R}\right)\right]}{\sqrt{\ggg +\dot{R}^{2}}},
\end{equation}
and so
\begin{equation}
\quad K^{\tau\pm}_{\ \tau} = \vert A_{\pm}\vert = \pm \left[\frac{\ddot{R} + \frac{m_{\pm}\e^{-\frac{a_{\pm}}{R}}}{R^{2}}\left(1-\frac{a_{\pm}}{R}\right)}{\sqrt{\gpm +\dot{R}^{2}}}\right].
\end{equation}

\noindent In summary, the extrinsic curvature components are given by
\begin{subequations}
\begin{align}
K\indices{^\theta_\theta^\pm} &= K\indices{^\phi_\phi^\pm} = \pm \frac{1}{R} \sqrt{\gpm + \dot{R}^2},		\label{curvtheta} \\
K\indices{^\tau_\tau^\pm} &= \pm \left[ \frac{m_\pm \, \e^{-a_\pm/R} (R-a_\pm) + R^3 \ddot{R}}{R^3 \sqrt{\gpm+\dot{R}^2}} \right],
\label{curvtau}
\end{align}
\end{subequations}
respectively.

\subsection{Surface stress--energy}

For our thin--shell analysis, the extrinsic curvature need not be continuous across the junction boundary \( \Sigma \).
Thus, we denote the discontinuity by \( \kappa_{ij} = K_{ij}^+ - K_{ij}^- \).
The surface stress--energy tensor on \( \Sigma \), \( S\indices{^i_j} \), can be calculated via the Lanczos equations:
\begin{equation}
S\indices{^i_j} = -\frac{1}{8\pi} \left( \kappa\indices{^i_j} - \delta\indices{^i_j} \kappa\indices{^k_k} \right).
\end{equation}
Due to spherical symmetry, the discontinuity can be represented by a diagonal matrix: \( \kappa\indices{^i_j} = \text{diag}(\kappa\indices{^\tau_\tau}, \kappa\indices{^\theta_\theta}, \kappa\indices{^\phi_\phi}) \), and so the surface stress--energy tensor simply reduces to \( S\indices{^i_j} = \text{diag}(-\sigma, \P, \P) \), where \( \sigma \) is the surface energy density and \( \P \) is the surface pressure.
Thus, with \( \kappa\indices{^k_k} = \kappa\indices{^\tau_\tau} + 2\kappa\indices{^\theta_\theta} \), the Lanczos equations imply:
\begin{subequations}
\begin{align}
\sigma &= -\frac{1}{4\pi}\kappa\indices{^\theta_\theta},	 \\
\P &= \frac{1}{8\pi}(\kappa\indices{^\tau_\tau} + \kappa\indices{^\theta_\theta}).
\end{align}
\end{subequations}
Using the extrinsic curvature components given in equations \eqref{curvtheta} and \eqref{curvtau}, the surface stress--energy at the junction throat \( \Sigma \) is finally found to be:
\begin{subequations}
\begin{equation}
\sigma = -\frac{1}{4\pi R} \left[ \sqrt{\gp + \dot{R}^2} \, + \ \sqrt{\gm + \dot{R}^2} \right],	\label{eq;surfE}
\end{equation}
\begin{multline}
\P = \frac{1}{8\pi R} \left[ \frac{1+\dot{R}^2 + R\ddot{R} - \frac{m_+\,\e^{-a_+/R}}{R^2}(R+a_+)}{\sqrt{\gp+\dot{R}^2}} \right. + \\
\left. \frac{1+\dot{R}^2 + R\ddot{R} - \frac{m_-\,\e^{-a_-/R}}{R^2}(R+a_-)}{\sqrt{\gm+\dot{R}^2}} \right]. 	
\label{eq;surfP}
\end{multline}
\end{subequations}
It can be seen from equation \eqref{eq;surfE} that negative energy is needed to keep the wormhole throat open, implying that exotic matter would be required.
This is in-keeping with every wormhole example we have encountered so far in this thesis.

An important ingredient explored in recent work~\cite{Lobo:2020, Garcia:2011aa} is the potential presence of an additional energy flux term, which arises from the conservation identity. This identity is obtained by combining the second contracted Gauss--Codazzi equation (or the ``ADM" constraint) $G_{\mu \nu}\,e^{\mu}_{(i)}n^{\nu}=K^j_{\ i|j}-K,_{i}$ with the Lanczos equations, and is given by $S^{i}_{\ j|i}=-\left[T_{\mu \nu}e^{\mu}_{(j)}n^{\nu}\right]^+_-$.
Here we have adopted the standard notation for the derivatives of tensors where \( X_{i|j} = \nabla_jX_i \) and \( X_{i\,,\,j} = \partial_jX_i \).
The momentum flux term in the right hand side corresponds to the net discontinuity in the momentum which impinges on the shell. 
Note that for the present geometry, this flux term vanishes:
\begin{align}
    \left[T_{\mu\nu}e^{\mu}_{(\tau)}n^{\nu}\right]^{+}_{-} 
    &= \left[T_{\mu\nu}U^{\mu}n^{\nu}\right]^{+}_{-} 
		\notag \\    
    &=  \left[\pm\left(-T^{t}{}_{t}+T^{r}{}_{r}\right)
    \frac{\dot{R}\sqrt{1-\frac{2m_{\pm}\e^{-a_{\pm}/R}}{R}+\dot{R}^{2}}}
    {1-\frac{2m_{\pm}\e^{-a_{\pm}/R}}{R}}
    \right]^{+}_{-} = 0,
\end{align}
where $T^{t}_{\ t}=-\rho$ and $T^{r}{}_{r}=p_r$, and equation (\ref{rhopr}) yields $-T^{t}{}_{t}+T^{r}{}_{r}=\rho+p_r=0$.
Thus, the conservation identity finally provides \(S^{i}_{\ \tau|i}=0=-\big[\dot{\sigma}+2\dot{a}(\sigma +{\cal P} )/a \big]\). That is:
\begin{equation}
\sigma'=-\frac{2}{a}\,(\sigma+{\cal P}).
\label{consequation2}
\end{equation}

\subsection{Stability analysis}\label{sec:stability}

\subsubsection{Equation of motion}

In order to force stability constraints on the mass of the thin-shell, $m_{s}(R)$, let us consider the thin--shell equation of motion, and write it in the form $\frac{1}{2}\dot{R}^{2}+V(R) = 0$. 
To obtain an explicit expression for the potential $V(R)$, taking into account $m_{s}(R) = 4\pi R^{2}\sigma(R)$, we rearrange equation~(\ref{eq;surfE}) to derive:
\begin{equation}
    V(R) = -\frac{1}{2}\dot{R}^{2} = \frac{1}{2}\left\lbrace 1+\frac{\bar{\Delta}(R)}{R}-\left[\frac{m_{s}(R)}{2R}\right]^{2}-\left[\frac{\Delta(R)}{m_{s}(R)}\right]^{2}\right\rbrace.
\end{equation}
Here $\bar{\Delta}(R)$ and $\Delta(R)$ are defined as:
\begin{equation}
    \bar{\Delta}(R) = m_{+}\e^{-a_{+}/R}+m_{-}\e^{-a_{-}/R} \,, \qquad 
    \Delta(R) = m_{+}\e^{-a_{+}/R}-m_{-}\e^{-a_{-}/R}\,.
\end{equation}
Having obtained this explicit form for $V(R)$, we may now recast the surface energy density $\sigma$ as a function of the effective potential:
\begin{multline}\label{sigmaV(R)}
    \sigma(R) = -\frac{1}{4\pi R} \left[ \sqrt{\gp - 2V(R)} \right. + \\
    \left. \sqrt{\gm - 2V(R)} \, \right].
\end{multline}

\subsubsection{Linearized equation of motion}

Let us assume there exists some static solution at $R=R_{0}$, and linearise around it accordingly. The equation of motion is $\frac{1}{2}\dot{R}^{2} + V(R) = 0$, which also directly yields that $\ddot{R} = -V'(R)$.
A second--order Taylor series expansion of $V(R)$ about $R_{0}$ yields:
\begin{equation}
    V(R) = V(R_{0})+V'(R_{0})\,(R-R_{0})+\frac{1}{2}V''(R_{0})\,(R-R_{0})^{2}+O[(R-R_{0})^{3}].
\end{equation}
Various simplifications ensue due to our solution being static, namely, 
$\dot{R}_{0} = \ddot{R}_{0} = 0$ and $V'(R_{0}) = -\ddot{R}_{0} = 0$.
Thus, our Taylor series for $V(R)$ reduces to:
\begin{equation}
    V(R) = \frac{1}{2}V''(R_{0})(R-R_{0})^{2}+O[(R-R_{0})^{3}].
\end{equation}
Now, the condition for our solution at $R_{0}$ to be stable is that $V(R_{0})$ is a local minimum; i.e. $V''(R_{0}) > 0$. Given our form for $\sigma$ as a function of $V(R)$ in equation~(\ref{sigmaV(R)}), we may now use this condition, along with $V(R_{0})=V'(R_{0})=0$, to force stability constraints on the mass of the thin--shell. 
It is in fact preferable to consider the effect of these constraints on the dimensionless quantity $\left[{m_{s}(R)}/{R}\right]$, rather than on $m_{s}(R)$ itself. 

In all generality, we have the following:
\begin{subequations}
\begin{multline}
    \frac{m_{s}(R)}{R} = 4\pi\sigma(R)R = -\Bigg[ \sqrt{\gp - 2V(R)} + \\
    \sqrt{\gm - 2V(R)} \Bigg],
\end{multline}
\begin{equation}    
    \left[\frac{m_{s}(R)}{R}\right]' = -\left[\frac{\frac{m_{+}\e^{-a_{+}/R}(R-a_{+})}{R^{3}}-V'(R)}{\sqrt{\gp - 2V(R)}} + \frac{\frac{m_{-}\e^{-a_{-}/R}(R-a_{-})}{R^{3}}-V'(R)}{\sqrt{\gm - 2V(R)}}\right],
\end{equation}
and
\begin{align}
    \left[\frac{m_{s}(R)}{R}\right]'' &= 
    \frac{\left[\frac{m_{+}\e^{-a_{+}/R}}{R^{2}}\left(1-\frac{a_{+}}{R}\right)-V'(R)\right]^{2}}{\left[1-\frac{2m_{+}\e^{-a_{+}/R}}{R}-2V(R)\right]^{\frac{3}{2}}} - \frac{\frac{m_{+}a_{+}\e^{-a_{+}/R}}{R^{4}}\left(4-\frac{a_{+}}{R}\right)-V''(R)}{\sqrt{1-\frac{2m_{+}\e^{-a_{+}/R}}{R}-2V(R)}} \notag \\
    &+\frac{\left[\frac{m_{-}\e^{-a_{-}/R}}{R^{2}}\left(1-\frac{a_{-}}{R}\right)-V'(R)\right]^{2}}{\left[1-\frac{2m_{-}\e^{-a_{-}/R}}{R}-2V(R)\right]^{\frac{3}{2}}} - \frac{\frac{m_{-}a_{-}\e^{-a_{-}/R}}{R^{4}}\left(4-\frac{a_{-}}{R}\right)-V''(R)}{\sqrt{1-\frac{2m_{-}\e^{-a_{-}/R}}{R}-2V(R)}}.
\end{align}
\end{subequations}

\subsubsection{Master equations}
Applying the stability constraints to these equations, we see that in order to have a stable solution at $R_{0}$, the thin-shell mass $m_s(R)$  must satisfy the following:

\begin{subequations}
\begin{equation}\label{stabilityineq0}
    \frac{m_{s}(R_{0})}{R_{0}} = -\left[\sqrt{1-\frac{2m_{+}\e^{-a_{+}/R_{0}}}{R_{0}}} + \sqrt{1-\frac{2m_{-}\e^{-a_{-}/R_{0}}}{R_{0}}}\right],
\end{equation}
\begin{equation}
    \left[\frac{m_{s}(R_{0})}{R_{0}}\right]' = -\left[\frac{m_{+}\e^{-a_{+}/R_{0}}(R_{0}-a_{+})}{R_{0}^{3}\sqrt{1-\frac{2m_{+}\e^{-a_{+}/R_{0}}}{R_{0}}}} + \frac{m_{-}\e^{-a_{-}/R_{0}}(R_{0}-a_{-})}{R_{0}^{3}\sqrt{1-\frac{2m_{-}\e^{-a_{-}/R_{0}}}{R_{0}}}}\right],
\end{equation}
and
\begin{multline}
    \left[\frac{m_{s}(R_{0})}{R_{0}}\right]'' \geq \frac{\left[\frac{m_{+}\e^{-a_{+}/R_{0}}}{R_{0}^{2}}\left(1-\frac{a_{+}}{R_{0}}\right)\right]^{2}}{\left[1-\frac{2m_{+}\e^{-a_{+}/R_{0}}}{R_{0}}\right]^{\frac{3}{2}}} - \frac{\frac{m_{+}a_{+}\e^{-a_{+}/R_{0}}}{R_{0}^{4}}\left(4-\frac{a_{+}}{R_{0}}\right)}{\sqrt{1-\frac{2m_{+}\e^{-a_{+}/R_{0}}}{R_{0}}}} \\
    + \frac{\left[\frac{m_{-}\e^{-a_{-}/R_{0}}}{R_{0}^{2}}\left(1-\frac{a_{-}}{R_{0}}\right)\right]^{2}}{\left[1-\frac{2m_{-}\e^{-a_{-}/R_{0}}}{R_{0}}\right]^{\frac{3}{2}}} - \frac{\frac{m_{-}a_{-}\e^{-a_{-}/R_{0}}}{R_{0}^{4}}\left(4-\frac{a_{-}}{R_{0}}\right)}{\sqrt{1-\frac{2m_{-}\e^{-a_{-}/R_{0}}}{R_{0}}}}.
    \label{stabilityineq}
\end{multline}
\end{subequations}
This final inequality gives us the stability regions for the thin--shell wormhole for various cases of the parameters $m_{\pm}$ and $a_{\pm}$.

\section{Examples}

Let us now analyse some of the more interesting specific sub--cases by fixing the parameters $a_{\pm}$ and $m_{\pm}$ and examining the corresponding stability criteria implied by equation~(\ref{stabilityineq}).

\subsection[Symmetrically vanishing $a$ parameter; asymmetric mass $m_+\neq m_-$.]{Symmetrically vanishing $a$ parameter; asymmetric\\ mass $m_+\neq m_-$}

In the bulk spacetime, we know that $a=0$ corresponds to the usual Schwarz\-schild solution. To fix $a_{+}=a_{-}=0$ in the wormhole construction while allowing asymmetric masses $m_{-}\neq m_{+}$ is to perform the thin--shell surgery exterior to two Schwarzschild spacetimes with distinct masses. By now, this particular thin--shell construction is rather well--known (see~\cite{Lobo:2020, Garcia:2011aa}). For the purposes of plotting the stability regions we define a dimensionless form for the stability constraint as follows.
First note that the equation~(\ref{stabilityineq}) reduces to:
\begin{equation}
    R_{0}^{2}\left[\frac{m_{s}(R_{0})}{R_{0}}\right]''\geq F_{1}(R_{0}, m_{\pm}) = \frac{m_{+}^{2}}{R_{0}^{2}\left(1-\frac{2m_{+}}{R_{0}}\right)^{\frac{3}{2}}} + \frac{m_{-}^{2}}{R_{0}^{2}\left(1-\frac{2m_{-}}{R_{0}}\right)^{\frac{3}{2}}}.
\end{equation}
Then, for the purposes of plotting the full domain of $R_{0}$, we shall consider the dimensionless definitions $(x=\frac{2m_{+}}{R_{0}}, \, y=\frac{2m_{-}}{R_{0}})$, so that the parameters $x$ and $y$ lie in the ranges $0<x<1$ and  $0<y<1$, respectively.
Hence,
\begin{equation}\label{F1}
    F_{1}(x, y) = \frac{1}{4}
    \left[\frac{x^{2}}{\left(1-x\right)^{\frac{3}{2}}}+\frac{y^{2}}{\left(1-y\right)^{\frac{3}{2}}}\right].
\end{equation}

\begin{figure}[t]
\centering
\includegraphics[scale=0.65]{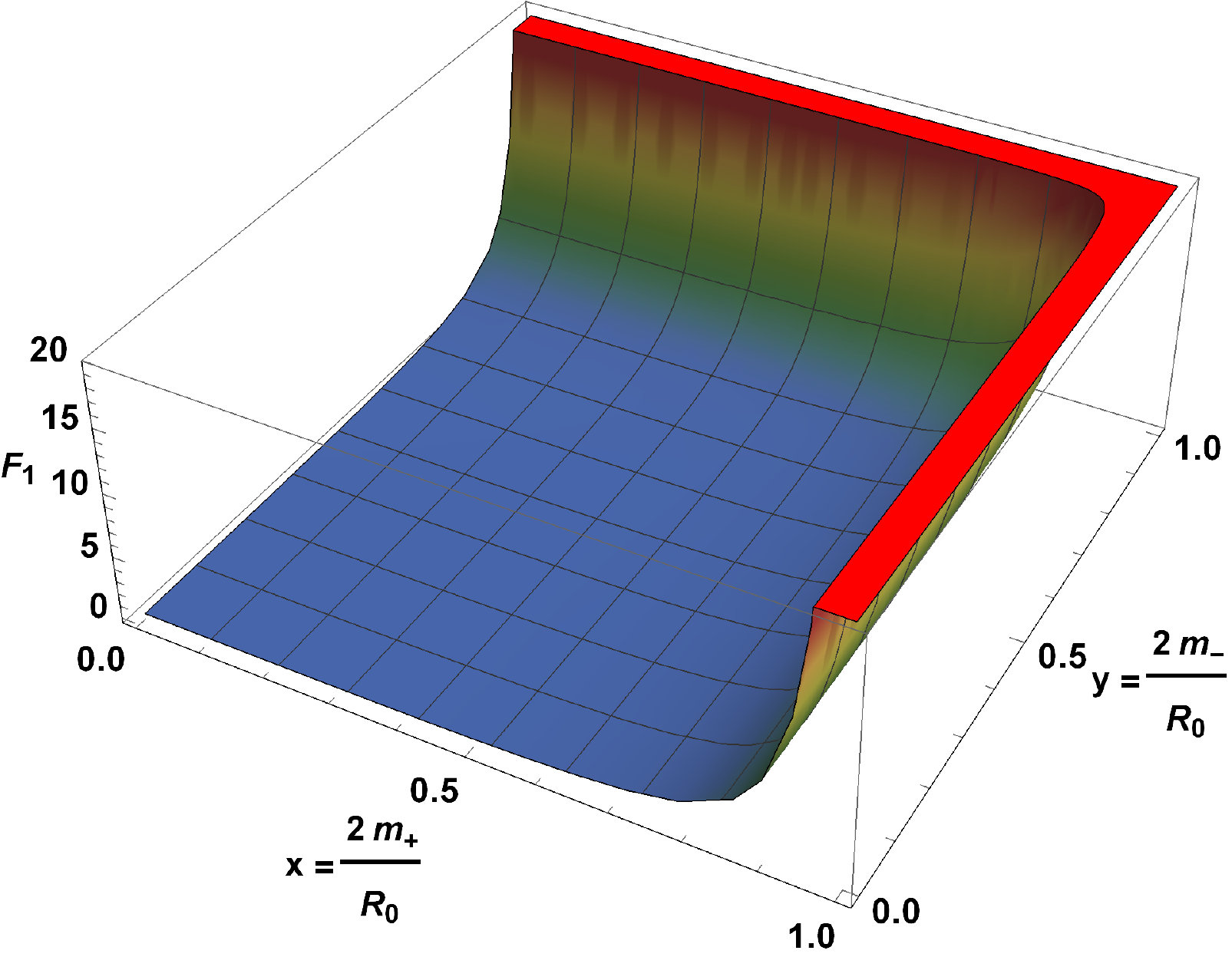}
\caption[Thin-shell wormhole stability analysis: \( a_\pm =0 \).]{Stability analysis for the $a_{\pm}=0$ case, which reduces to Schwarzschild surgery. The stability region lies above the surface $F_{1}(x, y)$, as given explicitly by equation~(\ref{F1}). The red region indicates where this function departs the specified range for $z$, and we can see that $F_{1}$ is increasing \emph{very steeply} within this region, as $x\rightarrow 1$ and/or $y\rightarrow 1$. Note that we chop off the plot  vertically once $F_1(x,y)=20$.
}
\label{F:1}
\end{figure}

We see from figure~\ref{F:1} that large stability regions exist for low values of $x$ and $y$, corresponding to $R_{0}\gg 2m_{\pm}$, while as $R_{0}\rightarrow 2m_{\pm}$ the size of the stability regions decreases steeply as we near the respective horizons.

The special case of equal masses $m_+=m_-$ simply corresponds to the diagonal $x=y$ in figure~\ref{F:1}.
Before proceeding to the next case of interest it is worth noting that, since our construction is formed from a spacetime which is strictly Minkowski in the $m\rightarrow 0$ limit, the case of symmetrically vanishing $m_{\pm}=0$ trivially reduces to Minkowski surgery. 
This corresponds to $x=0=y$ and $F_1(0,0)=0$. Thence in this specific situation the stability criterion simply reduces to
\begin{equation}
    \left[\frac{m_{s}(R_{0})}{R_{0}}\right]''\geq 0  .
\end{equation}
Similar logic is applied for asymmetric vanishing of parameters, say (without loss of generality) $m_+>0$ while $m_{-}=0$, as we are simply stitching Schwarzschild with Minkowski. This corresponds to $y=0$ but with $x>0$, and is represented by the $x$-axis in figure~\ref{F:1}.

\subsection{Mirror symmetry: Both $m_+=m_-$ and $a_+=a_-$}

For the specific case of mirror symmetry, let us fix both  $m_{+}=m_{-}=m$ as well as $a_{+}=a_{-}=a$. For this case the stability condition reduces to:
\begin{align}
    R_{0}^{2}\left[\frac{m_{s}(R_{0})}{R_{0}}\right]'' 
    &\geq F_{2}(R_{0}, m, a) \notag \\
    &= 2\left\lbrace\frac{\left[m\,\e^{-a/R_{0}}\left(1-\frac{a}{R_{0}}\right)\right]^{2}}
    {R_{0}^{2}\left[1-\frac{2m\,\e^{-a/R_{0}}}{R_{0}}\right]^{\frac{3}{2}}} -
    \frac{ma\,\e^{-a/R_{0}}\left(4-\frac{a}{R_{0}}\right)}
    {R_{0}^{2}\sqrt{1-\frac{2m\,\e^{-a/R_{0}}}{R_{0}}}}\right\rbrace.
\end{align}
In this case, we consider the two dimensionless parameters $(x=\frac{2m}{R_{0}}\, \e^{-a/R_0}$, $y=\frac{a}{R_{0}})$. 
Then the dimensionless function $F_{2}(x, y)$ is given by:
\begin{equation}\label{F2}
    F_{2}(x, y) = { x^2(1-y)^2\over2(1-x)^{3/2}} - {x y (4-y)\over(1-x)^{1/2}}.
\end{equation}
Notice that $x\in[0,1)$ to keep $F_{2}(x, y)$ real and finite. 
Furthermore, if the bulk spacetime contains horizons then $y\in(0,1]$; whilst if the bulk spacetime is horizonless, we are allowed to enter the region $y\in(1,\infty)$. 
It is worth noting that the parameter $x$ has a natural directly physical interpretation in terms of the gravitational redshift $z$ of the throat as seen from spatial infinity:
\begin{equation}
1+z = {1\over\sqrt{1-x}} = {1\over \sqrt{ 1 - \frac{2m}{R_{0}}\; \e^{-a/R_0}}}.
\end{equation}

\begin{figure}[t]
\begin{center}
\includegraphics[scale=0.5]{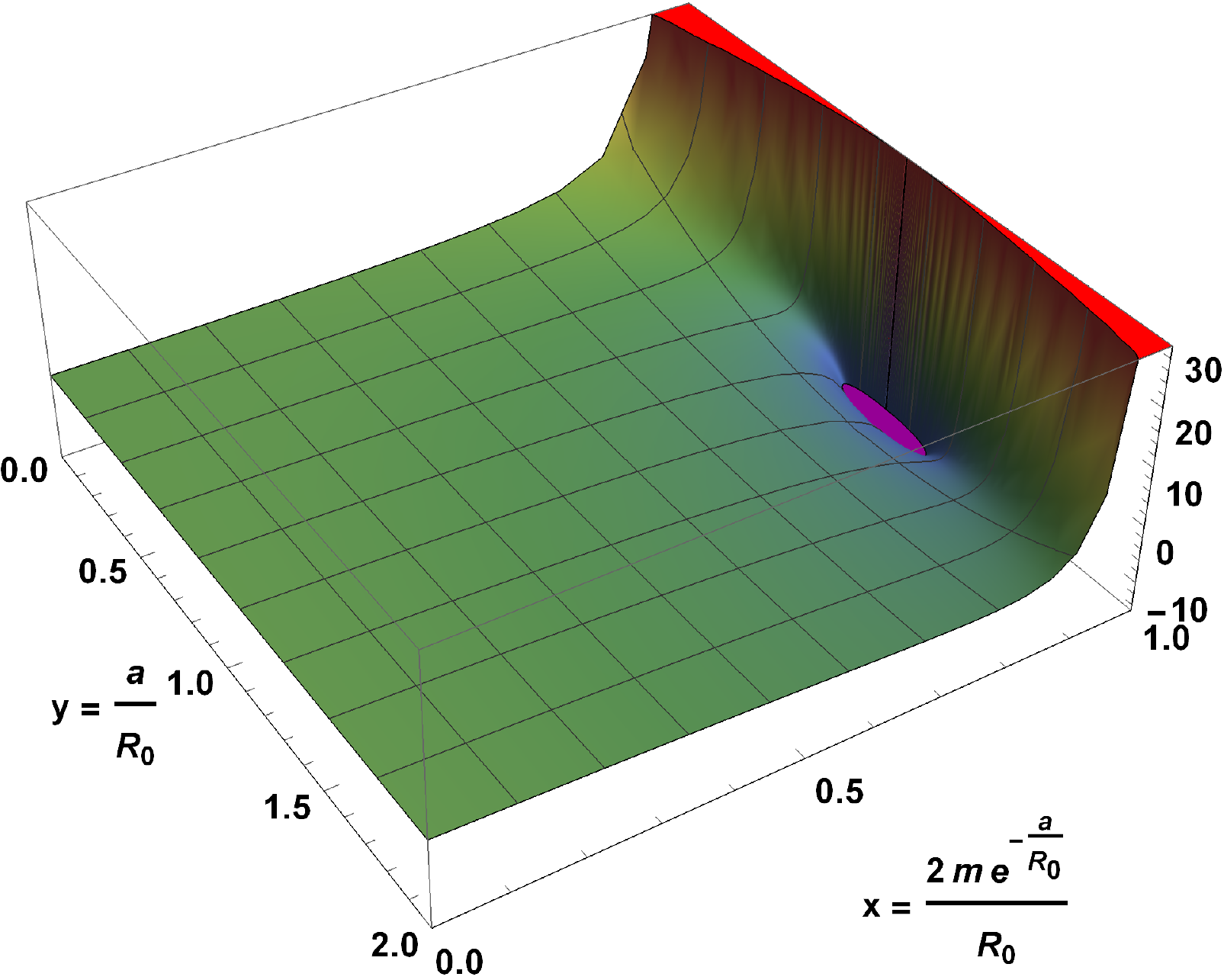}
\end{center}
\caption[Thin-shell wormhole stability analysis: Mirror symmetry.]{Stability analysis for the case of perfect mirror symmetry; $a_{+}=a_{-}$, and $m_{+}=m_{-}$. The stability region lies above the surface $F_{2}(x, y)$, given explicitly by equation~(\ref{F2}). The red and purple regions indicates where this function departs the specified range.
Note that we chop the graph vertically at $F_2(x,y)=30$ and at $F_2(x,y)=-10$.
}
\label{F:2}
\end{figure}

The point $(x,y)=(1,1)$ corresponds to the wormhole throat being located exactly at the degenerate horizon of an extremal bulk spacetime. 
The region $(x,y)\approx(1,1)$ corresponds to the wormhole throat being located near the almost degenerate horizon of a near-extremal bulk spacetime. 
It is easy to check that
\begin{equation}
\lim_{x\to1} F_{2}(x, y\neq 1) = +\infty \qq{and}
\lim_{x\to1} F_{2}(x, y=1) = -\infty. 
\end{equation}

\begin{figure}[t]
\centering
\includegraphics[scale=0.25]{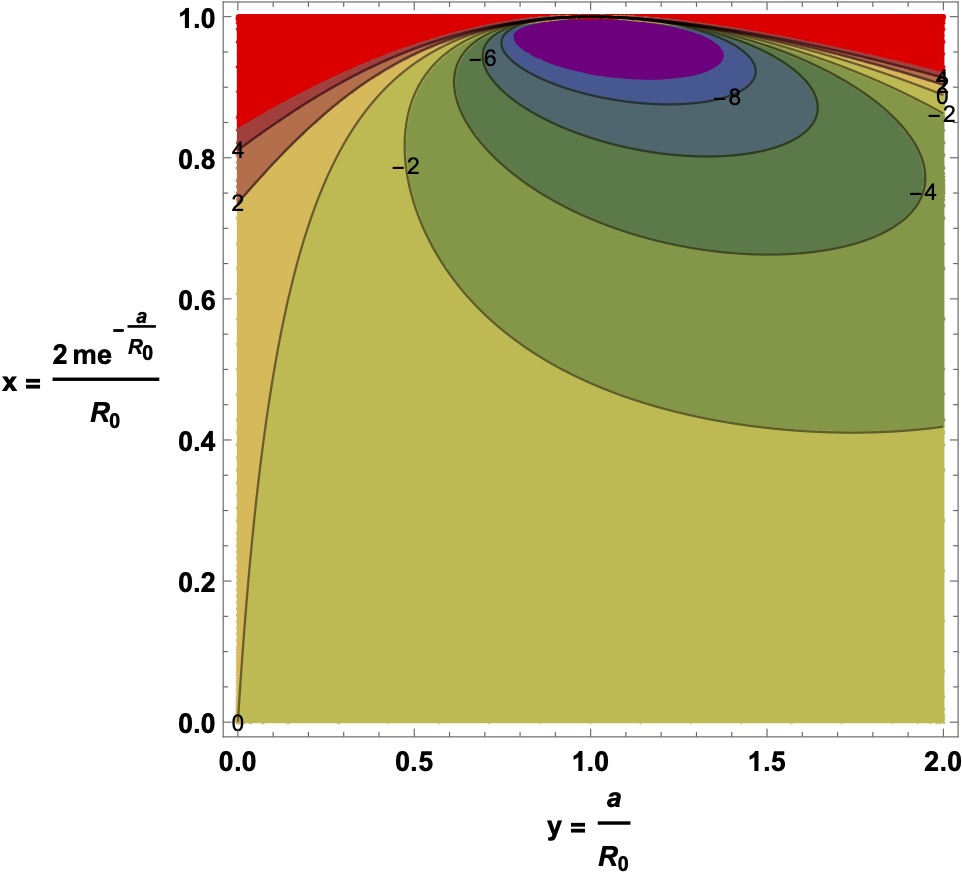}
\caption[Thin-shell wormhole stability analysis: Mirror symmetry contour plot.]{Contour plot: Stability analysis for the case of perfect mirror symmetry.
The purple region indicates the `pit' where  $F_2(x,y)<-10$. 
The red region indicates the region of lesser sability where  $F_2(x,y)>30$.
}
\label{F:2b}
\end{figure}

Inspecting figures \ref{F:2} and \ref{F:2b}, we observe relatively large stability regions.
An interesting feature of this plot is the presence of a  `pit' in the behaviour of $F_{2}(x,y)$ where the function is \emph{significantly} negative in the immediate vicinity of the extremal point $(x,y)=(1,1)$. 
This `pit' is a region which maximises the size of the stability region, and hence implies a preferred location for $R_{0}$ as a function of $m$ and $a$. 

\begin{figure}[t]
\centering
\includegraphics[scale=0.50]{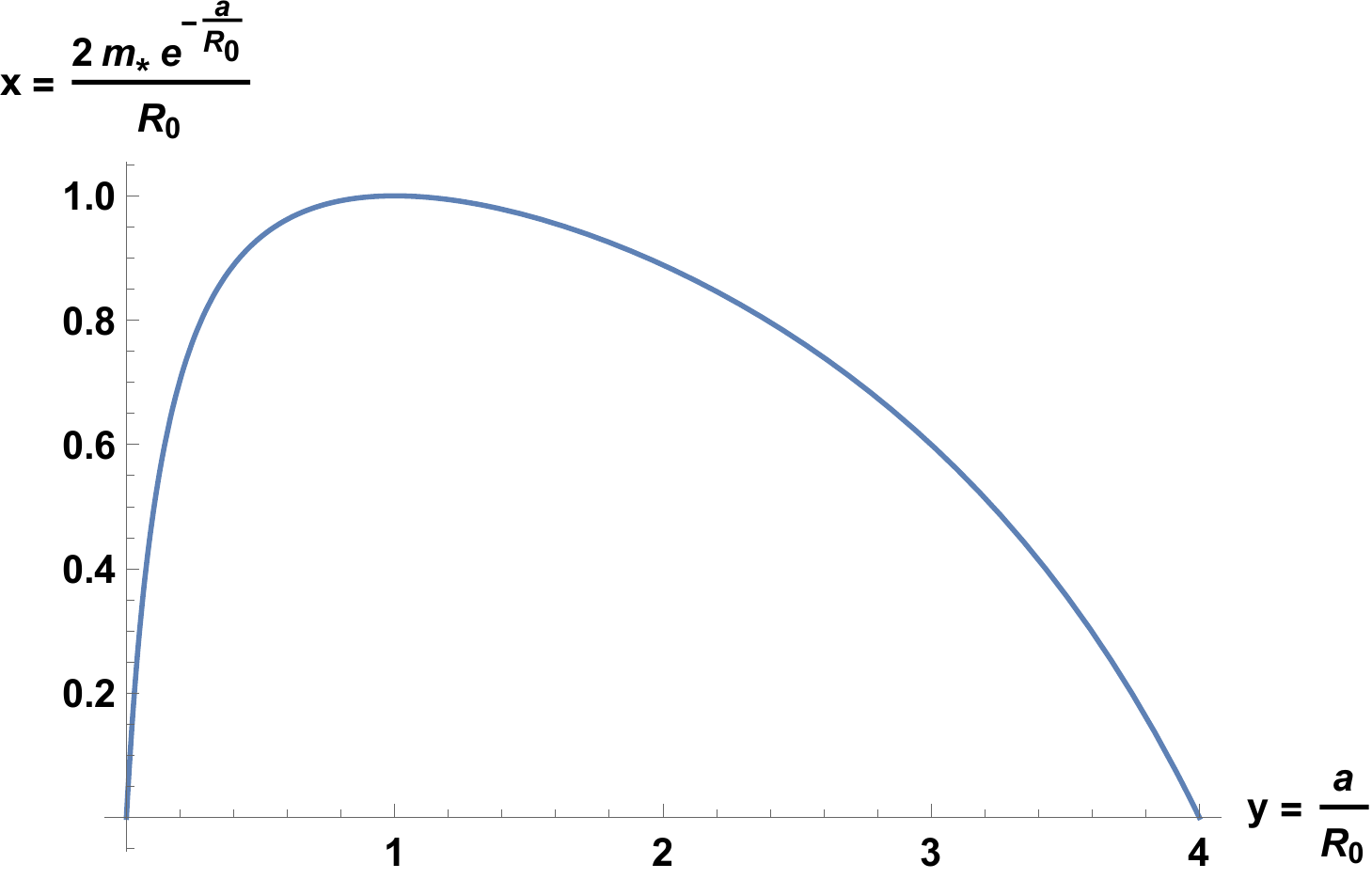}
\caption[Thin-shell wormhole stability analysis: \( F_2(x,y) < 0 \).]{Region in the $(x,y)$ plane where $F_2(x,y)$ flips sign.
}
\label{F:2c}
\end{figure}

\begin{figure}[t]
\centering
\includegraphics[scale=0.50]{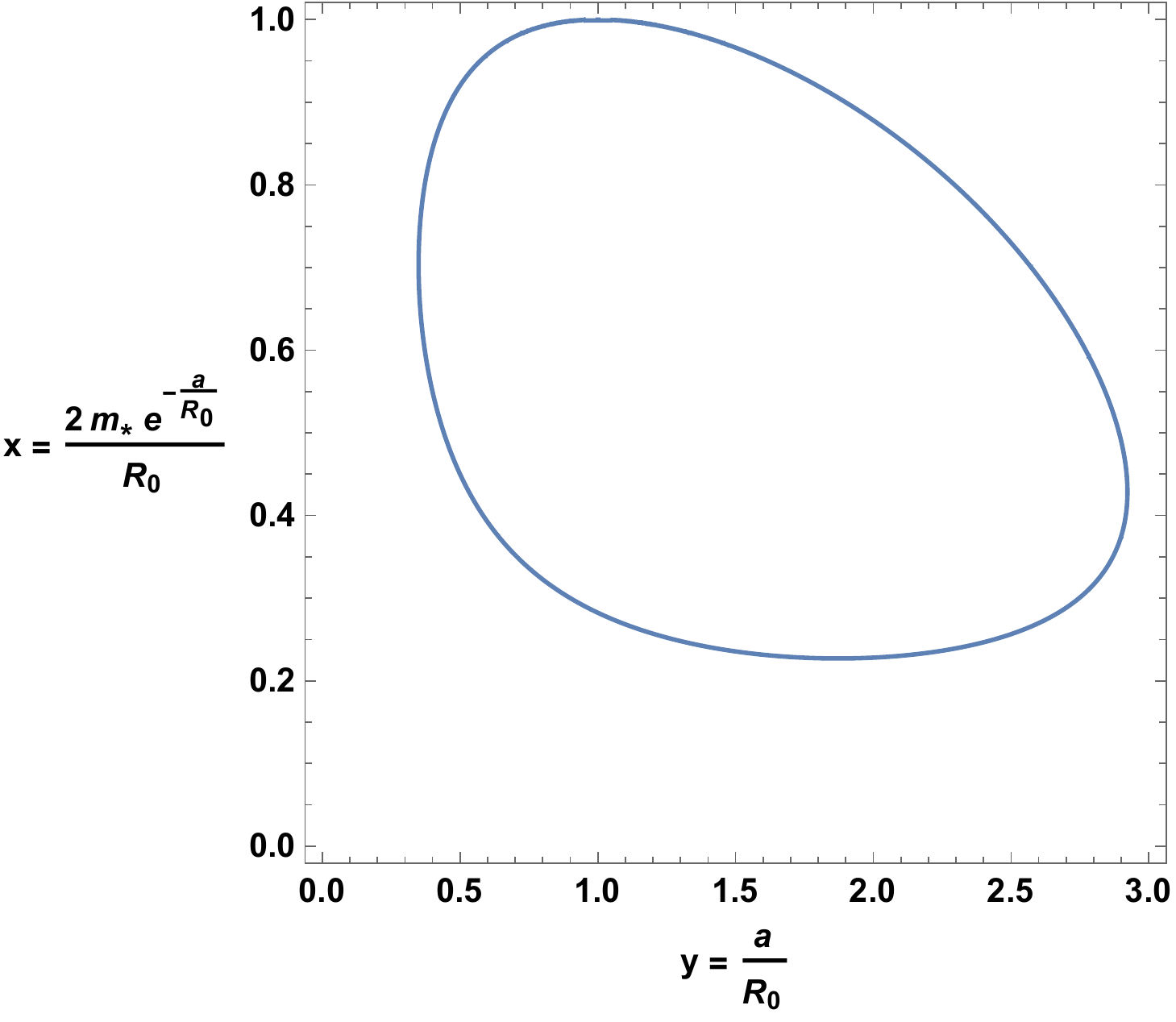}
\caption[Thin-shell wormhole stability analysis: \( F_2(x,y)< -1 \).]{Region in the $(x,y)$ plane where $F_2(x,y)<-1$.
}
\label{F:2d}
\end{figure}

The condition $F_{2}(x, y) = 0$, bounding the region where $F_{2}(x, y)$ changes sign, implicitly defines the curve 
\begin{equation}
x = {2y(4-y)\over1+6y-y^2}.
\end{equation}
In figure~\ref{F:2c} we plot the boundary of this region where $F_{2}(x,y)$ changes sign.
Then in figure~\ref{F:2d} we move deeper into the `pit' and plot the boundary of the region where $F_{2}(x,y)<-1$.

This `pit' in the stability plot is due to the wormhole throat getting close to where the extremal horizon would be in the bulk spacetime. 
It is well-known that having a wormhole throat get close to where a horizon would be in the bulk spacetime leads to interesting behaviour \cite{Kar:2004, visser-kar-dadhich}.
In particular, we note that in this symmetric situation
\begin{equation}\label{stabilityineq0b}
    \frac{m_{s}(R_{0})}{R_{0}} = -2\sqrt{1-\frac{2m\,\e^{-a/R_{0}}}{R_{0}}} = -2\sqrt{1-x},
\end{equation}
so that $x\approx 1$ corresponds to an arbitrarily small violation of the energy conditions \cite{Kar:2004, visser-kar-dadhich}. 
In terms of the redshift of the throat,
\begin{equation}\label{stabilityineq0c}
    \frac{m_{s}(R_{0})}{R_{0}} = -{2\over(1+z)^2}.
\end{equation}

\subsection{Specific asymmetry: $m_+\neq m_-$ while $a_+= a_-=a$}

Let us suppose $m_{+}\neq m_{-}$ while $a_{+}=a_{-}=a$. Hence we now have the case of surgery between two asymptotically Minkowski regular black holes with different masses but identical exponential suppression parameters. For a tractable analysis, 
let us define:
\begin{equation}
m_* = \max\{ m_+, m_-\}; \qquad 
\alpha = {\min\{m_+,m_-\} \over \max\{m_+,m_-\} } \leq 1,
\end{equation}

We may then re--express the stability condition of  equation~(\ref{stabilityineq}) as:
\begin{align}
    R_{0}^{2}\left[\frac{m_{s}(R_{0})}{R_{0}}\right]'' &\geq F_{3}(R_{0}, m_{*}, a, \alpha)    \notag \\ 
    &= \frac{\left[\alpha m_{*}\e^{-a/R_{0}}\left(1-\frac{a}{R_{0}}\right)\right]^{2}}{R_{0}^{2}\left[1-\frac{2\alpha m_{*}\e^{-a/R_{0}}}{R_{0}}\right]^{\frac{3}{2}}} - \frac{\alpha m_{*}a\,\e^{-a/R_{0}}\left(4-\frac{a}{R_{0}}\right)}{R_{0}^{2}\sqrt{1-\frac{2\alpha m_{*}\e^{-a/R_{0}}}{R_{0}}}} 	\notag \\
    &\quad+ \frac{\left[m_{*}\e^{-a/R_{0}}\left(1-\frac{a}{R_{0}}\right)\right]^{2}}{R_{0}^{2}\left[1-\frac{2m_{*}\e^{-a/R_{0}}}{R_{0}}\right]^{\frac{3}{2}}} - \frac{m_{*}a\,\e^{-a/R_{0}}\left(4-\frac{a}{R_{0}}\right)}{R_{0}^{2}\sqrt{1-\frac{2m_{*}\e^{-a/R_{0}}}{R_{0}}}}.
\end{align}

Now define the two dimensionless parameters, 
\begin{equation}
x=\frac{2m_{*}}{R_{0}}\, \e^{-a/R_0}, \qquad y=\frac{a}{R_{0}},
\end{equation}
so that the dimensionless function $F_{3}(x, y)$ takes the form
\begin{equation}
F_{3}(x, y) 
= \frac{\left[\alpha x (1-y)\right]^{2}}{4\left[1-\alpha x\right]^{\frac{3}{2}}} 
- \frac{\alpha x y (4-y)}{2\sqrt{1-\alpha x}} 
+ \frac{\left[x(1-y)\right]^{2}}{4\left[1-x\right]^{\frac{3}{2}}} 
- \frac{xy(4-y)}{2\sqrt{1-x}}.
\end{equation}
Note that the argument of the square root on the denominator forces our $x$--parameter to be less than unity, otherwise $F_{3}(x,y)$ will become complex. 
We therefore have $0<x<1$, while $0<y\leq1$ if the bulk spacetimes have horizons, and $y\in(1,\infty)$ is allowed if the bulk spacetimes are horizonless. 
By construction $\alpha\leq 1$, and so it is easy to check that
\begin{equation}
\lim_{x\to1} F_{3}(x, y\neq 1) = +\infty \qq{and} \lim_{x\to1} F_{3}(x, y=1) = -\infty. 
\end{equation}

\begin{figure}[t]
\begin{subfigure}{.5\textwidth}
\includegraphics[scale=0.450]{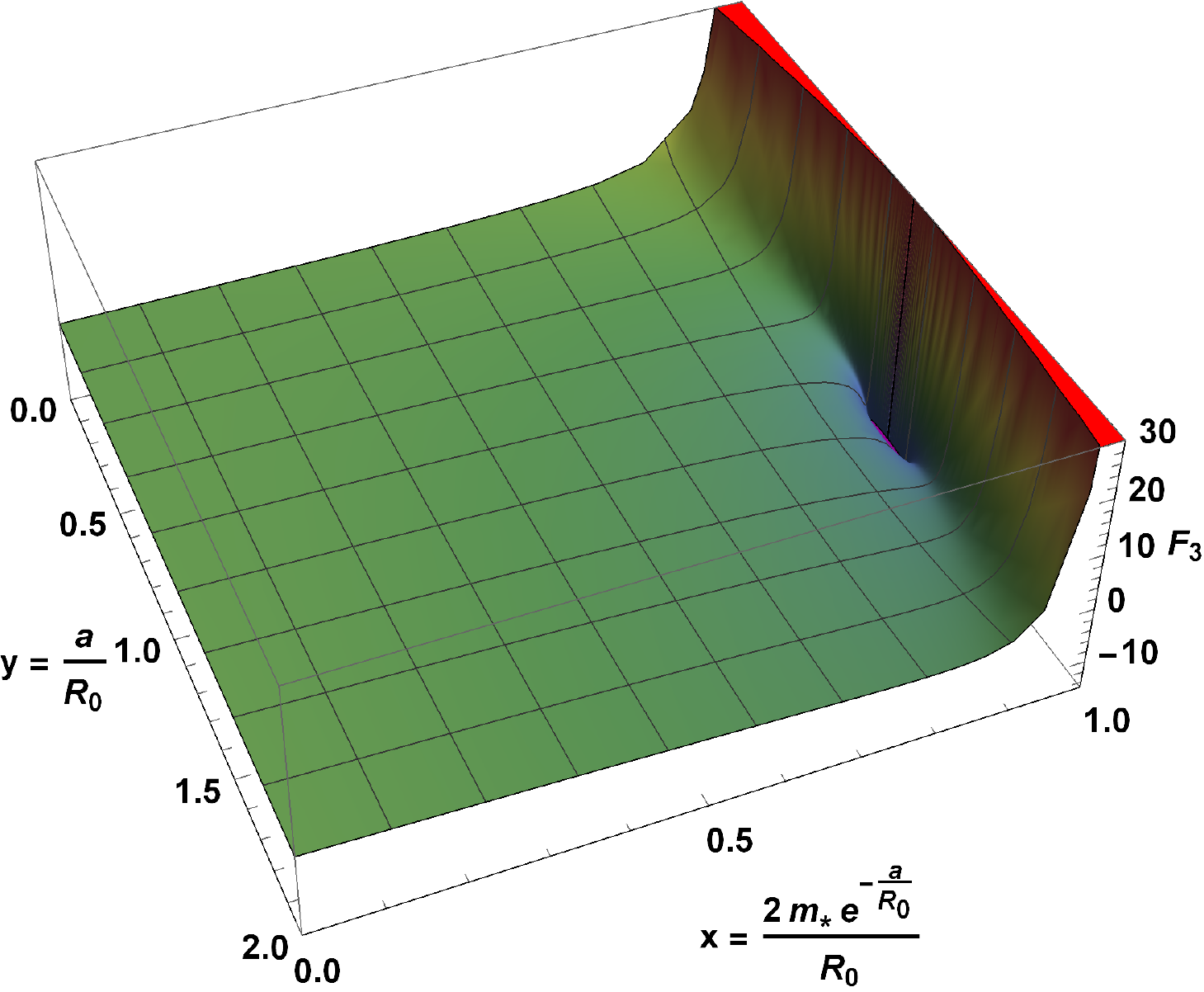}
\caption{$\alpha = 0.7$}
\end{subfigure}
\begin{subfigure}{.5\textwidth}
\includegraphics[scale=0.450]{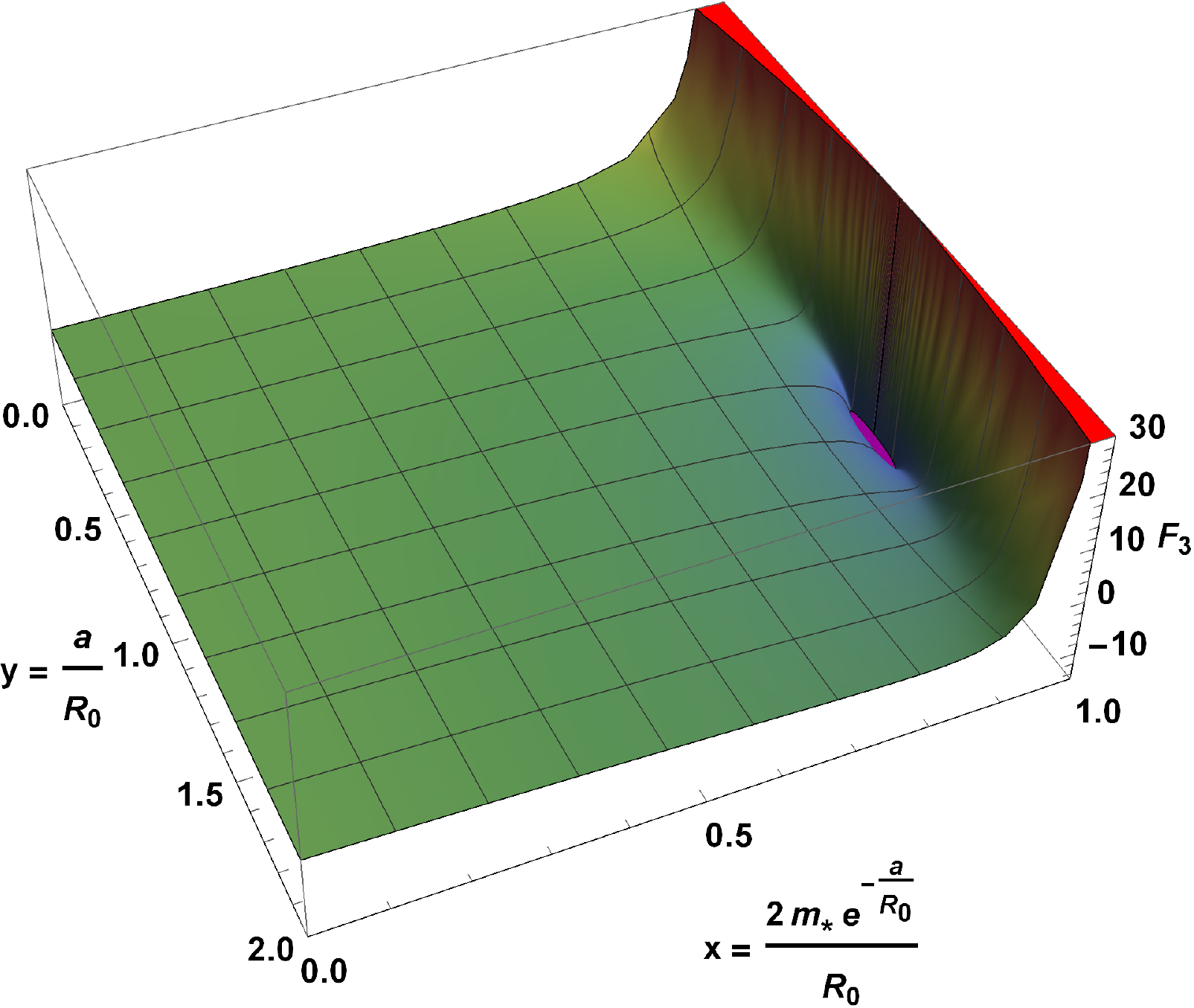}
\caption{$\alpha=0.9$}
\end{subfigure}
\caption[Thin-shell wormhole stability analysis: \( a_+ = a_- \) but \( m_+ \neq m_- \).]{Stability analysis for the specific asymmetry; $a_{+}=a_{-}=a$, while $m_{+}\neq m_{-}$. The stability region lies above the surface $F_{3}(x, y)$. The red and purple regions indicate where the function departs the specified range for $F_3(x,y)$.}
\label{F:3}
\end{figure}

\begin{figure}[t]
\begin{subfigure}{.5\textwidth}
\includegraphics[scale=0.215]{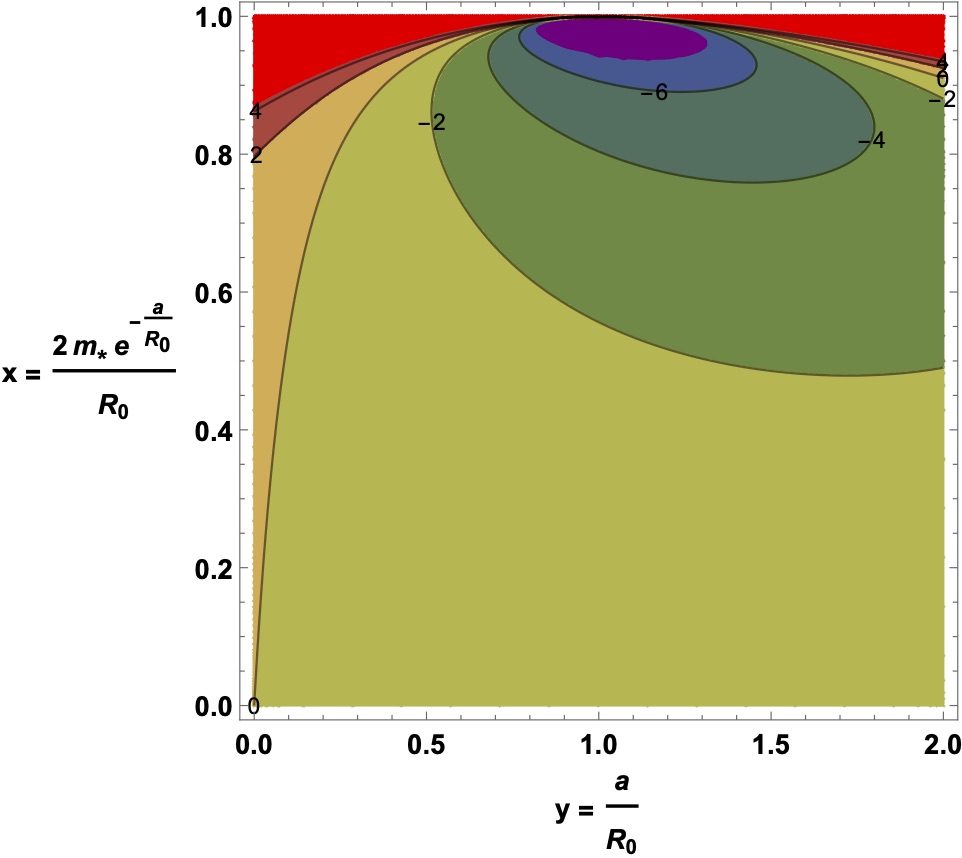}
\caption{$\alpha = 0.7$}
\end{subfigure}
\begin{subfigure}{.5\textwidth}
\includegraphics[scale=0.215]{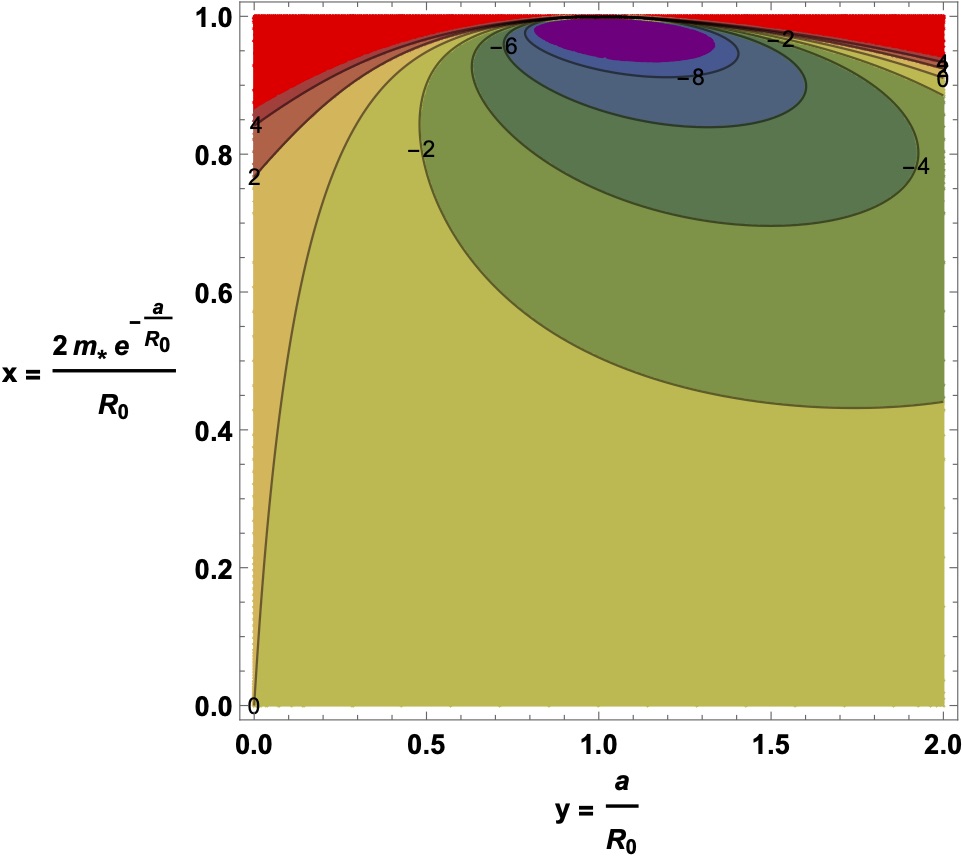}
\caption{$\alpha=0.9$}
\end{subfigure}
\caption[Thin-shell wormhole stability analysis: \( a_+ = a_- \) but\break \( m_+\neq m_- \) contour plots.]{Contour plots: Stability analysis for the specific asymmetry; $a_{+}=a_{-}=a$, while $m_{+}\neq m_{-}$. 
The purple region indicates `pit' where $F_3(x,y)<-10$.
The red region indicates region of lesser stability  where $F_3(x,y)$ is large and positive.}
\label{F:3b}
\end{figure}

We have chosen to illustrate two specific sub--cases, namely, $\alpha = 0.7$ and $\alpha = 0.9$. 
These correspond to the left--hand and right--hand plots of figures \ref{F:3} and \ref{F:3b} respectively. 
We observe that large stability regions exist, except in the limit $x\rightarrow1$ (with $y\neq1$). 
It appears that the difference between $\alpha=0.7$ and $\alpha=0.9$ is qualitatively negligible. 
However, of particular interest is the region \emph{very close} to the asymptote at $x=1$, where again, we have a `pit'. This leads to a preferred choice of the parameters $a, m_{\pm}$, which in turn leads to regions of maximal stability.
In this situation,
\begin{align}\label{stabilityineq00}
    \frac{m_{s}(R_{0})}{R_{0}} 
    &= -\sqrt{1-\frac{2\alpha m_{*}\e^{-a/R_{0}}}{R_{0}}} 
           - \sqrt{1-\frac{2m_{*}\e^{-a/R_{0}}}{R_{0}}} \notag \\
    &= -\sqrt{1-\alpha x} -\sqrt{1-x} = -\sqrt{1-\alpha} + \mathcal{O}(1-x). 
\end{align}
Thus, for $\alpha<1$, the energy condition violations are minimised (though no longer arbitrarily small) as the wormhole throat approaches the location of what would be a horizon in the bulk spacetime \cite{Kar:2004, visser-kar-dadhich}. 

\subsection{Specific asymmetry: $a_{+}\neq a_{-}$ while $m_{+}=m_{-}$}

Let us now suppose $a_{+}\neq a_{-}$ and $m_{+}=m_{-}=m$. 
Hence, we are now performing surgery between two asymptotically Minkowski black holes with identical masses, but different exponential suppression parameters. 

\enlargethispage{10pt}
To develop a tractable analysis, define
\begin{equation}
a_*=\min\{a_+,a_-\} \qq{and} \beta = {\max\{a_+,a_-\}\over \min\{a_+,a_-\} } \geq 1.
\end{equation}
We then have,
\begin{equation}
    R_{0}^{2}\left[\frac{m_{s}(R_{0})}{R_{0}}\right]^{''} \geq F_{4}(R_{0}, m, a_{-},\beta),
\end{equation}
where
\begin{multline}\label{asymm2}
    F_{4}(R_{0}, m, a_{-},\beta) = 
    \frac{\left[m\,\e^{-\frac{\beta a_*}{R_{0}}}\left(1-\frac{\beta a_{*}}{R_{0}}\right)\right]^{2}}{R_{0}^{2}\left[1-\frac{2m\,\e^{-\frac{\beta a_{*}}{R_{0}}}}{R_{0}}\right]^{\frac{3}{2}}} 
    - \frac{m\beta a_{*}\e^{-\frac{\beta a_{*}}{R_{0}}}\left(4-\frac{\beta a_{*}}{R_{0}}\right)}{R_{0}^{2}\sqrt{1-\frac{2m\,\e^{-\frac{\beta a_{*}}{R_{0}}}}{R_{0}}}}   \\
    + \frac{\left[m\,\e^{-\frac{a_{*}}{R_{0}}}\left(1-\frac{a_{*}}{R_{0}}\right)\right]^{2}}{R_{0}^{2}\left[1-\frac{2m\,\e^{-\frac{a_{*}}{R_{0}}}}{R_{0}}\right]^{\frac{3}{2}}} - \frac{ma_{*}\e^{-\frac{a_{*}}{R_{0}}}\left(4-\frac{a_{*}}{R_{0}}\right)}{R_{0}^{2}\sqrt{1-\frac{2m\,\e^{-\frac{a_{*}}{R_{0}}}}{R_{0}}}}.
\end{multline}
The stability analysis may now be simplified by employing the two dimensionless parameters $(x=\frac{2m}{R_0} \,\e^{-a_*/R_0}, \, y=\frac{a_{*}}{R_0})$, to re-express this stability condition as a function of these dimensionless parameters. 
Explicitly, 
\begin{multline}
\label{asymm3}
    F_{4}(x,y) = 
    \frac{\left[x\e^{(1-\beta)y}\left(1-\beta y\right)\right]^{2}}
    {4\left(1-x \e^{(1-\beta)y}\right)^{\frac{3}{2}}} 
    - \frac{\beta xy\,\e^{(1-\beta) y}(4-\beta y)}{2\sqrt{1-x\e^{(1-\beta) y}}} 	\\
    + \frac{\left[x(1-y)\right]^{2}}{4\left(1-x\right)^{\frac{3}{2}}} 
    - \frac{xy(4-y)}{\sqrt{1-x}}.
\end{multline}
Notice that the square root in the denominator implies $0<x<1$. 
We may, however, once again assert $0<y<1$ if the bulk spacetimes contain horizons, while $1<y<\infty$ is permitted if the bulk spacetimes are horizon-free.

\begin{figure}[t]
\begin{subfigure}{.5\textwidth}
\includegraphics[scale=0.4]{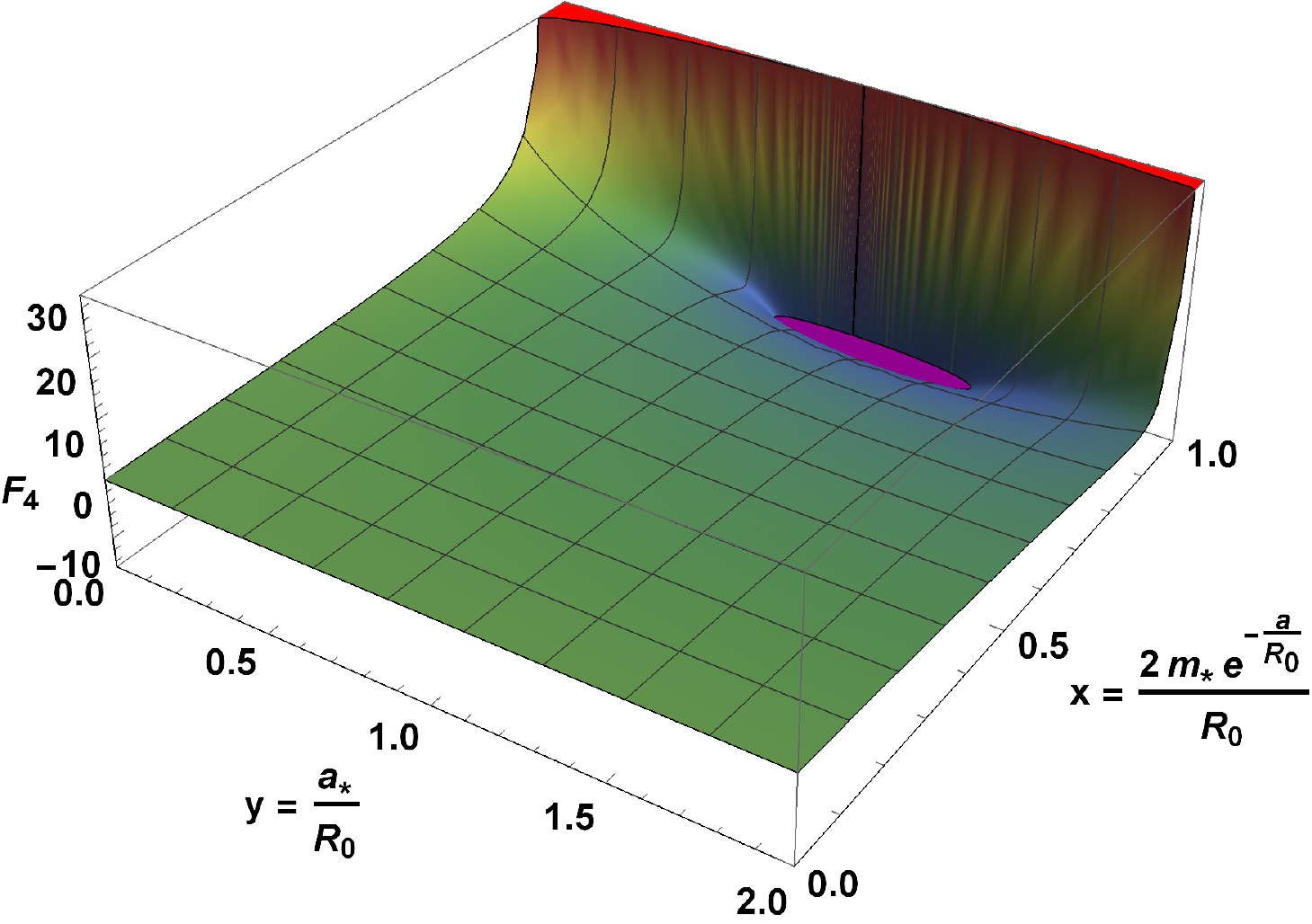}
\caption{$\beta = 1.2$}
\end{subfigure}
\begin{subfigure}{.5\textwidth}
\includegraphics[scale=0.4]{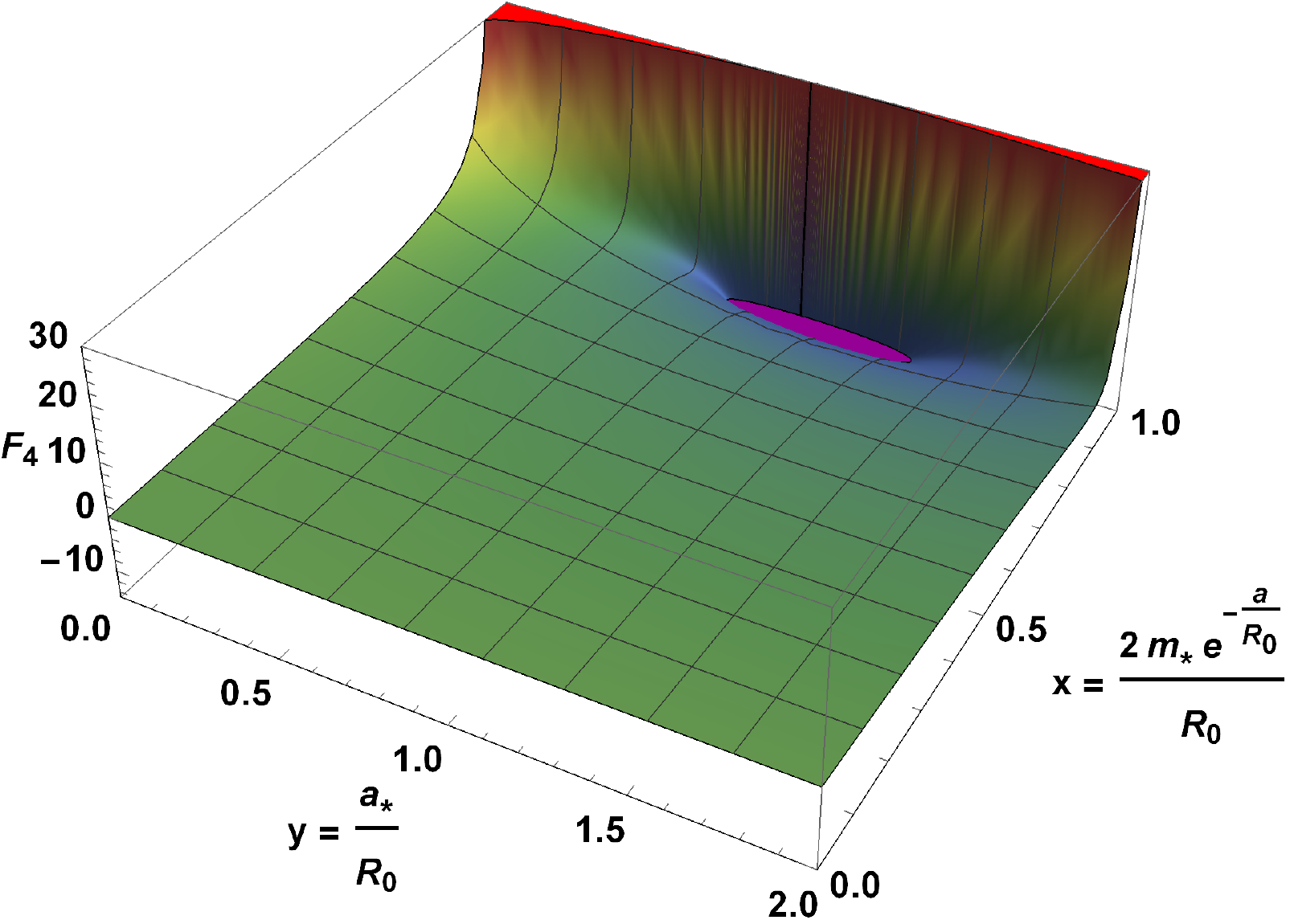}
\caption{$\beta=1.4$}
\end{subfigure}
\caption[Thin-shell wormhole stability analysis: \( a_+\neq a_- \) but \( m_+=m_- \).]{Stability analysis for the asymmetry $\max\{a_{+},a_-\} =\beta \min\{a_+,a_-\} = \beta a_{*}$, with $\beta >1$, and $m_{+} = m_{-}=m$. The stability region lies above the surface $F_{4}(x, y)$. The red and purple regions indicate where the function $F_4(x,y)$ departs the range $(-20,+30)$.}
\label{F:5a}
\end{figure}

\begin{figure}[t]
\begin{subfigure}{.5\textwidth}
\includegraphics[scale=0.2]{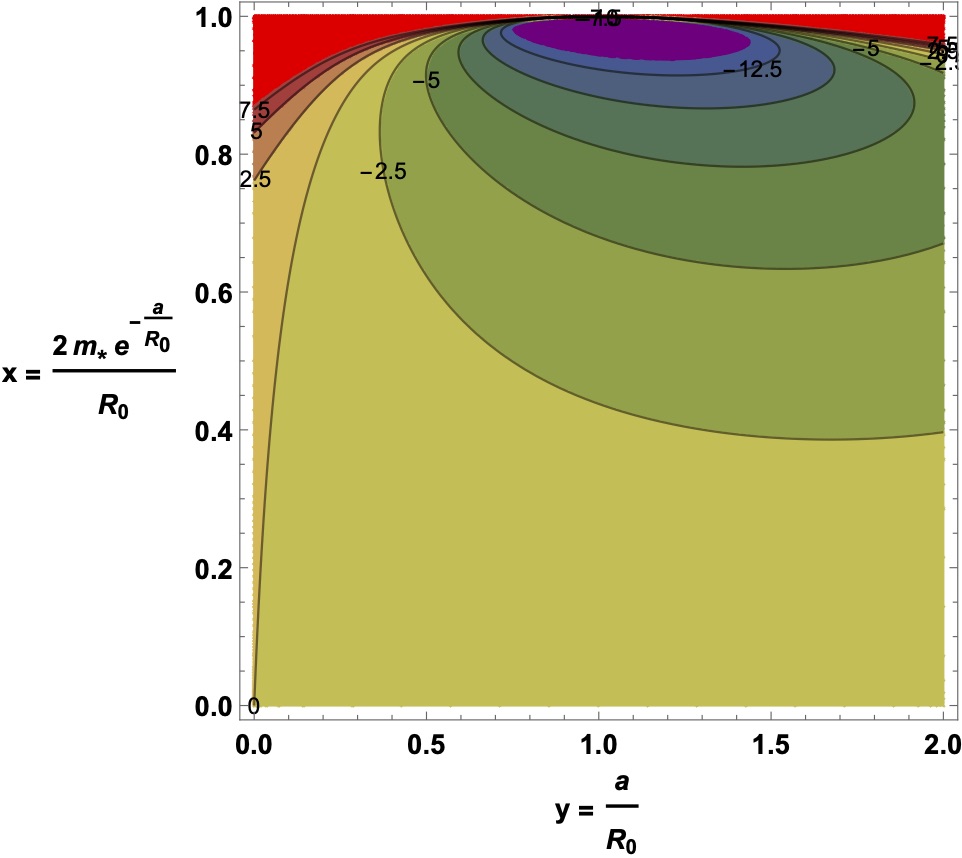}
\caption{$\beta = 1.2$}
\end{subfigure}
\begin{subfigure}{.5\textwidth}
\includegraphics[scale=0.2]{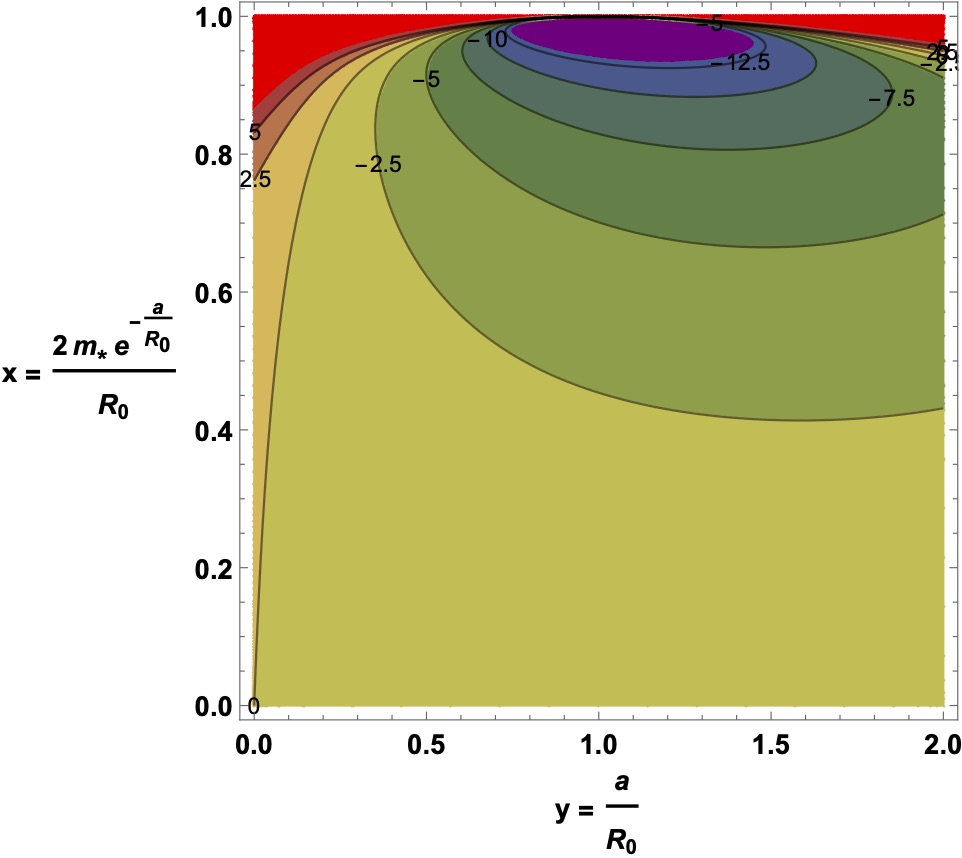}
\caption{$\beta=1.4$}
\end{subfigure}
\caption[Thin-shell wormhole stability analysis: \( a_+\neq a_- \) but\break \( m_+=m_- \) contour plots.]{Contour plots: Stability analysis for the specific asymmetry $\max\{a_{+},a_-\} =\beta \min\{a_+,a_-\} = \beta a_{*}$, with $\beta >1$, and $m_{+} = m_{-}=m$. These are contour plots for the function $F_{4}(x, y)$. The purple region indicates the `pit' where the function 
function $F_4(x,y)$ is strongly negative.
The red region indicates the region of decreased stability  where the function 
function $F_4(x,y)$ is strongly positive.}
\label{F:5b}
\end{figure}

Since by construction $\beta\geq 1$, it is easy to check that
\begin{equation}
\lim_{x\to1} F_{4}(x, y\neq 1) = +\infty \qq{and} \lim_{x\to1} F_{4}(x, y=1) = -\infty. 
\end{equation}
For illustrative purposes, we present the specific cases $\beta=1.2$ and $\beta=1.4$. These correspond to the left--hand and right--hand plots of figure \ref{F:5a} and figure \ref{F:5b}, respectively. 
We have large stability regions other than in the limit $x\to1$ (with $y\neq1$). 
There is again a `pit' in the vicinity of $(x,y)\approx(1,1)$. 

We note that, in this situation,
\begin{align}\label{stabilityineq0xxx}
    \frac{m_{s}(R_{0})}{R_{0}} &= -\sqrt{1-\frac{2m\,\e^{-\beta a_{*}/R_{0} }}{R_{0}}} - \sqrt{1-\frac{2m\,\e^{-a_{*}/R_{0}}}{R_{0}}} \notag \\
   &= - \sqrt{ 1 - \e^{(1-\beta) y} x} - \sqrt{1-x} \notag \\
   &= - \sqrt{ 1 - \e^{(1-\beta) y}} + \mathcal{O}(1-x).
\end{align}
Thus, for $\beta>1$, the energy condition violations are minimised (though no longer arbitrarily small) as the wormhole throat approaches the location of what would be a horizon in the bulk spacetime \cite{Kar:2004, visser-kar-dadhich}.

\section{Summary}

We have used a novel regular black hole model based on exponential mass suppression to construct a thin-shell wormhole using the cut-and-paste technique.
The construction under consideration provides an example of a black hole mimicker (the smaller the value of the mass suppression parameter $a$ and the closer the location of the wormhole throat to the Schwarzschild radius, the better this this model is to mimicking a standard Schwarzschild black hole). 
For suitable choices of parameters, the wormhole under consideration was found to violate the null energy condition in the bulk spacetime, whereas the strong energy condition is satisfied in this region.
The wormhole construction was analysed via the thin-shell formalism, allowing the four-velocity of the wormhole throat to be calculated along with the junction surface unit normal vectors, the extrinsic curvature, and the junction surface stress-energy.
The surface energy at the wormhole junction throat was found to be negative, and so, much like other traversable wormholes, exotic matter would be needed to keep the wormhole throat open.

We found that this class of wormholes permits a clean and quite general stability analysis, with wide swathes of stable behaviour. Furthermore the stability plateau exhibits a `pit' of enhanced stability when the wormhole throat is close to where a near-extremal horizon would have existed in the bulk spacetime before applying `cut-and-paste' surgery. 
Finally, we found that the quantity of exotic matter needed to support the wormhole throat could be minimised (and in some cases made arbitrarily small) by suitable choice of parameters. 
\chapter[General class of ``quantum deformed'' regular black holes]{General class of ``quantum deformed'' regular black holes}\label{C:QMS}
\chaptermark{\small{General class of ``quantum deformed'' RBH}}

In this chapter, we discuss the ``quantum deformed Schwarzschild spacetime'' as originally introduced by Kazakov and Solodukhin in 1993, and investigate the precise sense in which it does and does not satisfy the \emph{desiderata} for being a ``regular black hole''. We shall carefully distinguish (i) regularity of the metric components, (ii) regularity of the Christoffel components, and (iii) regularity of the curvature. 
We shall then embed the Kazakov--Solodukhin spacetime in a more general framework where these notions are clearly and cleanly separated. 
Finally we analyze aspects of the classical physics of these ``quantum deformed Schwarzschild spacetimes''. We shall discuss the surface gravity, the classical energy conditions, null and timelike geodesics, and the appropriate variant of Regge--Wheeler equation.

\section{Introducing the spacetime}\label{S:intro}

The unification of general relativity and quantum mechanics is of the utmost importance in reconciling many open problems in theoretical physics today. 
One avenue of exploration towards a fully quantised theory of gravity is to, on a case--by--case basis, apply various quantum corrections to existing black hole solutions to the Einstein equations, and thoroughly analyse the resulting geometries through the lens of standard general relativity.
As with the majority of theoretical analysis, to make progress one begins by applying quantum--corrections to the simplest case; the Schwarzschild solution~\cite{quantum-bh2}. 

Historically, various treatments of a quantum--corrected Schwarzschild metric have been performed in multiple different settings~\cite{quantum-bh3, quantum-bh1, Eslamzadeh:2020, Good:2020, Nozari:2020, Nozari:2021, Qi:2019, Md.:2018, Shahjalal:2019}. 
A specific example of such a metric is the ``quantum deformed Schwarzschild metric'' derived by Kazakov and Solodukhin in reference~\cite{quantum-bh2}. 
Much of the literature sees the original metric exported from the context of static, spherical symmetry into something dynamical, or else it invokes a different treatment of the quantum--correcting process to that performed in~\cite{quantum-bh2} (see,~\emph{e.g.}, reference~\cite{Burger:2018}). 

The metric derived in reference~\cite{quantum-bh2} invokes the following change to the line element for Schwarzschild spacetime in standard curvature coordinates:
\begin{eqnarray}
1-\frac{2m}{r} &\quad\longrightarrow\quad& \sqrt{1-\frac{a^{2}}{r^{2}}}-\frac{2m}{r},
\end{eqnarray}
so that
\begin{eqnarray}
\d s^{2} &=& -\left(\sqrt{1-\frac{a^{2}}{r^{2}}}-\frac{2m}{r}\right)\d t^{2} + \frac{\d r^{2}}{\sqrt{1-\frac{a^{2}}{r^{2}}}-\frac{2m}{r}} + r^{2}\,\d\Omega^{2}_{2}.
\label{k-s-metric}
\end{eqnarray}
To keep the metric components real, the $r$ coordinate must be  restricted to the range $r\in[a,\infty)$. So the ``centre'' of the spacetime at $r\to a$ is now a 2-sphere of finite area $A=4\pi a^2$.  The fact that the ``centre'' has now been ``smeared out'' to finite $r$ was originally hoped to render the spacetime regular.

\enlargethispage{40pt}
This metric was originally derived via an action principle which has its roots in the 2-D, (more precisely  (1+1)-D),  dilaton theory of gravity~\cite{quantum-bh2, Russo:1992yg}:
\begin{equation}
S = -\frac{1}{8} \int \d^2 z \sqrt{-g} \left[ r^2 R^{(2)} - 2 (\nabla r)^2 + \frac{2}{\kappa} \, U(r) \right].
\label{k-s-action}
\end{equation}
Here \( R^{(2)} \) is the two--dimensional Ricci scalar, \( \kappa \) is a constant with dimensions of length, and \( U(r) \) is the ``dilaton potential''.

The action \eqref{k-s-action} yields two equations of motion, one of which is then used to derive the general form of the metric:
\begin{equation}
\d s^2 = -f(r)\;\d t^2 + \frac{\d r^2}{f(r)} + r^2 \;\d\Omega^2_2, \qquad 
f(r) = -\frac{2m}{r} + \frac{1}{r}\int^r U(\rho)\,\d\rho.
\end{equation}
The dilaton potential \( U(r) \) is quantised within the context of the \( D=2 \) \( \sigma \)-model~\cite{quantum-bh2, Russo:1992yg}, resulting in the specific metric \eqref{k-s-metric}.
Specifically, Kazakov and Solodukhin choose
\begin{equation}
U(r) = {r\over\sqrt{r^2-a^2}}. 
\end{equation}
Note that generic metrics of the form 
\begin{equation}
\d s^2 = -f(r)\;\d t^2 + \frac{\d r^2}{f(r)} + r^2 \;\d\Omega^2_2,
\end{equation}
where one does not necessarily make further assumptions about the function $f(r)$, have a long and complex history~\cite{Kiselev2,Ted,Kiselev0,Kiselev1}.

In Kazakov and Solodukhin's original work~\cite{quantum-bh2}, they claim the metric \eqref{k-s-metric} is ``regular''. 
However, by this they just mean ``regular'' in the sense of the metric components (in this specific coordinate chart) being finite for all $r\in[a,\infty)$.
This is not the meaning of the word ``regular''  that is usually adopted in the GR community. 
We find it useful to carefully distinguish (i) regularity of the metric components, (ii) regularity of the Christoffel components, and (iii) regularity of the curvature. 
Indeed, within the GR community, the term ``regular'' means that the spacetime entirely is free of curvature singularities~\cite{rbhs-review, Bardeen:2014, Bardeen:1968, Borde:1996, Bronnikov:2000, Bronnikov:2005, Carballo-Rubio:2018pmi, Carballo-Rubio:2018jzw, Carballo-Rubio:2019fnb, Carballo-Rubio:2019nel, Carballo-Rubio:2021, frolov-rbh, hayward-rbh, Moreno:2002, Roman:1983}, with infinities in the curvature invariants being used as the typical diagnostic\footnote{Our technical definition of spacetime singularities from chapter \ref{C:BH-mimickers} is not very practical in most instances.}. 
While the metric \eqref{k-s-metric} is regular in terms of the metric components, it fails to be regular in terms of the Christoffel components, and has a Ricci scalar which is manifestly singular at \( r = a \):
\begin{equation}
R = \frac{2}{r^2} - \frac{2r^2-3a^2}{r (r^2-a^2)^{3\over2}}\,
 = {a\over(2a)^{3\over2} \, (r-a)^{3\over2}} - \frac{23}{4\left(2a\right)^{\frac{3}{2}}(r-a)^{\frac{1}{2}}} + \O(1).
\end{equation}

\enlargethispage{40pt}
The specific metric \eqref{k-s-metric} derived by Kazakov and Solodukhin falls in to a more general class of metrics given by
\begin{equation}
\d s_n^2 = -f_n(r) \d t^2 + \frac{\d r^2}{f_n(r)} + r^2\big(\d\theta^2 + \sin^2\theta \,\d\phi^2\big),
\label{E:g_n}
\end{equation}
where now we take
\begin{equation}
f_n(r) =   
\left(1-\frac{a^{2}}{r^{2}}\right)^{n\over2}-\frac{2m}{r}.
\label{E:f_n}
\end{equation}

Here \( n \in \{0\}\cup\{1,3,5,\dots\} \), \( r\in[a,\infty) \), and \( a\in(0,\infty) \). 
(Note, we include \( n = 0 \) as a special case since this reduces the metric to the Schwarz\-schild metric in standard curvature coordinates, which is useful for consistency checks).
We only consider odd values for \( n \) (excluding the \( n=0 \) Schwarz\-schild solution) as any even value of \( n \) will allow for the \( r \)-coordinate to continue down to \( r=0 \), and so produce a black-hole spacetime which is not regular at its core and hence not of interest in this work.

The class of metrics described by equations~\eqref{E:g_n}--\eqref{E:f_n} has the following regularity structure:
\begin{itemize}
        \item \( n = 0 \) (Schwarzschild): Not regular;
        \item \( n \geq 1 \): Metric--regular;
        \item \( n \geq 3 \): Christoffel--symbol--regular;
        \item \( n \geq 5 \): Curvature--regular.
\end{itemize}
We wish to stress that, unlike reference~\cite{quantum-bh2}, we make no attempt to \emph{derive} the class of metrics described by equations~\eqref{E:g_n}--\eqref{E:f_n}  from a modified action principle in this current work.
We feel that there are a number of technical issues requiring clarification in  the derivation presented in reference~\cite{quantum-bh2}, so instead, we shall simply use the results of Kazakov and Solodukhin's work as inspiration and motivation for the analysis of our general class of metrics.
As such, our extended class of Kazakov--Solodukhin models can be viewed as another set of black hole mimickers, arbitrarily closely approximating standard Schwarzschild black holes, and so potentially of interest to observational astronomers \cite{Barausse:2020}.

\section{Geometric analysis}

In this section, we shall analyse the metric \eqref{E:g_n}, its associated Christoffel symbols,  and the various curvature tensor quantities derived therefrom.
 %
 
\subsection{Metric components}
 
We immediately enforce $a\neq 0$ since $a=0$ is trivially the Schwarzschild spacetime, and in fact we shall specify $a>0$ since $a$ is typically to be identified with the Planck scale. 
At large $r$ and/or small $a$ we have:
\begin{equation}
f_n(r)= \left(1-\frac{a^{2}}{r^{2}}\right)^{n\over2}-\frac{2m}{r} = 1-\frac{2m}{r} - \frac{na^{2}}{2r^{2}} 
+ \O\left(\frac{a^4}{r^{4}}\right).
\end{equation} 
So the spacetime is asymptotically flat with mass $m$ for any fixed finite value of \( n \).
As $r\to a$, we note that for $n\geq1$ we have the finite limit
\begin{equation}
\lim_{r\to a} f_n(r) = - {2m\over a}.
\end{equation}
This is enough to imply metric--regularity.
Note however that for the radial derivative we have
\begin{equation}
f_n'(r) = {na^2\over r^3} \left(1-\frac{a^{2}}{r^{2}}\right)^{{n\over2}-1} +{2m\over r^2},
\end{equation}
and that only for $n\geq3$ do we have a finite limit
\begin{equation}
\lim_{r\to a} f_n'(r) =  {2m\over a^2}.
\end{equation}
Similarly for the second radial derivative
\begin{equation}
f_n''(r) = {na^2(na^2+a^2-3r^2)\over r^6} \left(1-\frac{a^{2}}{r^{2}}\right)^{{n\over2} -2} -{4m\over r^3},
\end{equation}
and only for $n\geq5$ do we have a finite limit
\begin{equation}
\lim_{r\to a} f_n''(r) =  -{4m\over a^3}.
\end{equation}
This ultimately is why we need $n\geq 3$ to make the Christoffel symbols regular, and $n\geq 5$ to make the curvature tensors regular.

\subsection{Event horizons}

Event horizons may be located by solving \( g_{tt}(r) = f_n(r) = 0 \), and so are implicitly characterized by 
\begin{equation}
r_H = 2m \left( 1 -{a^2\over r_H^2}\right)^{-{n\over2}}.
\label{E:horizon}
\end{equation}
This is not algebraically solvable for general \( n \), though we do have the obvious bounds that $r_H >2m$ and $r_H>a$. 

\enlargethispage{40pt}
Furthermore, for small \( a \) we can use (\ref{E:horizon}) to find an approximate horizon location 
by iterating the lowest-order approximation $r_H = 2m + \O(a^2/m)$ to yield
\begin{equation}
r_H = 2m \left\{ 1 +  {n a^2\over 8 m^2} + \O\left(a^4\over m^4\right) \right\}.
\end{equation}

Iterating a second time
\begin{equation}
r_H = 2m \left\{ 1 +  {n a^2\over 8 m^2} - {n(3n-2) a^4\over 128 m^4}   +\O\left({a^6\over m^6}\right) \right\}.
\end{equation}
We shall soon find that taking this second iteration is useful when estimating the surface gravity.
As usual, while event horizons are mathematically easy to work with, one should bear in mind that they are impractical for observational astronomers to deal with --- any physical observer limited to working in a finite region of space+time can at best detect apparent horizons or trapping horizons~\cite{observability}, see also reference~\cite{Hawking:2014}. In view of this intrinsic limitation, approximately locating the position of the horizon is good enough for all practical purposes.

\subsection{Christoffel symbols of the second kind}

Up to the usual symmetries, the non-trivial non-zero coordinate components of the Christoffel connection in this coordinate system are:
\begin{eqnarray}
\Gamma^t{}_{tr} &=&  -\Gamma^r{}_{rr}     =
{ 2m/r + n (a^2/r^2)(1-a^2/r^2)^{{n\over2} -1}
\over
2r  \{ (1-a^2/r^2)^{{n\over2}} - 2m/r \}
};
\nonumber\\
\Gamma^r{}_{tt} &=&
{ \{ 2m/r + n (a^2/r^2) (1-a^2/r^2)^{{n\over2} -1} \} \{(1-a^2/r^2)^{{n\over2}} - 2m/r \}
\over
2 r};
\nonumber\\
\Gamma^r{}_{\theta\theta} &=&  {\Gamma^r{}_{\phi\phi}\over \sin^2\theta}     =
2m - r(1-a^2/r^2)^{n\over2}.
\end{eqnarray}

The trivial non-zero components are 
\begin{eqnarray}
\Gamma^\theta{}_{r\theta} &=&  \Gamma^\phi{}_{r\phi}     = {1\over r};
\nonumber\\
\Gamma^\theta{}_{\phi\phi} &=& - \sin\theta\;\cos\theta;
\nonumber\\
\Gamma^\phi{}_{\theta\phi} &=&  \cot\theta.
\end{eqnarray}

Inspection of the numerators of \( \Gamma^{t}{}_{tr} \), \( \Gamma^{r}{}_{rr} \), and 
\( \Gamma^{r}{}_{\theta\theta} \) shows that (in this coordinate system) the Christoffel symbols are finite at \( r=a \) so long as \( n \geq 3 \).
Indeed as $r\to a$ we see
\begin{eqnarray}
\Gamma^t{}_{tr} &=&  -\Gamma^r{}_{rr}     \to -{1\over2a};
\qquad
\Gamma^r{}_{tt} \to -{2m^2\over a^3};
\nonumber\\
\Gamma^r{}_{\theta\theta} &=&  {\Gamma^r{}_{\phi\phi}\over \sin^2\theta}  \to2m ;
\quad\quad\;\;
\Gamma^\theta{}_{r\theta} \to  \Gamma^\phi{}_{r\phi}     = {1\over a}.
\end{eqnarray}

\subsection{Orthonormal components}

When a metric $g_{ab}$ is diagonal then the quickest way of calculating the orthonormal components of the Riemann and Weyl tensors is to simply set
\begin{equation}
R_{\hat a\hat b\hat c\hat d} = {R_{abcd}\over |g_{ac}|\,|g_{bd}|}; 
\qquad\qquad\qquad
C_{\hat a\hat b\hat c\hat d} = {C_{abcd}\over |g_{ac}|\,|g_{bd}|}.
\end{equation}
When a metric $g_{ab}$ is diagonal and a tensor $X_{ab}$ is diagonal then the quickest way of calculating the orthonormal components is to simply set
\begin{equation}
X_{\hat a\hat b} = {X_{ab}\over |g_{ab}|}.
\end{equation}
In both situations some delicacy is called for when crossing any horizon that might be present. 
Let us (using $-+++$ signature and assuming a diagonal metric) define
\begin{equation}
S = \sign(-g_{tt}) = \sign(g_{rr}).
\end{equation}
Then $S=+1$ in the domain of outer communication (above the horizon) and $S=-1$ below the horizon.

\subsection{Riemann tensor}

We shall now analyse what values of \( n \) result in non-singular components of various curvature tensors in an orthonormal basis \( (\t, \r, \th, \ph) \).
First, the non-zero orthonormal components of the Riemann tensor are:
\begin{align}
R_{\r\t\r\t} &= -\frac{2m}{r^3} - \frac{n a^2 \big[3 - (n+1)a^2/r^2\big](1-a^2/r^2)^{{n\over2} -2} }{2r^4},		\notag \\
R_{\r\th\r\th} &= R_{\r\ph\r\ph} = - R_{\th\t\th\t} = - R_{\ph\t\ph\t} =
-S\left\{ \frac{m}{r^3} + \frac{na^2 (1-a^2/r^2)^{{n\over2} -1} }{2r^{4}}\right\},	\notag \\
R_{\th\ph\th\ph} &=  \frac{2m}{r^3} + \frac{1-(1-a^2/r^2)^{n\over2}}{r^2}.
\end{align}

\enlargethispage{40pt}
Analysis of the numerator of \( R_{\r\t\r\t} \) shows that all of the orthonormal components of the Riemann tensor remain finite at \( r=a \) if and only if \( n\geq5 \).
Indeed as $r\to a$ (where $S\to-1$) we see
\begin{align}
R_{\r\t\r\t} &\to -\frac{2m}{a^3}; \qquad\qquad R_{\th\ph\th\ph} \to \frac{1}{a^2} + \frac{2m}{a^3}.
	\notag \\[5pt]
R_{\r\th\r\th} &= R_{\r\ph\r\ph} = - R_{\th\t\th\t} = - R_{\ph\t\ph\t} \to
+\frac{m}{a^3}.
\end{align}

Conversely at large $r$ (where  $S\to+1$) we see
\begin{align}
R_{\r\t\r\t} &= -\frac{2m}{r^3} + \O(a^2/r^4),		
\notag \\
R_{\r\th\r\th} &= R_{\r\ph\r\ph} = - R_{\th\t\th\t} = - R_{\ph\t\ph\t} =
-\frac{m}{r^3} +\O(a^2/r^4),	\notag \\
R_{\th\ph\th\ph} &=  \frac{2m}{r^3} +\O(a^2/r^4).
\end{align}
So, as it should, the spacetime curvature asymptotically approaches that of Schwarzschild.

\subsection{Ricci tensor}
The non-zero orthonormal components of the Ricci tensor are:
\begin{align}
R_{\t\t} &= - R_{\r\r} = - S \; \frac{na^2}{2r^4} \big[ 1 - (n-1)a^2/r^2 \big] (1-a^2/r^2)^{{n\over2} -2},	\notag \\
R_{\th\th} &= R_{\ph\ph} = \frac{1}{r^2} - \frac{1}{r^{2}} \big[1 + (n-1)a^2/r^2 \big] (1-a^2/r^2)^{{n\over2} -1}.
\end{align}

Analysis of the \( R_{\r\r} \) component shows that all of the components of the Ricci tensor remain finite at \( r=a \) so long as \( n\geq5 \).
Indeed as $r\to a$ we see
\begin{align}
R_{\t\t} &=  - R_{\r\r} \to 0,	
\qquad\qquad
R_{\th\th} = R_{\ph\ph} \to \frac{1}{a^2} .
\end{align}

Conversely at large $r$ we have
\begin{align}
R_{\t\t} &= - R_{\r\r} =R_{\th\th} = R_{\ph\ph}= -\frac{na^2}{2r^4} 
+\O(a^4/r^6).
\end{align}

\subsection{Ricci scalar}

As stated in Section \ref{S:intro}, our class of metrics is only curvature regular for \( n\geq5 \), where \( n \) is an odd integer.
Indeed, in general we have
\begin{equation}
R = \frac{2}{r^2} - (1-a^2/r^2)^{{n\over2}-2}  \bigg\{\frac{2 + (n-4)a^2/r^2 + (n-2)(n-1)a^4/r^4}{ r^{2} }\bigg\},
\end{equation}
and so the spacetime is non-singular at \( r=a \) if and only if \( n\geq5\).
Furthermore, any \( n\geq5 \) spacetime has positive scalar curvature at \( r=a \), where
$R \to  \frac{2}{a^2}$.

As an explicit example, 
\begin{equation}
R_{n=5} = \frac{2}{r^2} - \sqrt{r^2-a^2} \, \bigg\{\frac{2r^4 + a^2r^2 + 12a^4}{r^7}\bigg\},
\end{equation}
which is indeed singularity--free in the region \( r \in [a,\infty) \) and positive at \( r=a \).

\subsection{Einstein tensor}

The non-zero components of the Einstein tensor are
\begin{align}
G_{\t\t} &= - G_{\r\r} = \frac{S}{r^2}\left\{ 1
 - \left[1+(n-1){a^2\over r^2}\right] \left(1-{a^2\over r^2}\right)^{(n-2)/2}\right\},	\notag \\
G_{\th\th} &= G_{\ph\ph} = - \frac{na^2}{2r^4} \left[1-(n-1){a^2\over r^2}\right]
\left(1-{a^2\over r^2}\right)^{(n-4)/2}.
\end{align}

Analysis of the \( G_{\th\th} \) component reveals that the Einstein tensor remains finite in all of its orthonormal components if and only if \( n\geq5 \).
Indeed as $r\to a$ (where $S\to-1$) we see
\begin{align}
G_{\t\t} &=  - G_{\r\r} \to  - \frac{1}{a^2} ,	
\qquad\qquad
G_{\th\th} = G_{\ph\ph} \to 0.
\end{align}
At large $r$ (where $S\to+1$)  we have
\begin{equation}
G_{\t\t} =  - G_{\r\r} = G_{\th\th} = G_{\ph\ph} = -{na^2\over2r^4} + \O(a^4/r^6).
\end{equation}

\subsection{Weyl tensor}

The non-zero components of the Weyl tensor are
\begin{align}
C_{\r\t\r\t} &= 2S\, C_{\r\th\r\th} = 2 S\, C_{\r\ph\r\ph} 
= -2 S\,C_{\th\t\th\t} = -2 S\, C_{\ph\t\ph\t} = - C_{\th\ph\th\ph} 	\notag \\
&= - \frac{2m}{r^3} + {(1-a^2/r^2)^{{n\over2} -2} -1\over 3r^2}
\notag \\
&\qquad - a^2(1-a^2/r^2)^{{n\over2} -2} 
\left\{\frac{ (5n+4) - (n+2)(n+1)a^2/r^2}{6r^{4}}\right\}.
\end{align}
Thus, the components of the Weyl tensor remain finite at \( r=a \) so long as \( n\geq5 \).

Indeed as $r\to a$ (where $S\to-1$) we see
\begin{align}
C_{\r\t\r\t} &= -2C_{\r\th\r\th} = -2C_{\r\ph\r\ph} = +2 C_{\th\t\th\t} 
= +2 C_{\ph\t\ph\t} = - C_{\th\ph\th\ph}  \to - {1\over 3a^2} - {2m\over a^3}. 
\end{align}
At large $r$ (where $S\to+1$) we find
\begin{align}
C_{\r\t\r\t} &= 2C_{\r\th\r\th} = 2C_{\r\ph\r\ph} = -2 C_{\th\t\th\t} = -2 C_{\ph\t\ph\t} = - C_{\th\ph\th\ph}  \notag \\
&= - {2m\over r^3} -{na^2\over r^4} +\O(a^4/r^6). 
\end{align}

\subsection{Weyl scalar}
The Weyl scalar is defined by $C_{abcd}\, C^{abcd}$. In view of all the symmetries of the spacetime one can show that  $C_{abcd} \, C^{abcd} = 12 (C_{\r\t\r\t})^2$, so one gains no additional behaviour beyond looking at the Weyl tensor itself.
Thus, for purposes of tractability we will only display the result for \( n=5 \) at \( r=a \) in order to show that the \( n=5 \) spacetime is indeed regular at \( r=a \):
\begin{equation}
\left. (C_{abcd}\, C^{abcd})_{n=5}\right|_{r=a} = \frac{4(6m+a)^2}{3a^6}.
\end{equation}

\subsection{Kretschmann scalar}

The Kretschmann scalar is given by
\begin{equation}
K = R_{abcd}R^{abcd} = C_{abcd}C^{abcd} + 2R_{ab}R^{ab} - \frac{1}{3} R^2.
\end{equation}
The general result is rather messy and does not provide much additional insight into the spacetime.
Thus, for purposes of tractability we will only display the result for \( n=5 \) at \( r=a \) in order to show that the \( n=5 \) spacetime is indeed regular at \( r=a \):
\begin{equation}
\left. K_{n=5}\right|_{r=a} = \frac{4}{a^6} \big(a^2+4am+12m^2\big).
\end{equation}
The fact that the Kretschmann scalar is positive definite, and can be written as a sum of squares, is ultimately a due to spherical symmetry and the existence of a hypersurface orthogonal Killing vector~\cite{novel}.

\section{Surface gravity and Hawking temperature}\label{kappa}

Let us calculate the surface gravity at the event horizon for the generalised QMS spacetime. 
Because we are working in curvature coordinates we always have~\cite{DBH1}
\begin{equation}
\kappa_H = \lim_{r\to r_H} {1\over2} {\partial_r g_{tt}\over\sqrt{g_{tt}\; g_{rr}}}.
\end{equation}
For our general class of spacetimes,
\begin{equation}
\kappa_H = \left.{1\over2}\; \partial_r f_n(r)\right|_{r_H} 
= {m\over r_H^2} + {na^2\over 2 r_H^3} \left(1-{a^2\over r_H^2}\right)^{{n\over2}-1}.
\end{equation}
Using equation (\ref{E:horizon}) we can also rewrite this as
\begin{equation}
\kappa_H = 
 {m\over r_H^2} \left\{1 + {na^2\over r_H^2-a^2} \right\} .
\end{equation}
This result is, so far, exact.
Given that the horizon location is not analytically known for general $n$, we shall use the asymptotic result $r_{H}= 2m \left\{1+ \frac{n a^2}{8m^2}  - {n(3n-2) a^4\over 128 m^4} + \mathcal{O}({a^6\over m^6})\right\}$. 
This yields,
\begin{equation}
\kappa_H = 
 {1\over 4m}\left\{ 1 - {n(n-1)a^4\over32 m^4}  + \O(a^6/m^6)\right\}. 
\end{equation}
Note the potential $\O(a^2/m^2)$ term vanishes (which is why we estimated $r_H$ up to $\O(a^4)$).
As usual the Hawking temperature is simply $k_B T_H = {1\over2\pi} \,\hbar \,\kappa_H$.

\section{Stress-energy tensor}\label{stress}

Let us examine the Einstein field equations for this spacetime. 
Above the horizon, for $r > r_H$,  we have 
\begin{equation}
8\pi\,\rho = G_{\t\t};
\qquad\qquad
8\pi\,p_r = G_{\r\r}.
\end{equation}
Below the horizon, for $r < r_H$,  we have 
\begin{equation}
8\pi\,\rho = G_{\r\r};
\qquad\qquad
8\pi\,p_r = G_{\t\t}.
\end{equation}
But then regardless of whether one is above or below the horizon one has
\begin{align}
\rho &= - p_r= \frac{1}{8\pi r^2}\left\{ 1
 - \left[1+(n-1){a^2\over r^2}\right] \left(1-{a^2\over r^2}\right)^{(n-2)/2}\right\},	\notag \\
p_\perp&= - \frac{na^2}{16\pi r^4} \left[1-(n-1){a^2\over r^2}\right]
\left(1-{a^2\over r^2}\right)^{(n-4)/2}.
\end{align}

By inspection, for $n>1$ we see that $p_\perp(r)=0$ at $r= \sqrt{n-1} \;a$. Indeed we see that $p_\perp(r)>0$ for $r< \sqrt{n-1} \;a$ and $p_\perp(r)<0$ for $r> \sqrt{n-1} \;a$.
The analagous result for $\rho(r)$ is not analytically tractable (though it presents no numerical difficulty) as by inspection it amounts to finding the roots of 
\begin{equation}
(r^2-a^2) r^n - (r^2-a^2)^{n\over2} (r^2+ (n-1)a^2) = 0.
\end{equation}
We note that asymptotically
\begin{equation}
\label{E:rho}
\rho = - {na^2\over 16\pi r^4} + \O(a^4/ r^6),
\end{equation}
and 
\begin{equation}
p_\perp = - {na^2\over 16\pi r^4} + \O(a^4/ r^6).
\end{equation}

An initially surprising result is that the stress-energy tensor has no dependence on the mass \( m \) of the spacetime.
To see what is going on here, consider the Misner--Sharp quasi-local mass
\begin{equation}
1-{2m(r)\over r} = f_n(r) \qquad\implies \qquad 
m(r) = m +{r\over2} \left\{ 1- \left(1-{a^2\over r^2}\right)^{n\over2} \right\}.
\end{equation}
Then, noting that $m = m(r) + 4\pi \int_r^\infty \rho(\bar r) \bar r^2 \d\bar r$ above the horizon, we see
\begin{equation}
4\pi \int_r^\infty \rho(\bar r) \bar r^2 \d\bar r = -{r\over2} \left\{ 1- \left(1-{a^2\over r^2}\right)^{n\over2} \right\}.
\end{equation}
Here the right hand side of the equation is manifestly independent of $m$. 
Consequently, without need of any detailed calculation, $\rho(r)$ is manifestly independent of $m$. 
As an aside note that $m(r_H) = {r_H\over2}$, so we could also write
$ m(r) = {r_H\over2} + 4\pi \int_{r_H}^r \rho(\bar r) \bar r^2 \d\bar r$.

\section{Energy conditions}\label{S:ecs}

We will now analyse the classical energy conditions (as discussed in section \ref{subsubsec:energy}) for our general class of spacetimes. 
While it can be argued that the classical energy conditions are not truly fundamental~\cite{Barcelo:2002bv, LNP-survey, Visser:1996}, often being violated by semi-classical quantum effects, they are nevertheless extremely useful indicative probes, well worth the effort required to analyse them. 

\subsection{Null energy condition}

A necessary and sufficient condition for the null energy condition (NEC) to hold is that both $\rho + p_{r}\geq 0$ and $\rho+p_\perp\geq 0$  for all $r$, $a$, $m$.  
Since $\rho=-p_{r}$, the former inequality is trivially satisfied, and for all $r\geq a$ we may simply consider
\begin{multline}
\rho+p_\perp  = \frac{1}{8\pi r^2} \left\{ 1- \frac{(1-a^2/r^2)^{{n\over2}-2}}{2} \times \right. \\
\left. \bigg[ 2 + (3n-4)a^2/r^2 - (n+2)(n-1)a^4/r^4 \bigg] \right\}.
\end{multline}
Whether or not this satisfies the NEC depends on the value for \( n \).
Furthermore, for no value of \( n \) is the NEC \emph{globally} satisfied.

\begin{figure}[H]
\begin{center}
\includegraphics[width=\textwidth]{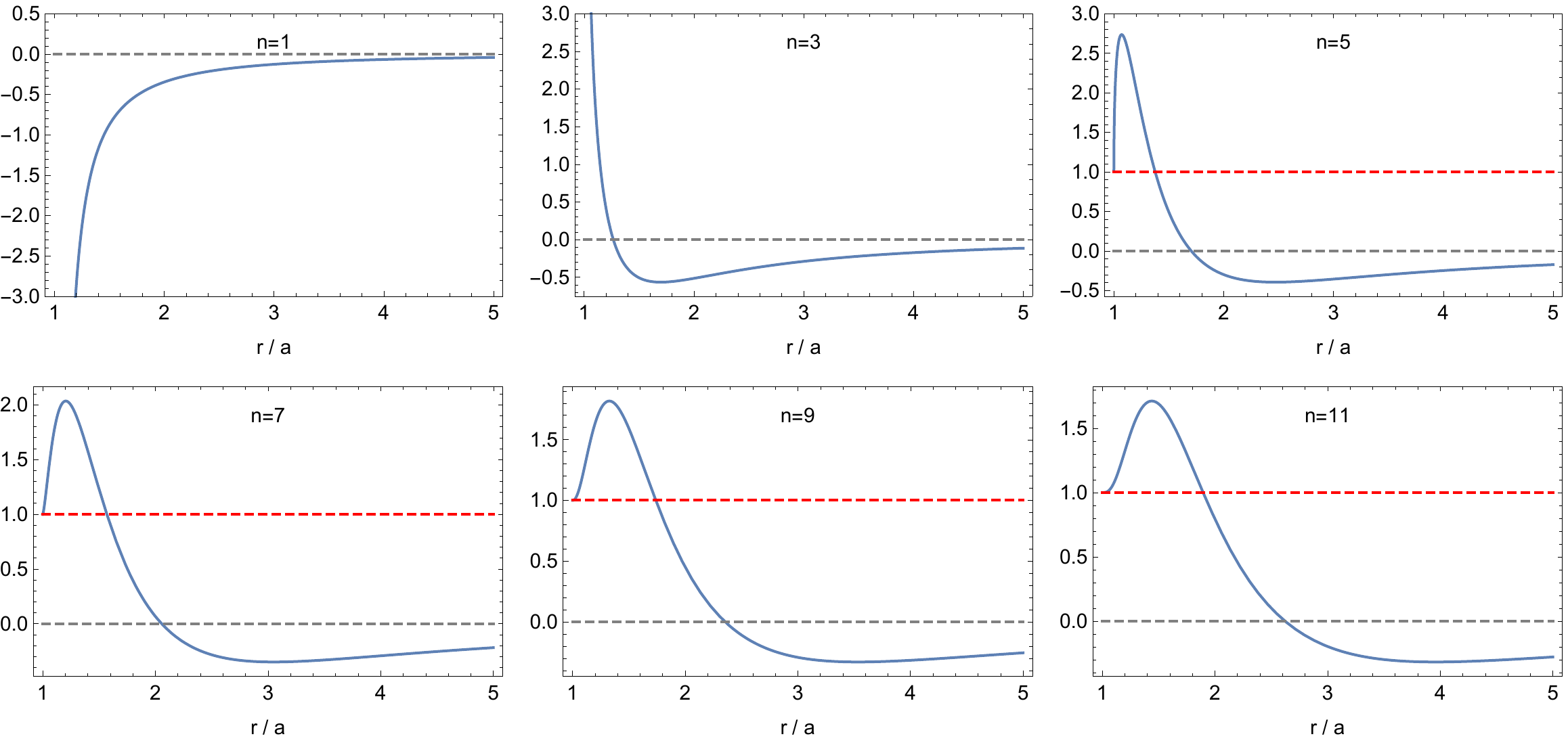}
\caption[Quantum deformed regular black hole: Plots of the NEC]{Plots of the NEC for several values of $n$. Here the $y$-axis depicts $8\pi r^2(\rho + p_{\perp})$, plotting against $r/a$ on the $x$-axis. Of particular interest are the qualitative differences in behaviour as $r/a\rightarrow 1$; we see divergent behaviour for the $n=1$ and $n=3$ cases, whilst for $n\geq 5$ we see $8\pi r^2(\rho + p_{\perp})\rightarrow 1$ as $r\rightarrow a$. This is ultimately due to the fact that the $n\geq 5$ cases are \emph{curvature regular}, with globally finite stress-energy components.}
\label{F:nec}
\end{center}
\end{figure}

Provided $n\geq 5$, so that the limits exist, we have
\begin{equation}
\lim_{r\to a} (\rho+p_\perp) = +{1\over8\pi a^2}.
\end{equation}
So the NEC is definitely satisfied deep in the core of the system.
Note that at asymptotically large distances
\begin{equation}
\rho+p_t  = - {na^2\over 8\pi r^4} + \O(a^4/ r^6).
\end{equation}
So the NEC (and consequently all the other classical point-wise energy conditions) are  always violated at asymptotically large distances.
However, for some values of \( n \), there are bounded regions of the spacetime in which the NEC is satisfied.
See figure \ref{F:nec}.

\subsection{Weak energy condition}
In order to satisfy the weak energy condition (WEC) we require the NEC be satisfied, and  in addition \( \rho \geq 0 \).
But in view of the asymptotic estimate (\ref{E:rho}) for $\rho$ we see that the WEC is always violated at large distances.
Furthermore, it can be seen from table~\ref{T:1} that the region in which the NEC is satisfied is always larger than that in which $\rho$  is positive (this would be as good as impossible to prove analytically for general \( n \)).
Thus, we can conclude (see table \ref{T:2})  that the WEC is satisfied for smaller regions than the NEC for all values of \( n \).

\vfil

\subsection{Strong energy condition}

In order to satisfy the strong energy condition (SEC) we require the NEC to be satisfied, and  in addition \( \rho + p_r + 2p_\perp = 2p_\perp \geq 0 \).
But regardless of whether one is above or below the horizon, the second of these conditions $p_\perp\geq0$ amounts to
\begin{equation}
0 < a < r \leq a \sqrt{n-1}.
\end{equation}
However, it can be seen from table \ref{T:1} that the region in which the NEC is satisfied is always smaller than that in which $p_\perp$ is positive (this would be as good as impossible to prove analytically for general \( n \)).
Thus, we can conclude (see table \ref{T:2}) that the SEC is satisfied in the same region as the NEC for all values of \( n \).

\subsection{Dominant energy condition}

The dominant energy condition is the strongest of the standard classical energy conditions.
Perhaps the best physical interpretation of the DEC is that for any observer with timelike 4-velocity $V^a$ the flux vector $F^a = T^{ab}\, V_b$ is non-spacelike (timelike or null). 
It is a standard result that in spherical symmetry (in fact for any type~I stress-energy tensor) this reduces to positivity of the energy density $\rho>0$ combined with the condition $|p_i| \leq \rho$. 
Since in the current framework for the radial pressure we always have $p_r = -\rho$, the only real constraint comes from demanding $|p_\perp| \leq \rho$.
But this means we want \emph{both} $\rho+p_\perp\geq 0$ \emph{and} $\rho-p_\perp\geq 0$.
The first of these conditions is just the NEC, so the only new constraint comes from the second condition. 
By inspection, it can be seen from table \ref{T:1} that the region in which the NEC is satisfied is always larger than that in which $\rho-p_\perp $  is positive (this would be as good as impossible to prove analytically for general \( n \)).
Thus, we can conclude (see table \ref{T:2})  that the DEC is satisfied for smaller regions than the NEC for all values of \( n \).

\begin{table}[H]
\centering
    \caption{Regions of the spacetime where the orthonormal components of the stress-energy tensor satisfy certain inequalities.}
    \tabulinesep=1.5mm
    \begin{tabu}{|c|c|c|c|c|}
        \hline
         $n$ & $\rho+p_\perp\geq0$ & $\rho\geq0$ & $p_\perp\geq0$ & $\rho-p_\perp\geq0$ \\
        \hline \hline
         0 & $a<r<\infty$ & $a<r<\infty$ & $a<r<\infty$ & $a<r<\infty$ \\
        \hline
         1 & globally violated & globally violated & globally violated & globally violated \\
        \hline
         3 & $a<r \lessapprox 1.26595a$ & $a<r \lessapprox 1.07457a$ & $a<r \lessapprox 1.41421a$ & globally violated\\
        \hline
        5 & $a<r \lessapprox 1.70468a$ & $a<r \lessapprox 1.37005a$ & $a<r \leq 2a$ & $a<r \lessapprox 1.00961a$ \\
        \hline
        7 & $a<r \lessapprox 2.05561a$ & $a<r \lessapprox 1.62933a$ & $a<r \lessapprox 2.44949a$ & $a<r \lessapprox 1.11129a$ \\
        \hline
        9 & $a<r \lessapprox 2.35559a$ & $a<r \lessapprox 1.85537a$ & $a<r \lessapprox 2.82843$ & $a<r \lessapprox 1.23076a$ \\
        \hline
        11 & $a< r\lessapprox 2.62173a$ & $a<r \lessapprox 2.05757a$ & $a<r \lessapprox 3.16228a$ & $a<r \lessapprox 1.34552a$ \\
        \hline
        \vdots & \vdots & \vdots & \vdots & \vdots \\
        \hline
    \end{tabu}
\label{T:1}
\end{table}
\begin{table}[H]
\centering
    \caption{Regions of the spacetime where the energy conditions are satisfied.}
    \tabulinesep=1.5mm
    \begin{tabu}{|c|c|c|c|c|}
        \hline
         $n$ & NEC & WEC & SEC & DEC \\
        \hline \hline
         0 & $a<r<\infty$ & $a<r<\infty$ & $a<r<\infty$ & $a<r<\infty$ \\
        \hline
         1 & globally violated & globally violated & globally violated & globally violated \\
        \hline
         3 & $a<r \lessapprox 1.26595a$ & $a<r \lessapprox 1.07457a$ & same as NEC & globally violated\\
        \hline
        5 & $a<r \lessapprox 1.70468a$ & $a<r \lessapprox 1.37005a$ & same as NEC & $a<r \lessapprox 1.00961a$ \\
        \hline
        7 & $a<r \lessapprox 2.05561a$ & $a<r \lessapprox 1.62933a$ & same as NEC & $a<r \lessapprox 1.11129a$ \\
        \hline
        9 & $a<r \lessapprox 2.35559a$ & $a<r \lessapprox 1.85537a$ & same as NEC & $a<r \lessapprox 1.23076a$ \\
        \hline
        11 & $a< r\lessapprox 2.62173a$ & $a<r \lessapprox 2.05757a$ & same as NEC & $a<r \lessapprox 1.34552a$ \\
        \hline
        \vdots & \vdots & \vdots & \vdots & \vdots \\
        \hline
    \end{tabu}
\label{T:2}
\end{table}

\vfil

\section{ISCO and photon sphere analysis}\label{ISCO}

We have the generalised quantum modified Schwarzschild metric

\begin{eqnarray}
\d s^{2} &=& -\left\{\left(1-\frac{a^{2}}{r^{2}}\right)^{\frac{n}{2}}-\frac{2m}{r}\right\}\d t^{2} + \frac{\d r^{2}}{\left(1-\frac{a^{2}}{r^{2}}\right)^{\frac{n}{2}}-\frac{2m}{r}} + r^{2}\,\d\Omega^{2}_{2}.
\label{k-s-metric-repeat}
\end{eqnarray}

Let us now find the location of both the photon sphere for massless particles, and the ISCO for massive particles, as functions of the parameters $m$, $n$, and $a$.
Consider the tangent vector to the worldline of a massive or massless particle, parameterized by some arbitrary affine parameter, $\lambda$: 
\begin{equation}
g_{ab}\frac{\d x^{a}}{\d\lambda}\frac{\d x^{b}}{\d\lambda}=-g_{tt}\left(\frac{\d t}{\d\lambda}\right)^{2}+g_{rr}\left(\frac{\d r}{\d\lambda}\right)^{2}+r^{2}\left\lbrace\left(\frac{\d\theta}{\d\lambda}\right)^{2}+\sin^{2}\theta \left(\frac{\d\phi}{\d\lambda}\right)^{2}\right\rbrace.
\end{equation}
As was done for the spacetime in chapter \ref{C:Mink-core}, we may without loss of generality separate the two physically interesting cases (timelike and null) by defining:
\begin{equation}
\epsilon = \left\{
\begin{array}{rl}
-1 & \qquad\mbox{massive particle, \emph{i.e.} timelike worldline} \\
0 & \qquad\mbox{massless particle, \emph{i.e.} null worldline}.
\end{array}\right. 
\end{equation}
That is, $\d s^{2}/\d\lambda^2=\epsilon$. Due to the metric being spherically symmetric we may fix $\theta=\frac{\pi}{2}$ arbitrarily and view the reduced equatorial problem:
\begin{equation}
g_{ab}\frac{\d x^{a}}{\d\lambda}\frac{\d x^{b}}{\d\lambda}=-g_{tt}\left(\frac{\d t}{\d\lambda}\right)^{2}+g_{rr}\left(\frac{\d r}{\d\lambda}\right)^{2}+r^{2}\left(\frac{\d\phi}{\d\lambda}\right)^{2}=\epsilon.
\end{equation}

The Killing symmetries yield the following expressions for the conserved energy $E$ and angular momentum $L$ per unit mass:
\begin{equation}
\left\{{\left(1-\frac{a^2}{r^2}\right)^{\frac{n}{2}}}-\frac{2m}{r}\right\}\left(\frac{\d t}{\d\lambda}\right)=E \ ; \qquad\quad r^{2}\left(\frac{\d\phi}{\d\lambda}\right)=L.
\end{equation}
Hence
\begin{equation}
\left\{{\left(1-\frac{a^2}{r^2}\right)^{\frac{n}{2}}}-\frac{2m}{r}\right\}^{-1}\left\lbrace -E^{2}+\left(\frac{\d r}{\d\lambda}\right)^{2}\right\rbrace+\frac{L^{2}}{r^{2}}=\epsilon,
\end{equation}
implying
\begin{equation}
\left(\frac{\d r}{\d\lambda}\right)^{2}=E^{2}+\left\{{\left(1-\frac{a^2}{r^2}\right)^{\frac{n}{2}}}
-\frac{2m}{r}\right\}\left\lbrace\epsilon-\frac{L^{2}}{r^{2}}\right\rbrace.
\end{equation}
This gives ``effective potentials" for geodesic orbits as follows:
\begin{equation}
V_{\epsilon}(r)=\left\{{\left(1-\frac{a^2}{r^2}\right)^{\frac{n}{2}}}-\frac{2m}{r}\right\}\left\lbrace -\epsilon+\frac{L^{2}}{r^{2}}\right\rbrace .
\end{equation}

\subsection{Photon orbits}
For a photon orbit we have the massless particle case $\epsilon=0$. Since we are in a spherically symmetric environment, solving for the locations of such orbits amounts to finding the coordinate location of the ``photon sphere''. These circular orbits occur at $V_{0}^{'}(r)=0$. That is:
\begin{equation}\label{V0r}
V_{0}(r)=\left\{\left(1-{a^2\over r^2}\right)^{\frac{n}{2}}-\frac{2m}{r}\right\}
\left\{\frac{L^{2}}{r^{2}}\right\},
\end{equation}
leading to:
\begin{equation}
    V_{0}^{'}(r) = \frac{L^2}{r^4}\left\lbrace 6m+r\left(1-\frac{a^2}{r^2}\right)^{\frac{n}{2}-1}\left[(n+2)\frac{a^2}{r^2}-2\right]\right\rbrace \ .
\end{equation}

Solving $V_{0}^{'}(r)=0$ analytically is intractable, but we may perform a Taylor series expansion of the above function about $a=0$ for a valid approximation (recall $a$ is associated with the Planck length). 

To fifth-order this yields:
\begin{equation}
    V_{0}^{'}(r) = \frac{2L^{2}}{r^4}(3m-r)+\frac{2L^{2}na^{2}}{r^5}-\frac{3na^{4}L^{2}(n-2)}{4r^{7}} + \mathcal{O}\left(L^{2}a^{6}/r^{9}\right) \ .
\end{equation}

Equating this to zero and solving for $r$ yields:

\begin{equation}\label{rphgenn}
    r_c = 3m \left\{1 + \frac{a^2n}{(3m)^2} 
    - {n(11n-6)a^4\over 8(3m)^4} + \mathcal{O}(a^6/m^6) \right\} .
\end{equation}

The $a=0$, (or $n=0$), Schwarzschild sanity check reproduces $r_c=3m$, the expected result.

To verify stability, we check the sign of $V_{0}^{''}(r)$:
\begin{multline}\label{V''}
    V_{0}^{''}(r) = -\frac{L^{2}}{r^{4}}\left\lbrace\frac{24m}{r}-\left(1-\frac{a^2}{r^2}\right)^{\frac{n}{2}-2} \times \right. \\
    \left. \left[6-(7n+12)\frac{a^{2}}{r^{2}}+(n+2)(n+3)\frac{a^4}{r^4}\right]\right\rbrace .
\end{multline}

We now substitute the approximate expression for $r_c$ into equation~(\ref{V''}) to determine the sign of $V_{0}''(r_c)$.
We find:
\begin{equation}
V_0''(r_c) = -{2L^2\over 81 m^4} \left\{ 1-{3n a^2\over (3 m)^2} + {n(67n-6)a^4 \over 8 (3 m)^4} 
+ \O(a^6/m^6) \right\}.
\end{equation}
Given that all bracketed terms to the right of the $1$ are strictly subdominant in view of $a\ll m$, we may conclude that $V_{0}^{''}(r_c)<0$, and hence the null orbits at $r=r_c$ are unstable.

Let us now recall the generalised form of equation~(\ref{V0r}), and specialise to $n=5$ (the lowest value for $n$ for which our quantum deformed Schwarzschild spacetime is \emph{regular}). We have:
\begin{align}
    V_{0}(r,n=5) &= \frac{L^2}{r^2}\left\lbrace\left(1-\frac{a^{2}}{r^2}\right)^{\frac{5}{2}}-\frac{2m}{r}\right\rbrace ; \\
    V_{0}'(r,n=5) &= \frac{L^2}{r^4}\left\lbrace 6m -\sqrt{r^2-a^2}\left(2-\frac{9a^2}{r^2}+\frac{7a^4}{r^4}\right)\right\rbrace .
\end{align}
Once again setting this to zero and attempting to solve analytically is an intractable line of inquiry, and we instead inflict Taylor series expansions about $a=0$. 
To fifth-order we have the following:
\begin{equation}
    V_{0}'(r,n=5) = -\frac{2L^2}{r^3} \left\{1-{3m\over r} - \frac{5a^2}{r^2}
    + {45 L^2 a^4\over 8 r^4}+\mathcal{O}(a^6/r^6) \right\}.
\end{equation}
Implying,
\begin{equation}
    \quad r_{c} = 3m\left\{ 1  + \frac{5a^2}{(3m)^2} -{245a^4\over 8(3 m)^4} + \mathcal{O}(a^6/m^6)\right\} ,
\end{equation}
which is consistent with the result for general $n$ displayed in equation~(\ref{rphgenn}).

\subsection{ISCOs}

For massive particles the geodesic orbit corresponds to a timelike worldline and we have the case that $\epsilon=-1$. Therefore:
\begin{equation}
V_{-1}(r) = \left\{\left(1-\frac{a^2}{r^2}\right)^{\frac{n}{2}}-\frac{2m}{r}\right\}
\left\{1+\frac{L^{2}}{r^{2}}\right\} ,
\end{equation}
and it is easily verified that this leads to:
\begin{equation}
V_{-1}^{'}(r) = \frac{2m(3L^2+r^2)}{r^4} + \frac{(1-a^2/r^2)^{{\frac{n}{2}-1}}}{r^{3}}
\left[n a^2 +L^2\left((n+2)\frac{a^2}{r^2}-2\right)\right] .
\end{equation}

For small $a$ we have
\begin{equation}
V_{-1}(r) = \left\{1+{L^2\over r^2}\right\} \left\{1 -{2m\over r}-{na^2\over2r^2}
+{n(n-2) a^4\over 8 r^4} +\O\left(a^6\over r^6\right)\right\},
\end{equation}
and
\begin{multline}
V_{-1}^{'}(r) = {2(L^2(3m-r) + mr^2)\over r^4} +{(2L^2+r^2)n a^2\over r^5} - \\
{(3L^2+2r^2)n(n-2)a^4\over 4r^7} 
+\O\left(a^6\over r^{{7}}\right).
\end{multline}
Equating this to zero and rearranging for $r$ presents an intractable line of inquiry. Instead it is preferable to assume a fixed circular orbit at some $r=r_{c}$, and rearrange the required angular momentum $L_{c}$ to be a function of $r_{c}$, $m$, and $a$. It then follows that the innermost circular orbit shall be the value of $r_{c}$ for which $L_{c}$ is minimised. It is of course completely equivalent to perform this procedure for the mathematical object $L_{c}^{2}$, and we do so for tractability.
Hence, if $V_{-1}^{'}(r_{c})=0$, we have:
\begin{equation}\label{Lcsq}
    L_{c}^{2} = \frac{na^2\left(1-\frac{a^2}{r^2}\right)^{\frac{n}{2}}+2mr\left(1-\frac{a^2}{r^2}\right)}{\left(1-\frac{a^2}{r^2}\right)^{\frac{n}{2}}\left[2-(n+2)\frac{a^2}{r^2}\right]-\frac{6m}{r}\left(1-\frac{a^2}{r^2}\right)}.
\end{equation}
For small $a$ we have
\begin{multline}\label{Lcsqa0}
L_c^2 = {mr^2\over r-3m} + {n r (r-m)a^2\over2(r-3m)^2} - \\
{n\{(2n+4)r^2 +(5n-18) mr -9(n-2) m^2 \}a^4\over 8 r (r-3m)^3}
+ \O(a^6).
\end{multline}

As a consistency check, for large $r_{c}$ (\emph{i.e.} $r_{c}\gg a, m$) we observe from the dominant term of equation~(\ref{Lcsqa0}) that $L_{c}\approx\sqrt{mr_{c}}$, which is consistent with the expected value when considering circular orbits in weak-field GR. 
Indeed it is easy to check that for large $r$ we have $L_{c}^{2} = mr_{c} + \mathcal{O}(1)$.
Recall that in classical physics the angular momentum per unit mass for a particle with angular velocity $\omega$ is $L_{c}\sim\omega r_{c}^2$, and Kepler's third law of planetary motion implies that $r_c^2 \omega^2\sim {G_{N}m}/{r_{c}}$. 
(Here $m$ is the mass of the central object, as above.) 
It therefore follows that $L_{c}\sim\sqrt{{G_Nm}/{r_{c}}}\; r_{c}$. 
That is $L_{c}\sim\sqrt{mr_{c}}$, as above.

Differentiating equation~(\ref{Lcsq}) and finding the resulting roots is not analytically feasible. We instead differentiate equation~(\ref{Lcsqa0}), obtaining a Taylor series for $\frac{\partial L_{c}^2}{\partial r_{c}}$ for small $a$:
\begin{multline}
\frac{\partial L_{c}^2}{\partial r_{c}} = {mr_c(r_c-6m)\over (r_c-3m)^2} 
- {mn(5r_c-3m)a^2\over2(r_c-3m)^3} - \\
{n\{16r_c^3 +(n-2) (4 r_c^3+21mr_c^2-36m^2r_c +27m^3)  \} a^4\over 8 r_c^2 (r_c-3m)^4}
+ \O(a^6) .
\end{multline}

Solving for the stationary points yields:
\begin{equation}
r_\ISCO = 6m 
\left\{ 1+ \frac{na^2}{8m^2} -{n(49n-22)a^4\over3456m^4}
+ \mathcal{O}\left(a^6\over m^6\right) \right\}  ,
\end{equation}
and the $a=0$ Schwarzschild sanity check reproduces $r_{c} = 6m$ as required.

\subsection{Summary}

Denoting $r_\H$ as the location of the horizon, $r_c$ as the location of the photon sphere, and $r_\ISCO$ as the location of the ISCO, we have the following summary:
\begin{itemize}
\item 
$r_{\scriptscriptstyle{H}}= 2m \times
\left\{1+ \frac{n a^2}{2(2m)^2}  - {n(3n-2) a^4\over 8(2m)^4} + \mathcal{O}\left({a^6\over m^6}\right)\right\}$;
\item 
$r_c= 3m \times
\left\{1 + \frac{a^2n}{(3m)^2} 
    - {n(11n-6)a^4\over 8(3m)^4} + \mathcal{O}\left({a^6\over m^6}\right) \right\}$ ;
\item 
$r_\ISCO= 6m \times
\left\{ 1+ \frac{na^2}{8m^2} -{n(49n-22)a^4\over3456m^4}+ \mathcal{O}\left({a^6\over m^6}\right) \right\} $.
\end{itemize}

\section{Regge--Wheeler analysis}\label{Regge}

Now considering the Regge--Wheeler equation, in view of the formalism developed in section \ref{subsection:RW-mink-core}, (see also references~\cite{expmetric, Fernando:2012, Flachi:2012}), we may explicitly evaluate the Regge--Wheeler potentials for particles of spin $S\in\lbrace 0,1\rbrace$ in our spacetime. 
Firstly define a tortoise coordinate as follows:
\begin{equation}
\d r_{*} = {\d r\over \left(1-{a^2\over r^2}\right)^{n\over2} - {2m\over r}}.
\end{equation}
This tortoise coordinate is, for general $n$, not analytically defined. 
However, this does not affect our analysis. 
This coordinate transformation yields the following expression for the metric:
\begin{equation}
\d s^2 = \left\{ \left(1-{a^2\over r^2}\right)^{n\over2} - \frac{2m}{r}\right\} \,
\bigg\lbrace -\d t^2+\d r_{*}^2\bigg\rbrace
+r^2\left(\d\theta^2+\sin^2\theta\; \d\phi^2\right)  .
\end{equation}
It is convenient to write this as:
\begin{equation}
\d s^2 = A(r_*)^2\bigg\lbrace -\d t^2+\d r_{*}^2\bigg\rbrace+B(r_*)^2\left(\d\theta^2+\sin^2\theta \;\d\phi^2\right)  .
\end{equation}
The Regge--Wheeler equation is~\cite{VisserRW, Fernando:2012, Flachi:2012}:
\begin{equation}
\partial_{r_{*}}^{2}\hat{\phi}+\lbrace \omega^2-\mathcal{V}_S\rbrace\hat\phi = 0 ,
\end{equation}
where, as in section \ref{subsection:RW-mink-core}, $\hat\phi$ is the scalar or vector field, $\mathcal{V}$ is the spin-depen\-dent Regge--Wheeler potential for our particle, and $\omega$ is some temporal frequency component in the Fourier domain.
For a scalar field ($S=0$) examination of the d'Alembertian equation quickly yields:
\begin{equation}
\mathcal{V}_{S=0} =   \left\lbrace{A^2 \over B^2} \right\rbrace \ell(\ell+1)
+ {\partial_{r_{*}}^2 B \over B}  .
\end{equation}
For a massless vector field, ($S=1$; \emph{e.g.} photon), explicit conformal invariance in 3+1 dimensions guarantees that the Regge--Wheeler potential can depend only on the ratio $A/B$, whence normalising to known results implies:
\begin{equation}
\mathcal{V}_{S=1} =   \left\lbrace{A^2 \over B^2} \right\rbrace \ell(\ell+1).
\end{equation}
Collecting results, for $S\in\{0,1\}$ we have:
\begin{equation}
\mathcal{V}_{S\in\{0,1\}} =   \left\lbrace{A^2 \over B^2} \right\rbrace \ell(\ell+1)
+ (1-S) {\partial_{r_{*}}^2 B \over B}  .
\end{equation}
The spin 2 axial mode is somewhat messier, and (for the purposes of this thesis) not of immediate interest. 

\noindent Noting that for our metric $\partial_{r_{*}} = \left\{\left(1-\frac{a^2}{r^2}\right)^{\frac{n}{2}} - \frac{2m}{r}\right\}\partial_{r}$ and $B(r)=r$, we have:
\begin{align}
\frac{\partial_{r_{*}}^2 B}{B} 
&= \frac{\partial_{r_{*}}\left\{\left(1-\frac{a^2}{r^2}\right)^{\frac{n}{2}} - \frac{2m}{r}\right\}}{r} \nonumber \\
&={1\over r^2} \left\{\left(1-\frac{a^2}{r^2}\right)^{\frac{n}{2}} - \frac{2m}{r}\right\}
\left\{n\left(1-\frac{a^2}{r^2}\right)^{\frac{n}{2}-1} {a^2\over r^2} + \frac{2m}{r}\right\}.
\end{align}
For small $a$:
\begin{multline}
\frac{\partial_{r_{*}}^2 B}{B} = {2m(1-2m/r)\over r^3} + {n(r-3m)\over r^5}a^2 + \\
 {n\{5(n-2) m - 4(n-1) r\}  \over 4 r^7} a^4+ \O\left(\frac{ma^6}{r^9}\right) .
\end{multline}
Therefore:
\begin{multline}
\mathcal{V}_{S\in\{0,1\}} = 
{1\over r^2} \left[\left(1-\frac{a^2}{r^2}\right)^{\frac{n}{2}}-\frac{2m}{r}\right] \times \\
\left\{\ell\left(\ell+1\right)+{\left(1-S\right)}\left[n\left(1-\frac{a^2}{r^2}\right)^{\frac{n}{2}-1} \,{a^2\over r^2} + \frac{2m}{r}\right]\right\}. 
\end{multline}
This has the correct behaviour as $a\to0$, reducing to the Regge--Wheeler potential for the Schwarzschild spacetime:
\begin{equation}
    \lim_{a\rightarrow 0}\mathcal{V}_{S\in\lbrace 0,1\rbrace} = \frac{1}{r^2}\left[1-\frac{2m}{r}\right]\left\lbrace\ell(\ell+1)+(1-S)\frac{2m}{r}\right\rbrace .
\end{equation}

In the small $a$ approximation we have the asymptotic result
\begin{align}
\mathcal{V}_{S\in\lbrace 0,1\rbrace} &= \frac{\left(1-\frac{2m}{r}\right)}{r^2}\left\lbrace\ell(\ell+1)+(1-S)\frac{2m}{r}\right\rbrace \notag \\
&-\frac{na^{2}}{2r^4}\left\lbrace\ell(\ell+1)+2(1-S)\left[\frac{3m}{r}-1\right]\right\rbrace \notag \\
&+\frac{na^4}{2r^6}\left\lbrace\frac{(n-2)}{4}\left[\ell(\ell+1)\right]
-\left(1-S\right)\left[2(n-1)+5\left(1-\frac{n}{2}\right)\frac{m}{r}\right]\right\rbrace  \notag\\
&+ \mathcal{O}\left(\frac{a^6}{r^8}\right).
\end{align}
As we saw in section \ref{subsection:RW-mink-core}, the Regge--Wheeler equation is fundamental to exploring the quasi-normal modes of the candidate spacetimes, an integral part of the ``ringdown'' phase of the LIGO calculation to detect astrophysical phenomena \emph{via} gravitational waves. 
However, exploring the quasi-normal modes is, for now, relegated to the domain of future research.

\section{Summary}

In this chapter we showed how the original Kazakov--Solodukhin ``quantum deformed Schwarz\-schild spacetime''~\cite{quantum-bh2} is slightly more ``regular'' than Schwarzschild spacetime, but it is not ``regular'' in the sense normally intended in the general relativity community.  
While the metric components are regular, both Christoffel symbols and curvature invariants diverge at the ``centre'' of the spacetime, a 2-sphere  where $r\to a$ with finite area $A=4\pi a^2$. 
The ``smearing out'' of the ``centre'' to $r\to a$ is not sufficient to guarantee curvature regularity.

We have generalised the original Kazakov--Solodukhin spacetime to a two-parameter class compatible with the ideas mooted in reference~\cite{quantum-bh2}.
Our generalised two-parameter class of ``quantum corrected'' Schwarz\-schild spacetimes contains exemplars which have  much better regularity properties, and we can distinguish three levels of regularity: metric regularity, Christoffel regularity, and regularity of the curvature invariants.  

Furthermore, our generalised two-parameter class of models distorts Schwarz\-schild spacetime in a clear and controlled way --- so providing yet more examples of black-hole ``mimickers'' potentially of interest for observational purposes.  
In this regard we have analysed the geometry, surface gravity, stress-energy, and classical energy conditions. 
We have also perturbatively analysed the locations of ISCOs and photon spheres, and set up the appropriate Regge--Wheeler formalism for spin-1 and spin-0 excitations.
\chapter{Conclusions}\label{C:conclusions}

This thesis explored many new and interesting results in the theory of relativity. 
The main focus of the work was to highlight the issue of singularities in black hole spacetimes and their classical resolution through black hole mimickers. 
Specific models of black hole mimicker spacetimes were thoroughly examined within the framework of classical general relativity. 
Each relevant chapter contains a summary of the results it contains, and so this chapter will only provide a brief summary of the key points from the thesis, and a general outlook on the future of the field.

\section{Combination of relativistic velocities using quaternions}
In chapter \ref{C:comb-veloc}, a new formula was presented for combining general velocities in special relativity using quaternions.
By identifying velocities \( \vec{v}_1,\vec{v}_2\in\RR^3 \) with pure quaternions via a natural isomorphism \( \hat{n}\mapsto\vu{n} \), it was shown that their combination could be expressed as
\begin{equation}
    \vb{w}_{1\oplus2} 
    = \vb{w}_1 \oplus \vb{w}_2 = (1 - \vb{w}_1\vb{w}_2)^{-1}(\vb{w}_1 + \vb{w}_2)
    = (\w{1}+\w{2}) (1 - \w{2}\w{1})^{-1},
\end{equation}
where the ``relativistic half-velocities'' \( \vb{w}_i \) are defined by \( \vb{v}_i = \vb{w}_i \oplus \vb{w}_i \).
In terms of the rapidity \( \xi \):
\begin{equation}
w = \tanh(\xi/2), \qquad v = \tanh(\xi) = {2w\over 1+w^2}.
\end{equation}
That is,
\begin{equation}
w = \tanh\left(\frac{1}{2}\tanh^{-1}(v)\right) = \frac{v}{1+\sqrt{1-v^2}}.
\end{equation}

This result was further extended to obtain novel, elegant and compact formulae for both the associated Wigner angle \( \Omega \) and the direction of the combined velocities:
\begin{equation}
\e^{\vb{\Omega}} = \e^{\Omega\,\vb{\hat\Omega}} = (1-\w{1}\w{2})^{-1}  (1-\w{2}\w{1}),
\qquad
\vb{\hat{w}}_{1\oplus2} = \mathrm{e}^{\vb{\Omega}/2} {\vb{w}_1+\vb{w}_2\over |\vb{w}_1+\vb{w}_2|}.
\end{equation}

Finally, the formalism was used discuss the conditions under which the relativistic composition of 3-velocities is associative.

The central argument of the chapter was that many key results that are ultimately due to the non-commutativity of non-collinear boosts can be easily rephrased in terms of the non-commutative algebra of quaternions.

\section[Regular black hole with asymptotically Minkowski core]{Regular black hole with asymptotically \\ Minkowski core: Photon sphere and \\ timelike circular orbits}
Chapter \ref{C:Mink-core} analysed the existence of null and timelike circular orbits for the regular black hole with line element 
\begin{equation}
    \dd s^2 
    = -\left(1-\frac{2m\,\e^{-a/r}}{r}\right)\dd{t}^2 + \frac{\dd{r}^2}{1-\frac{2m\,\e^{-a/r}}{r}} + r^2\left(\dd{\theta}^2 + \sin^2\theta \dd{\phi}^2\right).
\end{equation}

It was found that the photon spheres exist a locations described by the implicit equation
\begin{equation}
    r_c^2 = m\,\e^{-a/r_c}(3r_c-a).
\end{equation}
In contrast, extremal timelike circular orbits (ESCOs) can only exist in the region described implicitly by
\begin{equation}
    r_c > a, \qquad {ma-3mr_c+r_c^2\,\e^{a/r_c}}>0.
\end{equation}
The analysis of the photon sphere and ESCO was extended to horizonless compact massive objects, leading to the surprising results that for fixed values of $m$ and $a$, up to two possible photon sphere and up to two possible ESCO locations exist in our model spacetime; and that the very existence of the photon sphere and ESCO depends explicitly on the ratio $a/m$. 
Somewhat unexpectedly, due to the effectively repulsive nature of gravity in the region near the core, it was found that there are some situations in which the photon orbits are stable, and some situations where the ESCOs are outer-most stable circular orbits (OSCOs) rather than the inner-most stable circular orbits (ISCOs).
This provided a rich phenomenology that is significantly more complex than for the Schwarzschild spacetime.

\section[Regular black hole to thin-shell wormhole]{From regular black hole to thin-shell\\ wormhole}

In chapter \ref{C:thin-shell-wormhole}, the same regular black hole analysed in chapter \ref{C:Mink-core} was used to construct a spherically symmetric thin-shell traversable wormhole via the ``cut-and-paste'' method, thereby resulting in yet another black hole mimicker.
The smaller the value of the mass suppression parameter $a$, and the closer the location of the wormhole throat is to the Schwarzschild radius, the better this model mimics a standard Schwarzschild black hole.
The surface energy at the wormhole junction throat was found to be negative, and so, much like other traversable wormholes, exotic matter would be needed to keep the wormhole throat open.

This class of wormholes permits a clean and quite general stability analysis, with wide swathes of stable behaviour. 
Furthermore, crucial to the central theme of this thesis, it was shown the stability plateau exhibits a `pit' of enhanced stability when the wormhole throat is close to where a near-extremal horizon would have existed in the bulk spacetime before applying cut- and-paste surgery. 
Thus, the quantity of exotic matter needed to support the wormhole throat could be minimised (and in some cases made arbitrarily small) by suitable choice of parameters.
Hence, this model of traversable wormhole can be considered a potential astrophysical alternative to classical black holes.

\section{General class of ``quantum deformed''\\ regular black holes}

Chapter \ref{C:QMS} analysed a general class of regular black holes inspired by a (non-regular) metric proposed by Kazakov and Solodukhin in 1993.
The class of spacetimes are described by the line element
\begin{equation}
\d s_n^2 = -\left[\left(1-\frac{a^{2}}{r^{2}}\right)^{n\over2}-\frac{2m}{r}\right] \d t^2 
+ \frac{\d r^2}{\left(1-\frac{a^{2}}{r^{2}}\right)^{n\over2}-\frac{2m}{r}} 
+ r^2\big(\d\theta^2 + \sin^2\theta \,\d\phi^2\big),
\end{equation}
where \( n \) is an odd integer. 
This general class contains spacetimes which have different notions of `regularity', dependent on the single parameter \( n \).
We showed that the spacetimes are metric regular for \( n\geq1 \), Christoffel regular for \( n\geq3 \), and curvature regular for \( n\geq5 \).

The spacetime was found to have an event horizon located implicitly at
\begin{equation}
r_H = 2m \left( 1 -{a^2\over r_H^2}\right)^{-{n\over2}}.
\end{equation}
For small \( a \), we found an approximate but \emph{explicit} horizon location given by
\begin{equation}
r_{H}= 2m \left\{1+ \frac{n a^2}{8m^2}  - {n(3n-2) a^4\over 128 m^4} + \mathcal{O}\left({a^6\over m^6}\right)\right\}.
\end{equation}

Locations of circular null and timelike geodesics (the ``photon sphere'' and ``ISCO'') were calculated.
The locations of these orbits could not be solved analytically \emph{and} explicitly, however in the small \( a \) limit they were calculated to be located at
\begin{equation}
r_c= 3m \left\{1 + \frac{a^2n}{(3m)^2} - {n(11n-6)a^4\over 8(3m)^4} + \mathcal{O}\left({a^6\over m^6}\right) \right\},
\end{equation}
and
\begin{equation}
r_\ISCO= 6m \left\{ 1+ \frac{na^2}{8m^2} -{n(49n-22)a^4\over3456m^4}+ \mathcal{O}\left({a^6\over m^6}\right) \right\},
\end{equation}
for the photon sphere and ISCO, respectively. 

The surface gravity was calculated to be approximately given by
\begin{equation}
\kappa_H = 
 {1\over 4m}\left\{ 1 - {n(n-1)a^4\over32 m^4}  + \O\left(\frac{a^6}{m^6}\right)\right\},
\end{equation}
and the Hawking temperature is simply $k_B T_H = {1\over2\pi} \,\hbar \,\kappa_H$.

The classical energy conditions were analysed, and it was found that the regions in which these conditions are satisfied is dependent on the value of the parameter \( n \).
It was should that for no value of \( n \) (barring the trivial \( n=0 \) Schwarzs\-child spacetime) was there satisfaction of any of the energy conditions globally in the whole of the spacetime. 

In addition to these results, the Regge--Wheeler potential was calculated for spin 0 and spin 1 excitations. 

Somewhat importantly, the spacetimes in this general class could be made arbitrarily close to the Swarz\-schild spacetime, and so are indeed able to be classified as ``black hole mimickers''.

\section{Overall summary and outlook}

It was the aim of this thesis to show that the issue of black hole singularities may be resolved in a tractable,  classical manner by the concept of black hole mimickers. 
Indeed, the results presented in this thesis are far cleaner and simpler than proposed resolutions of singularities via modified theories of gravity or quantum gravity candidate theories. 

There is still a significant amount of open problems on the subject of black hole mimickers.
A small list of problems are listed below as possible avenues of future research. 
\begin{itemize}
    \item Construct a regular black hole spacetime model which violates the classical energy conditions in a `minimal' way.
    \item Construct a regular black hole spacetime model which is charged and rotating, but still allows for tractable analysis of key astrophysical observables (photon spheres, ISCOs, etc.).
    \item Analyse the possibility of regular black holes as explanation for some of the currently unexplainable phenomena in the universe (mass-gap black holes, etc.).
    \item Analyse specific black hole mimicker spacetime models through the lens of quantum field theory in curved spacetimes.
    Specifically, analyse the effect that regularising black hole singularities has on quantum field theoretic phenomena, such as Hawking radiation.
\end{itemize}

In addition to these problems, there are major un-resolved questions in physics which relate directly to black hole mimickers and the study of black hole singularities in general:
\begin{itemize}
    \item Do traversable wormholes really exist, or are they just mathematical curiosities?
    \item Will a consistent theory of quantum gravity regularise the singularities at the core of black holes?
    \item Is there any black hole remnant left over when a black hole Hawking evaporates that could be detected astronomically?
\end{itemize}

Of course, there are many more open problems in this field of research, and this discussion is in no way exhaustive.
However, it is hoped that this thesis has shown the power in the simplicity of black hole mimickers, and that future research will continue to be directed towards their analysis.
I would once again like to thank everyone who has helped me in the completion of this thesis.



\bibliographystyle{acm}
\bibliography{Thesis-master-bib}

\begin{appendices}

\chapter*{\vspace{-3cm} Appendix \\ \vspace{0.5cm} List of publications}\label{C:app-papers}
\addcontentsline{toc}{chapter}{List of publications} 
\chaptermark{List of publications}

\noindent Publications in journals:
\begin{itemize}
    \item \textit{``Thin-shell traversable wormhole crafted from a regular black hole with asymptotically Minkowski core,''} 
    \textbf{T. Berry}, A. Simpson, and M. Visser. \\
    Phys. Rev. D \textbf{102}, {\#6}, 064054 (2020). \\
    DOI:\href{https://doi.org/10.1103/PhysRevD.102.064054}{10.1103/PhysRevD.102.064054}, 
    [arXiv:2008.07046 [gr-qc]].
    \item \textit{``Relativistic combination of non-collinear 3-velocities using quaternions,''} 
    \textbf{T. Berry} and M. Visser. 
    Universe \textbf{6}, \#12, 237 (2020). \\
    DOI:\href{https://doi.org/10.3390/universe6120237}{10.3390/universe6120237},
    [arXiv:2002.10659 [gr-qc]].
    \item \textit{``Photon spheres, ISCOs, and OSCOs: Astrophysical observables for\break regular black holes with asymptotically Minkowski cores,''} \\
    \textbf{T. Berry}, A. Simpson, and M. Visser.
    Universe \textbf{7}, \#1, 2 (2020). \\
    DOI:\href{https://doi.org/10.3390/universe7010002}{10.3390/universe7010002}, 
    [arXiv:2008.13308 [gr-qc]].
    \item \textit{``Darboux diagonalization of the spatial 3-metric in Kerr spacetime,''} \\
    J. Baines, \textbf{T. Berry}, A. Simpson, and M. Visser. \\
    Gen. Rel. Grav. \textbf{53}, \#1, 3 (2021).
    DOI:\href{https://doi.org/10.1007/s10714-020-02765-0}{10.1007/s10714-020-02765-0}, \\
    {[arXiv:2009.01397 [gr-qc]].}
    \item \text{``Unit-lapse versions of the Kerr spacetime,''} \\
    J. Baines, \textbf{T. Berry}, A. Simpson, and M. Visser. \\
    Class. Quant. Grav. \textbf{38}, \#5, 055001 (2021). \\
    DOI:\href{https://doi.org/10.1088/1361-6382/abd071}{10.1088/1361-6382/abd071},
    [arXiv:2008.03817 [gr-qc]].
\end{itemize}
\vfil

\newpage
\noindent Articles under review:
\begin{itemize}
    \item \textit{``Painlev\'e--Gullstrand form of the Lense--Thirring spacetime,''} \\
    J. Baines, \textbf{T. Berry}, A. Simpson, and M. Visser. \\ 
    {[arXiv:2006.14258 [gr-qc]]} (2020).
    \item \textit{``Lorentz boosts and Wigner rotations: self-adjoint complexified\\ quaternions,''}
    \textbf{T. Berry} and M. Visser. \\
    {[arXiv:2101.05971 [gr-qc]]} (2021).
    \item \textit{``General class of `quantum deformed' regular black holes,''} \\
    \textbf{T. Berry}, A. Simpson, and M. Visser. \\
    {[arXiv:2102.02471 [gr-qc]]} (2021).
\end{itemize}

\end{appendices}

\end{document}